\documentclass[final,3p,times,twocolumn]{elsarticle}
\newcommand{\nuebar}{\ensuremath{\overline{\nu}_{e}}}
\newcommand{\g}{\gamma}
\newcommand{\degrees}{$^{\circ}$}
\newcommand{\ptwo}{PROSPECT2}
\newcommand{\NIST}{National Institute of Standards and Technology (NIST) \renewcommand{\NIST}{NIST}}
\newcommand{\NBSR}{National Bureau of Standards Reactor (NBSR) \renewcommand{\NBSR}{NBSR}}
\newcommand{\INL}{Idaho National Laboratory (INL) \renewcommand{\INL}{INL}}
\newcommand{\ATR}{Advanced Test Reactor (ATR) \renewcommand{\ATR}{ATR}}
\newcommand{\ORNL}{Oak Ridge National Laboratory (ORNL) \renewcommand{\ORNL}{ORNL}}
\newcommand{\HFIR}{High Flux Isotope Reactor (HFIR) \renewcommand{\HFIR}{HFIR}}
\newcommand{\LLNL}{Lawrence Livermore National Laboratory (LLNL) \renewcommand{\LLNL}{LLNL}}

 \usepackage{graphicx}
\usepackage{float,amssymb,multirow,dcolumn,amsmath} 
\usepackage{url}
        \usepackage{breakurl} 
        \usepackage[breaklinks]{hyperref}
\newcolumntype{d}[1]{D{.}{.}{#1} }

\begin{document}

\title{Background Radiation Measurements at High Power Research Reactors}

\author[yale]{J.~Ashenfelter}
\author[uwp]{B.~Balantekin}
\author[ornl]{C.~X.~Baldenegro}
\author[yale]{H.~R.~Band}
\author[ornl]{G.~Barclay}
\author[lm]{C.~D.~Bass}
\author[temple]{D.~Berish}
\author[llnl]{N.~S.~Bowden\corref{cor1}}
\ead{nbowden@llnl.gov}
\cortext[cor1]{Corresponding author}
\author[ornl]{C.~D.~Bryan}
\author[uw]{J.~J.~Cherwinka}
\author[ornl,ut]{R.~Chu}
\author[llnl]{T.~Classen}
\author[wm]{D.~Davee}
\author[ornl]{D.~Dean}
\author[ornl]{G.~Deichert}
\author[drexel]{M.~J.~Dolinski}
\author[bnl]{J.~Dolph}
\author[lbnl]{D.~A.~Dwyer}
\author[ornl,ut]{S.~Fan}
\author[drexel]{J.~K.~Gaison}
\author[ornl,ut]{A.~Galindo-Uribarri}
\author[iit]{K.~Gilje}
\author[llnl]{A.~Glenn}
\author[ornl]{M.~Green}
\author[yale]{K.~Han}
\author[bnl_c]{S.~Hans}
\author[yale]{K.~M.~Heeger}
\author[ornl,ut]{B.~Heffron}
\author[bnl]{D.~E.~Jaffe}
\author[bnl]{S.~Kettell}
\author[yale]{T.~J.~Langford}
\author[iit]{B.~R.~Littlejohn}
\author[iit]{D.~Martinez}
\author[wm]{R.~D.~McKeown}
%\author[nist]{M.~P.~Mendenhall}
\author[inl]{S.~Morrell}
\author[ornl]{P.~E.~Mueller}
\author[nist]{H.~P.~Mumm}
\author[temple]{J.~Napolitano}
\author[yale]{D.~Norcini}
\author[waterloo]{D.~Pushin}
\author[ornl,ut]{E.~Romero}
\author[bnl_c]{R.~Rosero}
\author[yale]{L.~Saldana}
\author[llnl]{B.~S.~Seilhan}
\author[bnl]{R.~Sharma}
\author[yale]{N.~T.~Stemen}
\author[iit]{P.~T.~Surukuchi}
\author[inl]{S.~J.~Thompson}
\author[ornl]{R.~L.~Varner}
\author[wm]{W.~Wang}
\author[inl]{S.~M.~Watson}
\author[ornl]{B.~White}
\author[iit]{C.~White}
\author[temple]{J.~Wilhelmi}
\author[ornl]{C.~Williams}
\author[yale]{T.~Wise}
\author[wm]{H.~Yao}
\author[bnl_c]{M.~Yeh}
\author[drexel]{Y.-R.~Yen}
\author[bnl]{C.~Zhang}
\author[iit]{X.~Zhang}
\author[]{\protect\\(The PROSPECT Collaboration)}

\address[bnl_c]{Chemistry Department, Brookhaven National Laboratory, Upton, NY 11973}
\address[bnl]{Physics Department, Brookhaven National Laboratory, Upton, NY 11973}
\address[drexel]{Department of Physics, Drexel University, Philadelphia, PA 19104}
\address[inl]{Nuclear Nonproliferation Division, Idaho National Laboratory, Idaho Falls, ID 83401}
\address[iit]{Department of Physics, Illinois Institute of Technology, Chicago IL 60616}
\address[lbnl]{Physics Division, Lawrence Berkeley National Laboratory, Berkeley, CA 94720}
\address[llnl]{Nuclear and Chemical Sciences Division, Lawrence Livermore National Laboratory, Livermore, CA~94550}
\address[lm]{Department of Chemistry and Physics, Le Moyne College, Syracuse, NY 13214}
\address[nist]{National Institute of Standards and Technology, Gaithersburg, MD 20899}
\address[ornl]{Oak Ridge National Laboratory, Oak Ridge, TN 37831} 
\address[temple]{Department of Physics, Temple University, Philadelphia, PA 19122}
\address[ut]{Department of Physics, University of Tennessee, Knoxville, TN 37996} 
\address[waterloo]{Institute for Quantum Computing and Department of Physics, University of Waterloo, Waterloo, ON N2L 3G1, Canada}
\address[wm]{Department of Physics, College of William and Mary, Williamsburg, VA 23187} 
\address[uwp]{Department of Physics, University of Wisconsin, Madison, WI 53706} 
\address[uw]{Physical Sciences Laboratory, University of Wisconsin, Madison, WI 53706} 
\address[yale]{Wright Laboratory, Department of Physics, Yale University, New Haven, CT 06520} 

\begin{abstract}

 %\linenumbers

Research reactors host a wide range of activities that make use of the intense neutron fluxes generated at these facilities. Recent interest in performing measurements with relatively low event rates, e.g. reactor antineutrino detection, at these facilities necessitates a detailed understanding of  background radiation fields. Both reactor-correlated and naturally occurring background sources are potentially important, even at levels well below those of importance for typical activities. Here we describe a comprehensive series of  background assessments at three high-power research reactors, including $\gamma$-ray, neutron, and muon measurements. For each facility we describe the characteristics and identify the sources of the background fields encountered. The general understanding gained of background production mechanisms and their relationship to facility features will prove valuable for the planning of any sensitive measurement conducted therein.

\end{abstract}

%\pacs{14.60.Lm, 14.60.Pq, 14.60.St,  28.50.Dr, 29.40.Mc}% PACS, the Physics and Astronomy
                             % Classification Scheme.
%\keywords{Research Reactors; Background Measurements; Reactor Antineutrino Detection.}%Use showkeys class option if keyword
                              %display desired
\maketitle

%\newpage
%
% \linenumbers
%
%\newpage

\section{Introduction}
\label{sec:intro}

\begin{table*}[tb]
\begin{center}
\begin{tabular}{l|l|d{4.1}|d{1}|d{1}|d{1}|d{4.1}}
\hline
Location&Reactor& \multicolumn{1}{c|}{Thermal power (MW)}&\multicolumn{1}{c|}{Latitude} & \multicolumn{1}{c|}{Longitude} & \multicolumn{1}{c|}{Altitude (m)} &\multicolumn{1}{c}{Fast Neutron Flux}\\
\hline
NIST&NBSR & $20$&$39.13\degrees$~\text{N} & $77.22\degrees$~\text{W} & 1$05$& $1.0$ \\
ORNL &HFIR& $85$&$35.93\degrees$~\text{N} & $84.31\degrees$~\text{W} & $259$& $1.1$ \\
INL&ATR  & $110$&$43.59\degrees$~\text{N} & $112.96\degrees$~\text{W} & $1435$ & $3.1$ \\
\hline
\end{tabular}
\caption{Facility parameters, including reactor thermal power, geographic location, and predicted fast neutron fluxes relative to NBSR.}
\label{tab:reactors}
\end{center}
\end{table*}%

Research reactors have for decades been important facilities for an enormous variety of activities including, but by no means limited to, isotope production, transmutation, materials and reactor studies, teaching and training, and fundamental physics investigations~\cite{IAEA_RR_report}. More than 250 research reactor facilities are operational or planned in 57 countries~\cite{IAEA_RR_site}. The large neutron flux generated by a controlled fission chain reaction enables such activities. Typically experiments are conducted within or close to the reactor core, or using neutron beams generated through the thermal moderation and collimation of fission neutrons.  With the addition of specialized moderators at cryogenic temperatures, neutron beams with flux rates on the order of 10$^9$~cm$^{-2}$s$^{-1}$ can be produced and efficiently guided tens of meters away from the reactor to both reduce reactor related backgrounds, i.e. high-energy $\gamma$-rays and fast neutrons, and provide more space for experiment deployment~\cite{Nico2005}.

Recently, there has been renewed interest in using research reactors as a source for another product of the fission process: electron antineutrinos (\nuebar{}). On average approximately six \nuebar{} result from each fission reaction in a reactor via the beta decay of neutron rich daughter nuclei. 
The ability to site a  \nuebar{} detector close to a research reactor would enable a sensitive search for additional sterile neutrino states suggested as an explanation for anomalous results in several neutrino oscillation experiments~\cite{Mention:2011rk,VSBL}, the observation of coherent neutrino nuclear scattering (CNNS)~\cite{Hagmann:2004uv,Akimov:2009ht}, and testing the hypothesis of neutrino-induced decay-rate modulation~\cite{deMeijer2011320,deMeijer:2014kxa}. A measurement of the reactor \nuebar{} energy spectrum performed at a research reactor fueled by $^{235}$U would help constrain uncertainties in the prediction of reactor \nuebar{} emissions and may provide additional information on short-lived daughter states that contribute to decay heat uncertainties~\cite{Ashenfelter:2013oaa}. Using measurements of  research reactor \nuebar{} emissions is also of interest for nuclear safeguards and non-proliferation applications, allowing verification of operator declarations~\cite{Bernstein:2001cz,Christensen:2013eza,Christensen:2014pva}.

Conducting a reactor \nuebar{} measurement in a research reactor facility, however, carries a significant challenge. In contrast to typical activities performed at such facilities, the expected signal event rates are low (100s to 1000s of events per day for ton-scale detectors). Therefore strong suppression and an excellent understanding of all background sources is required. To obtain the broadest sensitivity to the possible existence of additional neutrino states~\cite{VSBL} and to maximize the event rate for a \nuebar{} energy spectrum measurement, such an experiment should be placed as close to the reactor core as practical. At this close proximity $\gamma$-rays and neutrons produced by reactor operation can not be neglected, and indeed may be the dominant background source. This is in contrast to most reactor \nuebar{} experiments~(e.g.~\cite{bugey:1996,Bowden:2006hu,An:2012eh}), which are sited 10s to 1000s of meters from the reactor core(s) of interest. Similar considerations apply to searches for CNNS and nuclear decay-rate modulation experiments. 

In preparation for PROSPECT~\cite{Ashenfelter:2013oaa,prospectURL}, we have therefore conducted a comprehensive background radiation survey at three reactor facilities in the US. Our goal is to characterize the background radiation fields generally encountered at research reactor facilities, to understand the sources of those backgrounds, and to develop background mitigation strategies appropriate for low-background experiments generally. While obviously essential for the planning and execution of PROSPECT, we expect this study to provide valuable insight into background sources, intensities, and mitigation strategies for other research reactor facility users. 

The outline of this paper is as follows. In Sec.~\ref{sec:facilities} we describe the research reactor facilities examined in this study, highlighting features that influence the background fields encountered. In Sec.~\ref{sec:Measurements} we describe the instruments used to perform background measurements at each of the selected facilities and the results obtained in Sec.~\ref{sec:results}.  In Sec.~\ref{sec:rxCharacteristics} we use the measurements to illustrate characteristics of reactor-correlated backgrounds at these facilities. Finally, using the understanding of the background radiation fields gained, we describe the steps taken to mitigate reactor-correlated backgrounds in a prototype detector deployment in Sec.~\ref{sec:prospect@hfir}. 

\section{Reactor Facilities}
\label{sec:facilities}

The \NIST{} \cite{NIST}, the \ORNL{} \cite{HFIR}, and the \INL{} \cite{ATR}  operate powerful, highly compact research reactors and have identified potential sites for the deployment of compact $\nuebar{}$ detectors at distances between $4$--$20$~m  from the reactor cores. Important parameters for these facilities are summarized in Table~\ref{tab:reactors}. While designed for a variety of purposes, all three are research reactors with active user programs.  All use  similar Highly Enriched Uranium (HEU) fuel and operate at typical peak thermal powers of $20$~MW, $85$~MW, and $110$~MW respectively.  While having much lower power than typical commercial reactors, the availability of sites at short baselines roughly compensates in terms of available $\nuebar{}$ flux.  Importantly, these facilities operate single cores with  refueling and maintenance outages of significant length, thus allowing precise characterization of natural background during reactor off periods.  Nonetheless, placing extremely sensitive detectors at such short baseline locations requires careful assessment of both natural and reactor generated background radiation.

The potential deployment sites at these facilities include locations as close as practical to the reactor core (``near'', $5$--$10$~m) and at slightly greater separation (``far'', $15$--$20$~m). A wide variety of measurements have been performed at each location, as will be described in later sections. In the following, we describe the general features of these locations with a focus on those relevant to the background measurements performed.   Broadly speaking, backgrounds at the near locations exhibit significant reactor correlations since they are as close to the reactor as practical, while at the far locations that have more intervening shielding and typically greater separation from plant systems there is little or no reactor correlation observed.  

\begin{figure}[tb]
\centering
\includegraphics*[clip=true, trim=20mm 89mm 5mm 70mm,width=0.45\textwidth]{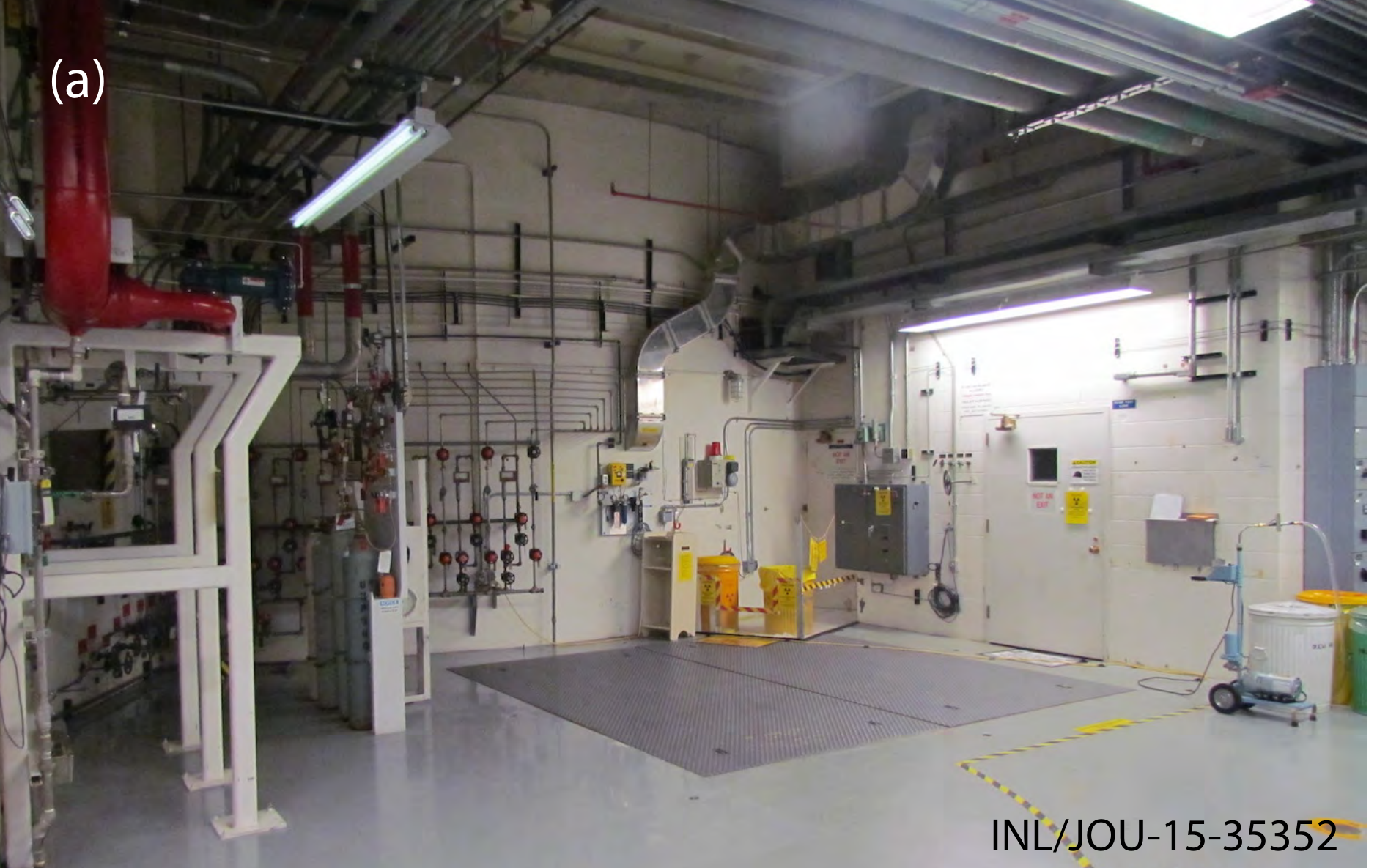}
\includegraphics*[clip=true, trim=15mm 89mm 10mm 60mm,width=0.45\textwidth]{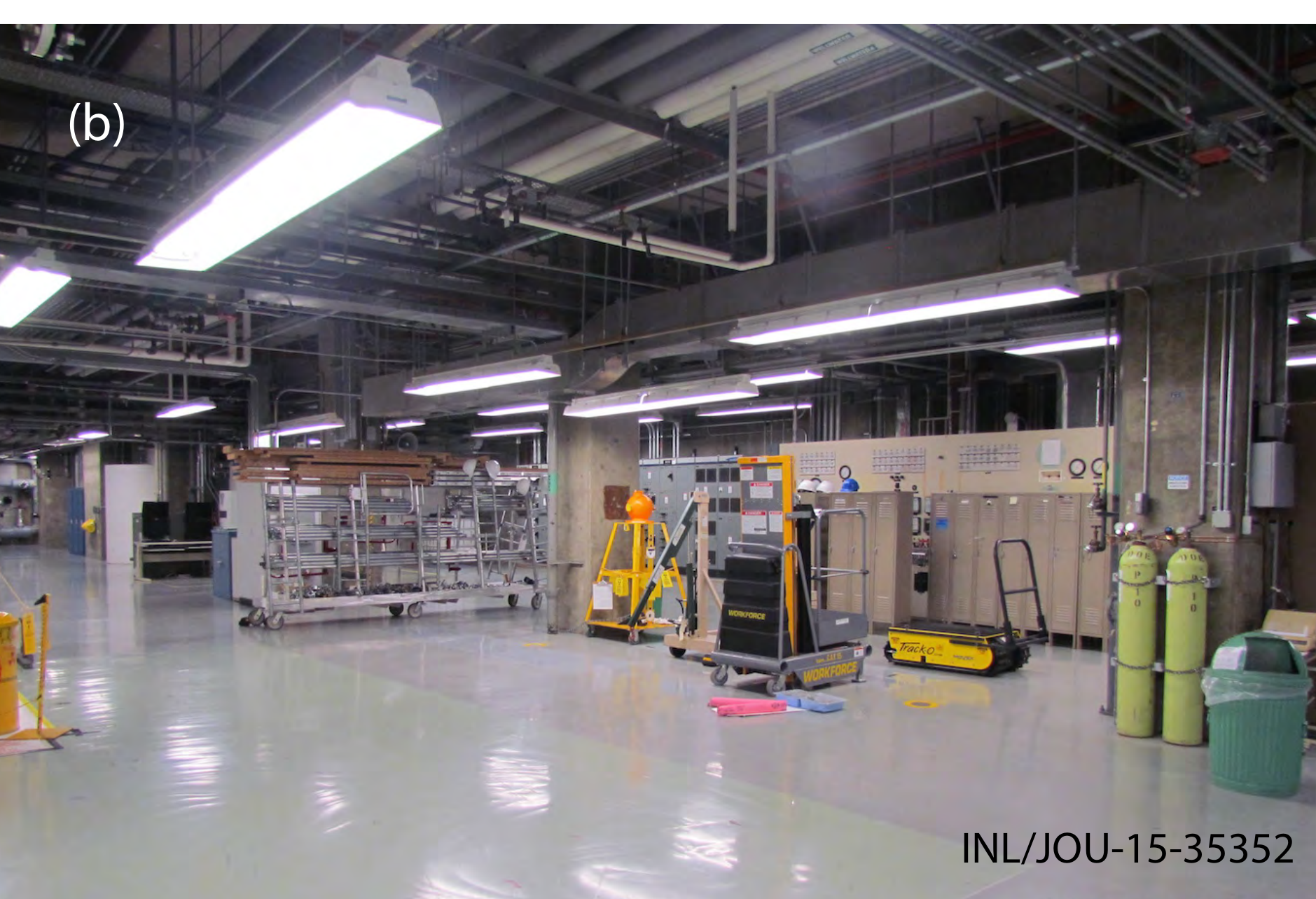}
\caption{Photographs of the near (a) and far (b) locations studied at ATR.}
\label{fig:atrPhotos}
\end{figure}

\begin{figure}[tb]
\centering
\includegraphics*[clip=true, trim=30mm 43mm 55mm 18mm,width=0.45\textwidth]{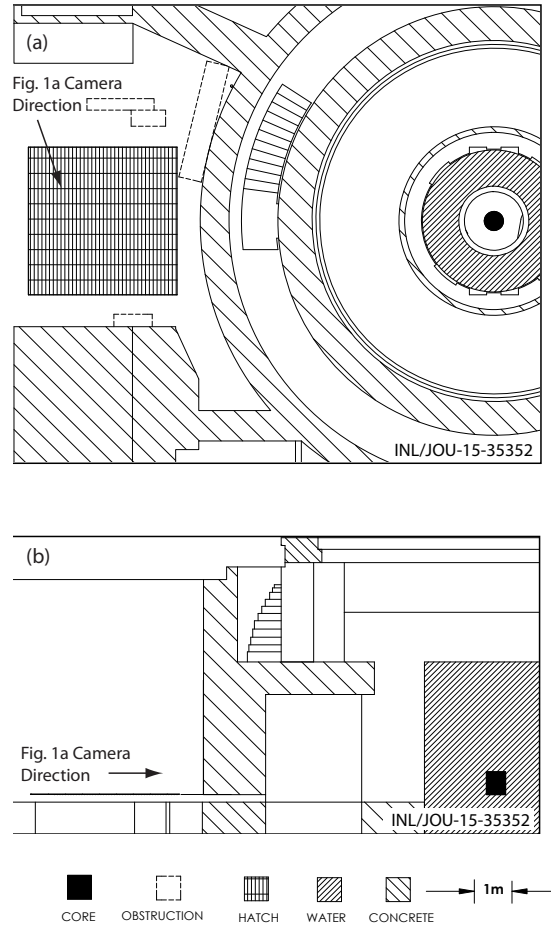}
\caption{(a) Plane and (b) elevation views of the ATR near location.}
\label{fig:atrLayout}
\end{figure}

%\clearpage

At all facilities considered here the thermal neutron flux at the periphery of the reactor vessel varies by less than 10\% over the course of a cycle with the reactor held at nominal power. While this variation may result in background rate variations within the facility, other background generation mechanisms described later (e.g. scattering from beam lines) will also result in time dependent variations.

In addition to variations in background due to site design, variation in cosmogenic background rates are expected due to differences in facility location and elevation (Table~\ref{tab:reactors}). Tools developed for Single Event Upset (SEU) predictions can be used to estimate the relative fast neutron flux at each location, relative to a reference location~\cite{SEUTest}. These estimates predict a minor cosmogenic background difference between the \NBSR{} at \NIST{} and the \HFIR{} at \ORNL{}, while the higher altitude of the \ATR{} at \INL{} leads to a significantly higher cosmogenic neutron flux, absent any shielding or enhancement effects due to the local surroundings or potential overburden. The actual overburden available at these facilities will depend upon the precise location and size of a deployed detector since it will depend in detail upon the facility layout (floor and roof thicknesses, wall locations and thicknesses, etc). The muon measurements presented in Sec.~\ref{sec:muon} provide an indication of the relative cosmogenic flux at each location, accounting for construction details and altitude effects.

\subsection{ATR Locations}

\begin{figure}[tb]
\centering
\includegraphics[clip=true, trim=155mm 115mm 10mm 10mm,width=0.45\textwidth]{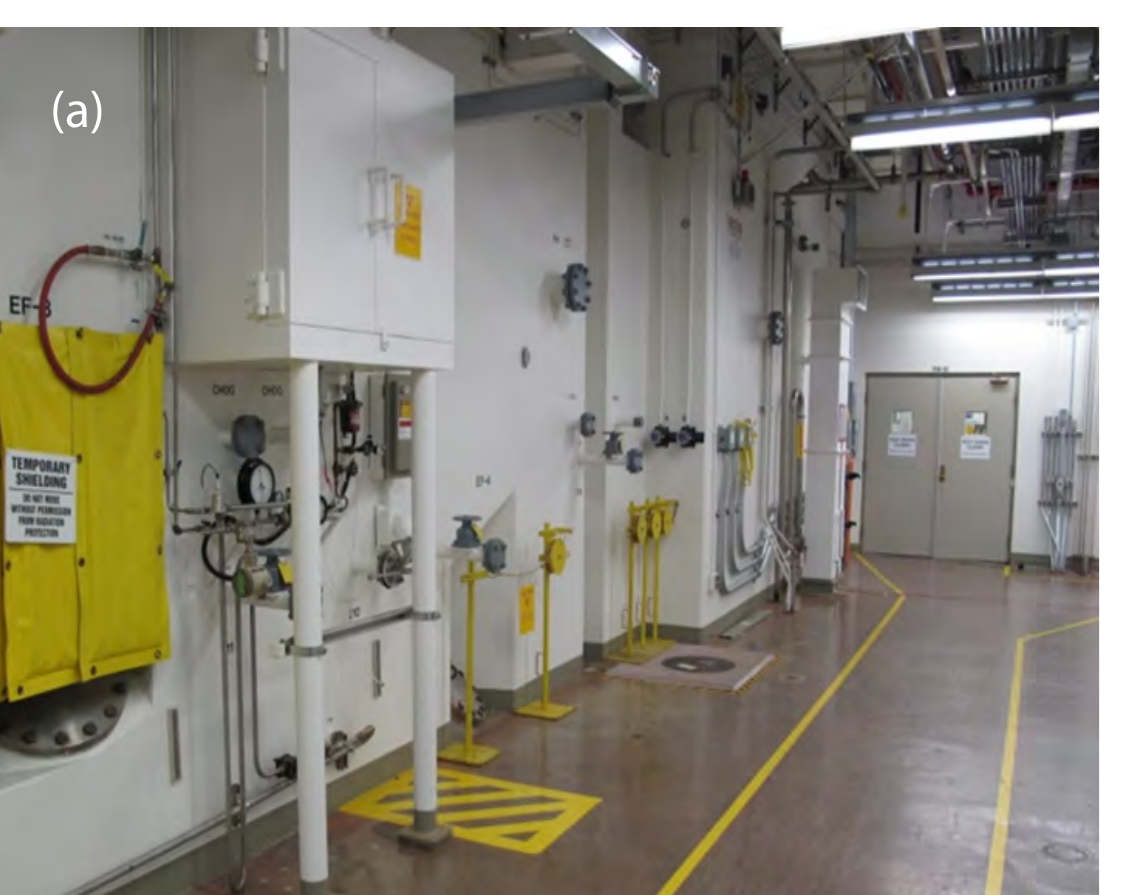}
\includegraphics[clip=true, trim=140mm 20mm 20mm 104mm,width=0.45\textwidth]{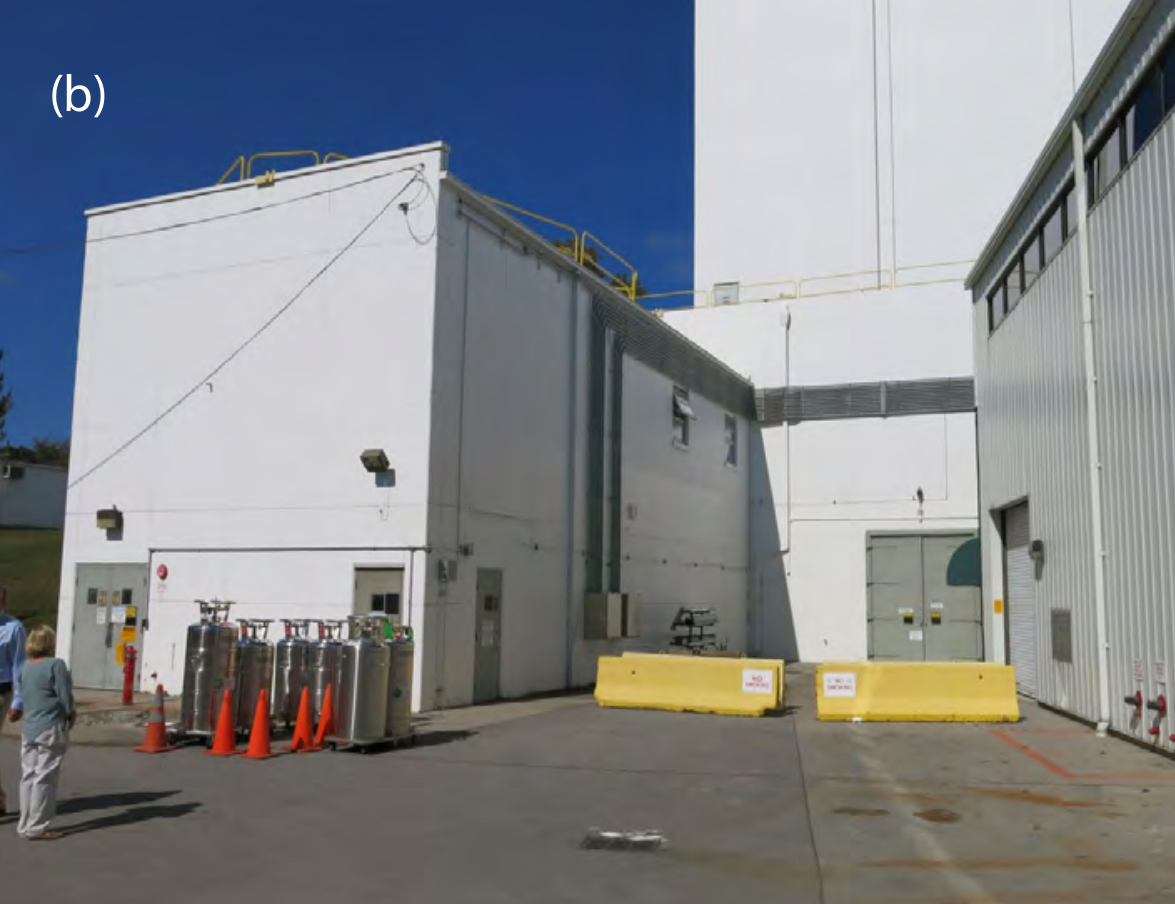}
\caption{ Photographs of the near (a) and far (b) locations studied at HFIR.}
\label{fig:hfirPhotos}
\end{figure}

\begin{figure}[tb!b]
\centering
\includegraphics*[clip=true, trim=40mm 00mm 40mm 10mm,width=0.45\textwidth]{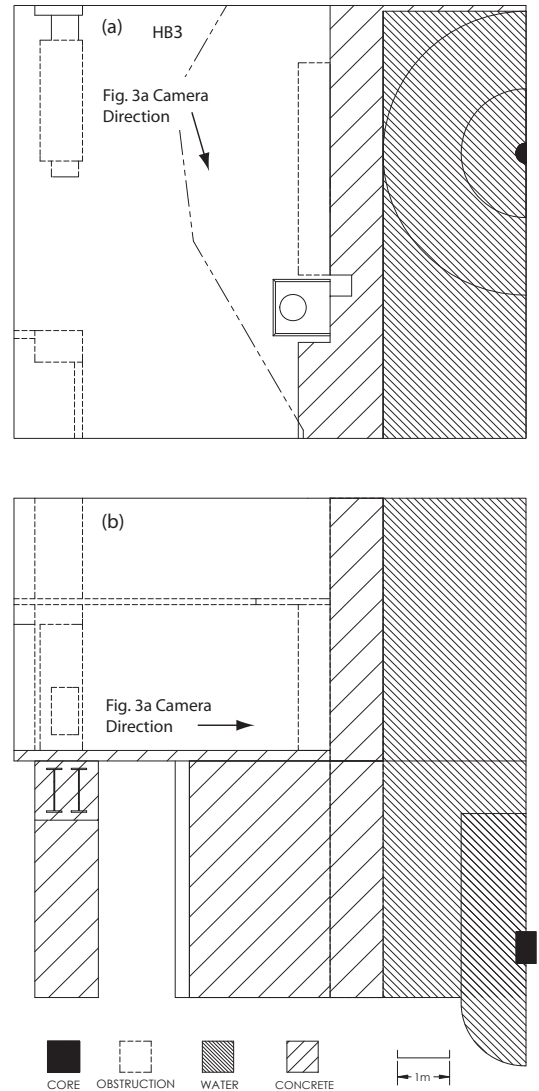}
\caption{(a) Plan and (b) elevation views of the HFIR near location. The location of the HB3 beamline on the floor below is indicated in the top panel.}
\label{fig:hfirLayout}
\end{figure}

ATR is a light water moderated reactor that uses fuel made from a U$_3$O$_8$-Al dispersion clad in and  burnable $^{10}$B poison aluminum and has reactivity control elements composed of beryllium and hafnium. While the data presented here were collected during a typical cycle with a thermal power of $\approx 110$~MW, this parameter and the power distribution within the core can be varied based upon the needs of in-core experiments. Some cycle-to-cycle variation in background can therefore be expected, but given limited available measurement time this possibility is not explored in this work. The  \ATR{} core is sited below-grade and adjacent basement levels contain the possible detector deployment locations. While this below-grade siting provides some cosmogenic attenuating overburden, this is offset by the  higher cosmic ray rate encountered at  $1435$~m elevation of the facility. Cosmogenic background rate measurements were therefore of particular interest at this site. 

Background measurements were performed in the following locations at \ATR{} (Figs.~\ref{fig:atrPhotos}~and~\ref{fig:atrLayout}):
\begin{itemize}
\item \textbf{Near: First sub-basement hatch area}  

Located $\approx6$~m below-grade, this site is directly beneath a large service hatch that provides crane access to the lower levels of the facility. Therefore, despite being below-grade, there is relatively little overburden provided by the facility structure directly above. Several plant systems containing small amounts of reactor primary coolant are located in this area.
\item \textbf{Far: Second sub-basement storage area}  

This below-grade location is a possible far detector location. It is located $\approx12$~m below-grade, and is a relatively open location used for equipment storage and staging. There is one significant plant system that passes through this area: a ceiling mounted pipe that returns a small amount of primary coolant from a $^{16}$N power monitoring system to the main coolant loop.
\end{itemize}

\begin{figure}[tb]
\centering
\includegraphics[width=0.45\textwidth]{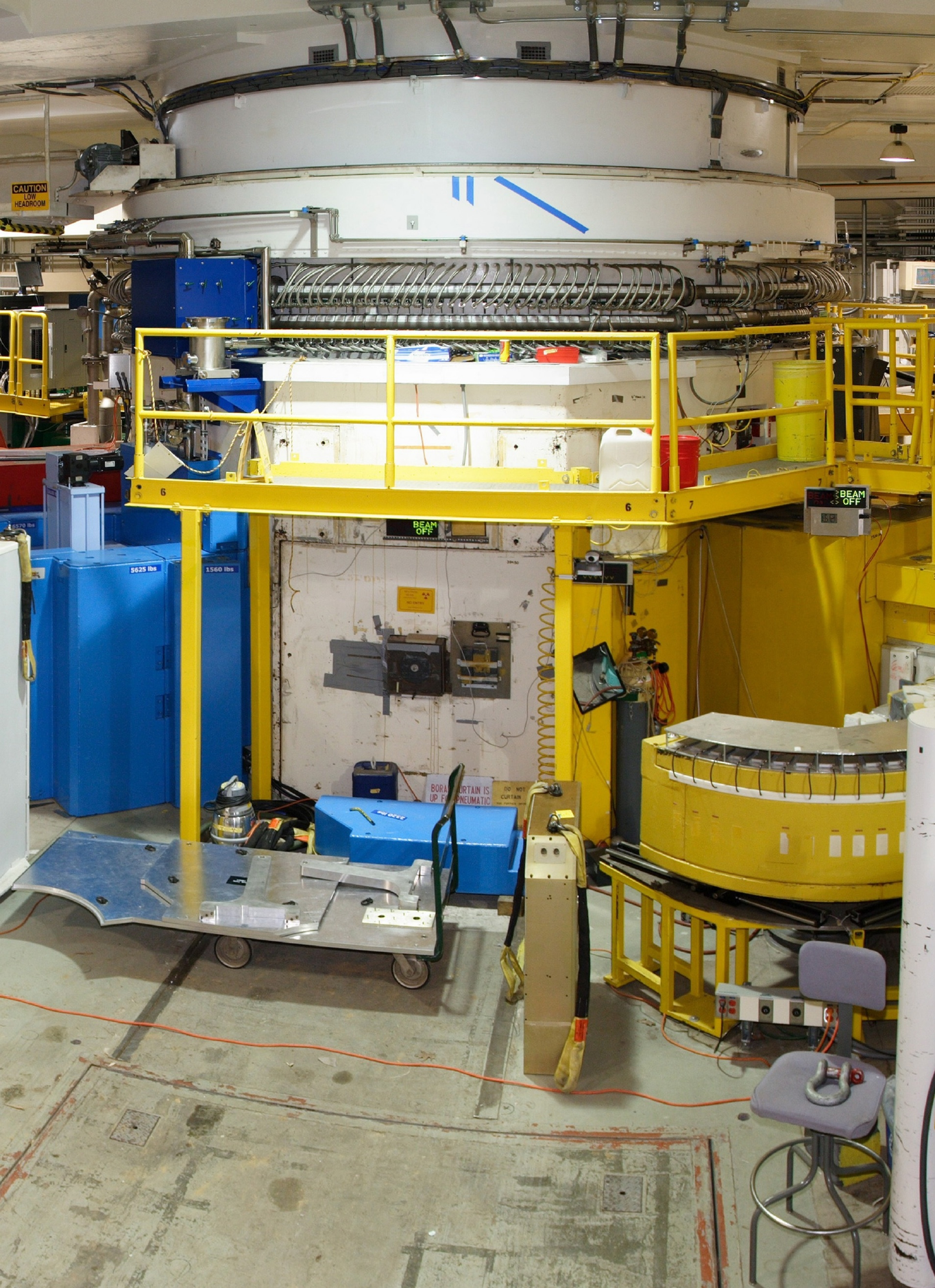}
\caption{Photograph of the near location studied at \NBSR{}.}
\label{fig:nistNearLocation}
\end{figure}

\begin{figure}[tb]
\centering
\includegraphics*[clip=true, trim=50mm 30mm 35mm 32mm,width=0.45\textwidth]{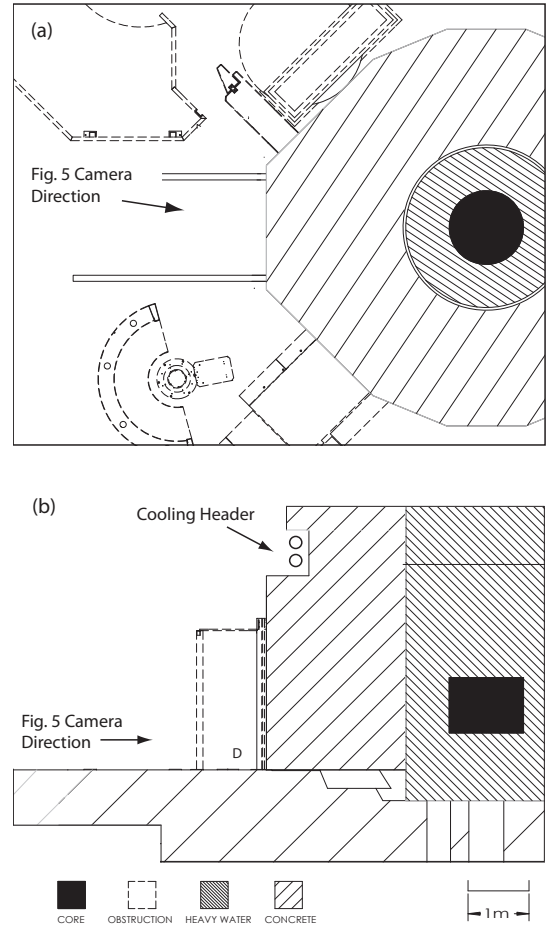}
\caption{(a) Plane and (b) elevation views of the \NBSR{} near location.}
\label{fig:nistLayout}
\end{figure}

\subsection{\HFIR{} Locations}

HFIR is a light water moderated reactor that uses fuel made from a U$_3$O$_8$-Al dispersion and  burnable $^{10}$B poison clad in aluminum. The fuel elements are surrounded by a beryllium neutron reflector. Thin cylindrical control elements containing europium, tantalum, and aluminum are used to maintain a constant thermal power of $85$~MW throughout each reactor cycle. %The thermal neutron flux at the periphery of the reactor vessel increases by approximately 10\% over the course of a reactor cycle, which may result in background rate variations within the facility. 
The \HFIR{} reactor core is located near grade at an elevation of $259$~m.  Background measurements were performed in two locations at \HFIR{} (Figs.~\ref{fig:hfirPhotos}~and~\ref{fig:hfirLayout}):
\begin{itemize}
\item \textbf{Near: Experiment hall}  

The potential location is in a broad corridor on the building level above the core. Large concrete blocks on the level below provide substantial shielding, but the shielding wall at the deployment level is considerably thinner and contains a number of penetrations. In addition, there is a shielded enclosure at this location used intermittently for measurements of activated gas. Measurements were taken at several positions throughout this location.
\item \textbf{Far: Loading area}  

The potential far location is located outdoors on the level below the near location, adjacent to an area in which neutron beam experiments are conducted. Measurements were taken either in the outdoor location or, during inclement weather, in an adjacent steel clad building that supplied little overburden.
\end{itemize}

\subsection{\NBSR{} Locations}
NBSR is a heavy water moderated reactor that uses fuel made from a U$_3$O$_8$-Al dispersion clad in aluminum. The \NBSR{} building is located slightly above grade at an elevation of $106$~m.  Background measurements were performed in two locations at \NBSR{}:
\begin{itemize}
\item \textbf{Near: Thermal column}  

This location is an area within the \NBSR{} confinement building immediately adjacent to the biological shield surrounding the core.   This location was designed to provide high-flux thermal neutron beams (thermal column), but is currently decommissioned.   The moderator and shielding for these sources are still in place however, and result in very low neutron penetration from the core.  Photographs and cross-sections of the thermal column shielding and possible detector location are shown in Figs.~\ref{fig:nistNearLocation}~and~\ref{fig:nistLayout}.  Unique to the \NBSR{} site, this location has  neutron scattering instruments to either side.  These are sources of both thermal neutrons and prompt $\gamma$-rays arising from neutron capture.  Above the thermal column area is a cooling water manifold that is part of the biological shield cooling system.  As described in the sections to follow, this is a source of $\gamma$-rays from $^{16}$N that illuminates roughly half of the potential detector area.  Measurements have been taken under a variety of reactor and adjacent instrument operating conditions.   

\item \textbf{Far: Loading area and high-bay}  

There are two potential far locations at \NBSR{}: outside the confinement building and in a high-bay area adjacent to the confinement building.  Since Health Physics surveys indicated no reactor-correlated background in these locations, representative measurements of naturally occurring radiogenic and cosmogenic background were carried out in a laboratory space nearby.
\end{itemize}

\section{Background Measurement Techniques}
\label{sec:Measurements}

All of the locations studied have low radiation fields from a Health Physics perspective (that are typically of greatest concern to reactor operators and users), i.e. personnel dose rates are low.  For context, an area is typically designated a ``radiation area" and will need controlled access given a 50~$\mu$Sv/h (5~mrem/h) $\gamma$-ray field. At 2.5~MeV this corresponds to roughly $1\times10^3$ cm$^{-2}$s$^{-1}$, which represents a significant flux relative to expected neutrino interaction rates.  In many cases, the background levels of interest in this study are beneath the sensitivity of common radiation survey instruments. Therefore, we have assembled a suite of measurement instruments with greater sensitivity, the ability to provide spectral information, and the ability to separately measure the most important backgrounds for low-background experiments. %: $\g$-rays, fast and thermal neutrons, and cosmogenics (primarily muons).

Where possible, compact portable instruments were selected that could be easily transported between the three facilities, so as to provide a robust relative background comparison. These relative measurements are augmented by higher resolution or higher precision absolute flux measurements using  devices available for use at only one or two locations.  Such measurements proved valuable in determining the sources of particular backgrounds.  The types of measurement performed during this survey were high and low resolution $\g$-ray spectroscopy, fast and thermal neutron flux measurements, muon flux measurements, and fast neutron spectroscopy. Where possible and appropriate, angular and spatial variations of the background fields have been measured. In particular, we have sought to localize $\g$-ray sources corresponding to particular site features like piping or shield wall penetrations, since in principle it is possible to substantially reduce such sources using localized shielding.

\subsection{$\g$-ray Measurements}
\label{sec:gamma}

Interaction of $\g$-rays is likely to dominate the singles rate in a $\nuebar{}$ detector in the locations examined. Typically,  low-background experiments are most concerned with $\g$-rays with energies up to $2.614$~MeV emitted by small amounts of radioisotopes such as $^{60}$Co, $^{40}$K, and $^{208}$Tl found in construction materials. These  $\g$-ray emissions can be effectively controlled through shielding and careful material selection and screening. For operation at a research reactor, there are several important differences. First, the relatively compact spaces available and floor loading limitations at a research reactor facility constrain the shielding that can be used. Second, short-lived radioisotopes with high-energy $\g$-ray emissions can be present in reactor facilities due the higher neutron fluxes present. Of course, at all sites under consideration there is significant shielding incorporated in the facility design that eliminates direct transport of radiation from the reactor core. Nonetheless, there are several mechanisms that can produce elevated $\g$-ray background rates.  These include:
\begin{itemize}
\item \textbf{Local neutron interactions.}  

Fast and thermal neutrons transported through shielding or scattered from beams can interact with material in the local environment. In particular, neutron interactions with water and iron in structural steel can result in high-energy $\g$-ray emissions. 
\item \textbf{Activation product transport in plant piping.}  

Short-lived radioisotopes produced in water exposed to high neutron fluxes near the reactor core can be transported in plant piping to locations close to the measurement locations.
\item \textbf{$\g$-ray transport through shielding.}  

The shielding between a location of interest and a high intensity background source (e.g. pipe carrying activated water) may not attenuate the emitted $\g$-ray flux to levels comparable with natural background. Seams or piping penetrations in shielding walls may allow a scattering path for $\g$-ray (and neutrons) that results in a localized ``hot-spot''.
\end{itemize}

Because each of these sources would be expected to depend on specific features of a site, detailed characterization of the $\g$-ray flux present in each location is necessary. Spectroscopic studies identifying particular radioisotopes present and surveys indicating emission locations can aid in determining the production mechanism.   

A variety of $\g$-ray spectroscopy instruments were used for these measurements. The same easily-transported NaI(Tl) moderate-resolution crystal spectrometer was used at each site, with measurements providing a robust relative comparison of the overall rate and the general shape of the energy spectrum at each location. The NaI(Tl) detector used was a Bicron~\cite{disclaimer} model 2M2 (2" right cylindrical crystal size). A Bridgeport qMorpho~\cite{qmorpho} Data Acquisition (DAQ) system was used in a Multi-Channel Analyzer (MCA) mode to collect measured spectra in the $0$--$12$~MeV electron equivalent energy range.

Higher resolution instruments supplied by the host institutions were used for more complete characterization of  the $\g$-ray fields at each location.  While the differences in efficiency between these instruments precluded direct comparisons between collected spectra, the higher energy resolution allows specific $\g$-ray lines to be identified. 

At \ATR{}, a 2" LaBr$_3$(Ce) scintillation detector  (St. Gobain BrilLanCe~\cite{labr}) was used. This material has good detection efficiency and a very good resolution for an inorganic crystal ($\approx3$\% FWHM at $662$~keV). An Ortec DigiBASE~\cite{digibase} MCA was used for spectra collection.

At \NBSR{}, a Canberra High-Purity Germanium detector (Model CPHA7.5-37200S) was used.  The crystal is a closed-end coaxial geometry of $55$~mm length and $62.5$~mm diameter.  The health-physics group at the \NIST{} Center for Neutron Research (NCNR) carries out regular energy calibrations of the detector and these data are combined with Monte Carlo calculations to determine absolute efficiency as a function of energy.  The photo-peak efficiency at 6 MeV is roughly 0.025\%.  

At \HFIR{}, a standard n-type High-Purity Germanium detector with $15\%$ intrinsic efficiency was used to perform high-resolution background measurements. 

In addition, several centimeters of configurable lead shielding was used at both \HFIR{} and \NBSR{} to make directional measurements of $\g$-ray fields.   As will be discussed in subsequent sections, collimated measurements were useful in locating specific hot spots at these sites.

\begin{figure}[tb!pb]
\begin{center}
\includegraphics[width=.48\textwidth]{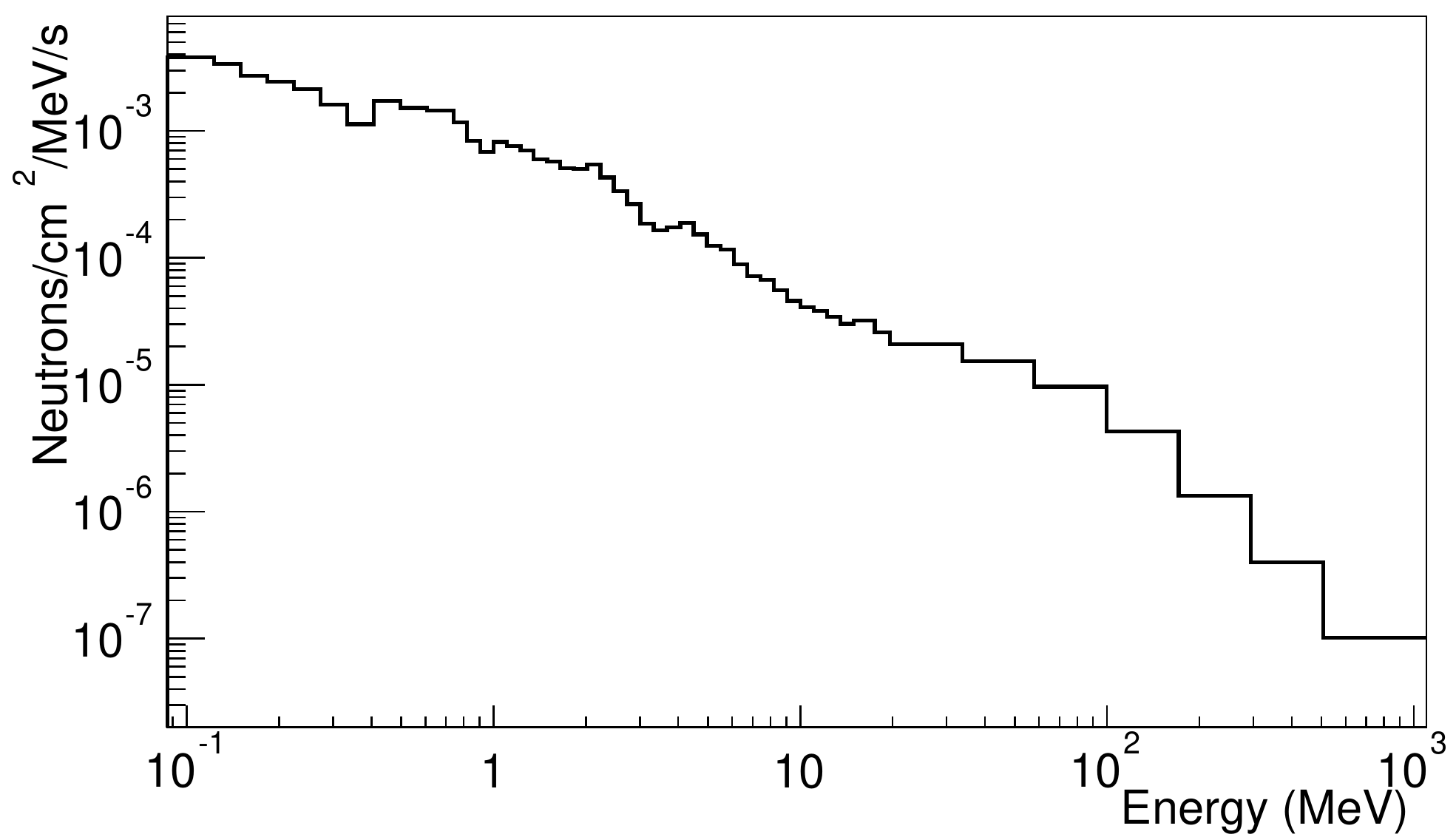}
\caption{The JEDEC standard fast neutron spectrum recorded at sea level in New York~\cite{JESD89a}. }
\label{fig:cosmogenicSpec}
\end{center}
\end{figure}

\subsection{Neutron Measurements}
\label{sec:fn}

Neutron backgrounds at each site were also measured with several instruments. Dose measurements were recorded primarily as a simple method of characterizing thermal neutrons. Particular attention was focussed on fast neutrons as these are an important background for a reactor $\nuebar{}$ measurement that can mimic the signature of  inverse-beta decay (IBD) events. Fast neutron backgrounds at each site predominantly fall into two categories: cosmogenic fast neutron and reactor-related fast neutrons.  

Cosmogenic neutrons can be produced in the atmospheric column above the detector, in structures (e.g. buildings) surrounding the detector, or within the detector itself, especially in high-Z passive shielding materials. These neutrons range in energy from thermal to many GeV with a spectral shape that is reasonably well known. Figure~\ref{fig:cosmogenicSpec} shows the most recent JEDEC standard spectrum for fast neutrons (JESD89a) at sea-level~\cite{JESD89a}, based upon measurements by Gordon {\it et al.} using an array of Bonner spheres~\cite{Gordon2004}.

Work by Kowatari \textit{et al.} has shown that the shape of the cosmogenic neutron spectrum varies little between different locations~\cite{Kowatari2005}. However, the total flux of these neutrons, particularly the thermal part of the spectrum, depends on the local conditions present, including altitude, geomagnetic cutoff, solar activity, weather, and the presence of high-Z material that may cause spallation from high-energy particles in cosmic ray showers including muons and fast neutrons.    

Reactor-correlated fast neutron backgrounds can be expected to have different characteristics. Neutrons produced in the reactor will follow a fission spectrum, with very few neutrons expected at energies $>$10~MeV. Since the reactors are surrounded by moderating material and, of course, no line of sight exists between active fuel elements and the locations under consideration, any reactor-correlated neutron can be expected to have undergone multiple scattering interactions and therefore to have degraded energy. This fast neutron source is of less concern with regard to \nuebar{}-mimicking correlated background but could still be a significant source of singles background from neutron capture in an antineutrino detector. Similar to the preceding discussion of localized $\g$-ray sources, reactor-correlated neutron background can be expected to correspond to penetrations, beam instruments, or other shielding leakage paths. 

\begin{figure}[tb]
\centering
\includegraphics[width=0.48\textwidth]{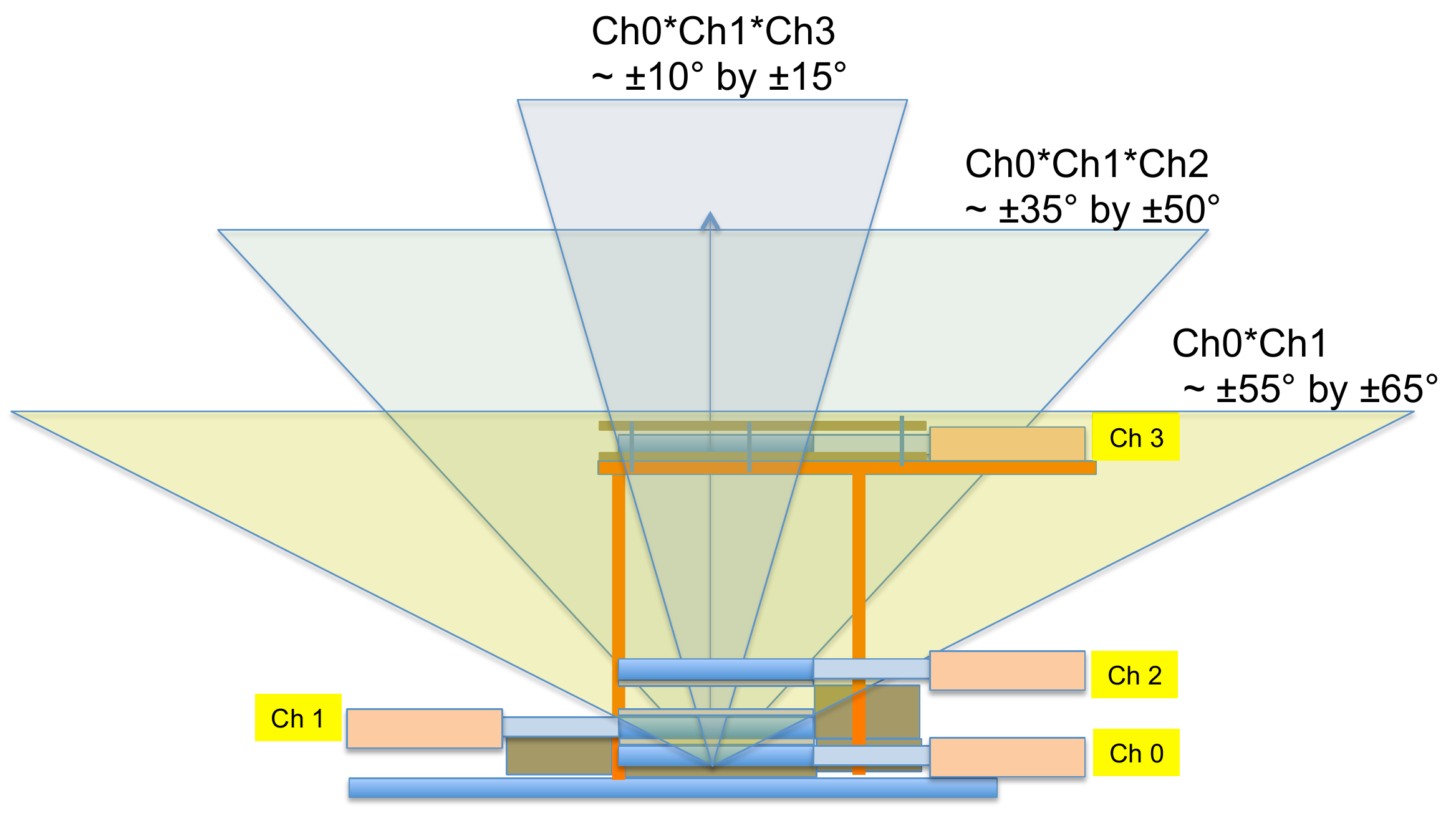}
\caption{The angular acceptances for the muon telescope instrument used at all sites is determined  by the coincidence requirement enforced between the 4 plastic scintillator paddles.
}
\label{fig:telescope}
\end{figure}

\subsubsection{Fast Neutron Measurements}

As with the $\g$-ray background measurements, two classes of instruments were used to assess the fast neutron backgrounds at the potential reactor sites. First, a well characterized fast neutron spectrometer, FaNS-1, was used at \NBSR{} and \HFIR{} to validate the assumption that the higher-energy ($> $1 MeV) portion of the cosmogenic  neutron energy spectrum varies little between locations and to measure the absolute flux at those sites. This device was difficult to transport, so it was not used at \ATR{}. In addition, measurements were taken at the three facilities using a small portable fast neutron recoil counter. Due to its small size and operating method, this device could not readily provide absolute flux and spectral information but, analogous to the use of the NaI(Tl) $\g$-ray detector, it provides a robust basis for a relative site comparison. 

The FaNS-1 spectrometer is a fast neutron detector consisting of segments of BC-400 plastic scintillator with $^3$He proportional counters positioned between~\cite{Langford201578, Langford2013}. 
Each of the six optically-decoupled plastic scintillator segments are 9~cm~$\times$~18.5~cm$~\times$~15~cm, for a total active volume of 15~liters.  Light from each segment is collected by pairs of Photo-Multiplier Tubes (PMTs) attached to cylindrical light-guides.   
The signal from each PMT passes through an asymmetric splitter circuit that produces two signals, a delayed full-amplitude signal and one attenuated by a factor of nine.  Each pulse pair was waveform digitized.  This approach allows for the construction of a linear response over a large dynamic range.  The six $1$" diameter $^3$He proportional counters are filled with 4.0 bar $^3$He and 1.1 bar of natural krypton and have high thermal neutron capture efficiency.  All six helium-counter signals were combined in one fan in/out module and digitized.

FaNS-1 operates via the concept of capture-gated spectroscopy. A fast neutron enters the detector, where it thermalizes through multiple (n,p) scatters. After thermalizing, it randomly walks until it is captured by a $^3$He counter or leaves the volume. Thus the signature of a fast neutron is a scintillator signal followed by a delayed neutron capture. The neutron energy is determined by the quantity of light detected in the PMTs. Segmentation reduces the effect of non-linear light yield and improves the energy resolution of the spectrometer.  The energy response of the device was determined via irradiation with well calibrated $^{252}$Cf, $2.5~$MeV  and $14~$MeV generator sources and detailed MCNP models of detector response.  By examining the time separation between a scatter-like event and a capture-like event, it is possible to differentiate accidental coincidences, which are uniform in time, and correlated coincidences, which have a distinct exponential distribution. 
The rate of accidental coincidences is driven by the product of the $^3$He trigger rate and the scintillator signal rate. If either of these sustains a substantial increase from non-fast neutron interactions, e.g. $\g$-ray interactions in the scintillator or thermal neutron captures in the $^3$He tubes, it will degrade the ability for FaNS-1 to determine the fast neutron rate.   Backgrounds such as these limited the measurements that could be carried out at the \NBSR{} near location.

Additionally, a small stilbene detector was taken to all three reactor sites. This device comprised a $2$" trans-stilbene crystal, a $2$" PMT, and a Bridgeport eMorph DAQ system packaged in a small aluminum tube. 
Power and readout were supplied via a USB connection to a laptop computer. Trans-stilbene is an organic crystal with good Pulse Shape Discrimination (PSD) properties that allow for the identification of fast neutron recoil events. The crystal used for these measurements was grown in a materials development laboratory at the \LLNL{}~\cite{Zaitseva20158}.

\subsubsection{Neutron Dose and Thermal Measurements}
\label{sec:thermalNeutron}

Neutron dose measurements were performed at \NBSR{} and \HFIR{} using neutron survey instruments. 
The  detector used at \NBSR{} consisted of a $23~$cm diameter cadmium-loaded polyethylene sphere surrounding a BF$_3$ tube.   The cadmium loading is designed to create a neutron response such that the instruments directly reads  dose rates in units of rem/hr ($\approx 2.7 \times 10^5$ cm$^{-2}$s$^{-1}$ for thermal energies~\cite{nrc}).   A thermal spectrum was assumed in the conversion to flux.  The \HFIR{} detector was similar, except that it was calibrated to read in s$^{-1}$ and an absolute efficiency was provided by the instrument manufacturer.  Since the two  similar instruments were primarily used for assessing spatial variations, no attempt was made to perform relative or absolute response calibrations.  Calibrated bare  BF$_3$ tubes were also used to measure the approximate thermal neutron flux at several locations.

\subsection{Muon Measurements}
\label{sec:muon}

Cosmic ray measurements are important as they indicate the amount of overburden provided by reactor buildings and other structures at these shallow sites. The cosmic muon rate and angular dependence was measured using a telescope detector comprised of 4 scintillation paddles spaced at varying distances vertically as shown in Fig.~\ref{fig:telescope}. Requiring coincidences between the paddles is equivalent to restricting the muon angular acceptance of the telescope. Up to three angular ranges (approximately $\pm10^{\circ} \times  \pm15^{\circ}$, $\pm35^{\circ} \times  \pm50^{\circ}$, and $\pm55^{\circ} \times  \pm65^{\circ}$) can be measured at the same time.

The $25$~cm $\times$ $15$~cm $\times$ $2.5$~cm scintillator paddles were constructed from Eljen EJ-200 plastic scintillator. Each paddle was connected to a 5~cm ADIT B51D01 PMT by a trapezoidal shaped acrylic light guide. 
A Bridgeport Instruments hvBase-P-B14D10 provided the high voltage and served as the voltage divider for the 10-dynode chain. Data acquisition was provided by a Bridgeport Instruments qMorpho-2010 ADC. The DAQ had the capability to record  individual waveforms and could individually control the gain settings of each PMT channel.  The qMorpho contains four 20 MHz multichannel analyzers with 10 bit resolution and was controlled through a USB interface. 

This simple telescope  cannot discriminate between particle type as it can only identify particles creating time coincident hits in 
the separated scintillator paddles. Cosmic rays near sea level are an admixture of muons, hadrons, electrons, photons and neutrons. According to the Review of Particle Properties (Particle Data Group) the integral rate of muons $\ge 1$ GeV/c is  $\approx60$--$70~\rm{m^{-2} sr^{-1}s^{-1}} $ and follows a roughly $\rm{cos}^2\theta$ angular distribution~\cite{PDG}. The number of electrons and positrons is very energy dependent with rates of $30~\rm{m^{-2} sr^{-1}s^{-1}} $ above 10~MeV,  $6~\rm{m^{-2} sr^{-1}s^{-1}}$ above 100 MeV, and $ 0.2~\rm{m^{-2} sr^{-1}s^{-1}}$ above 1 GeV.  Protons and neutrons $ \ge 1$ GeV/c add $\approx 0.9~\rm{m^{-2} sr^{-1}s^{-1}}$. Even the small amount of material in the roof over a typical laboratory space can reduce the observed rate by a few percent consistent with low energy electron fraction quoted above. Adding 3 mm of lead between paddles reduces the overall coincidence rate in the lab by roughly 5\%. The roofs over the confinement buildings at both the \NBSR{} and \HFIR{} are $\approx0.5$~m concrete and reduced the coincident rate by $\approx17$--$19$\%. Our data show that  the scintillator energy spectra are consistent with minimum-ionizing tracks and follow a roughly $\rm{cos}^2\theta$  angular distribution as expected from muons.  Therefore we will assume that the rates measured with the telescope near the reactor are due to muons with a small  $\le5$\% contribution from other particles.

\begin{table*}[tb]
\begin{center}
\begin{tabular}{c | c | c | d{4.1} | d{1}}
    Isotope    & Reaction    & Source Material    &\multicolumn{1}{c|}{ Energy (keV)} & \multicolumn{1}{c}{$t_{1/2}$}\\ \hline
    $^{187}$W & $^{186}$W(n,$\gamma$)$^{187}$W    &Unknown& 479.5       &  23.9\text{h}\\ 
     - & annh.             && 511.0     &  \\
    $^{82}$Br &    &Fission Product& 554.3       & 35.3\text{h}\\
    $^{208}$Tl &             &Structural Material& 583.2    &   \\
    $^{214}$Bi &    &Radon& 609.3       &  \\%19.7\text{m}\\
    $^{82}$Br &    &Fission Product& 619.0       &  35.3\text{h}\\
    $^{137}$Cs &    &Fission Product& 661.6     &    \\%30y\\
    $^{187}$W & $^{186}$W(n,$\gamma$)$^{187}$W   & Unknown & 685.8      & 23.9\text{h}  \\
    $^{82}$Br &  & Fission Product& 776.5      &  35.3\text{h} \\
    $^{27}$Mg & $^{27}$Al(n,p)$^{27}$Mg   &Fuel Cladding, Structural Material& 843.8      &  9.5\text{m} \\
    $^{27}$Mg & $^{27}$Al(n,p)$^{27}$Mg  &Fuel Cladding, Structural Material& 1014.5     &  9.5\text{m} \\
    $^{60}$Co & $^{59}$Co(n,$\gamma$)$^{60}$Co  &Stainless Steel& 1173.2       & \\%5.3y \\
    $^{41}$Ar & $^{40}$Ar(n,$\gamma$)$^{41}$Ar   & Air & 1293.6       &  1.8\text{h}\\
    $^{60}$Co & $^{59}$Co(n,$\gamma$)$^{60}$Co   &Stainless Steel& 1333.2       &  \\%5.3y\\
    $^{24}$Na & $^{27}$Al(n,$\alpha$)$^{24}$Na   &Fuel Cladding, Structural Material& 1368.6       & 15.0\text{h} \\
    $^{40}$K &              &Structural Material& 1460.9      &  \\%$1.3 \times 10^9$y \\ 
    $^{214}$Bi &    &Radon& 1764.5       &  \\%19.7\text{m}\\
    $^{2}$H & $^{1}$H(n,$\gamma$)$^{2}$H &Water, HDPE& 2223.2  &   \\
    $^{55}$Fe & $^{54}$Fe(n,$\gamma$)$^{55}$Fe &Steel& 2469.9  & \text{fs}   \\
    $^{208}$Tl &             &Structural Material& 2614.5     &  \\
    $^{24}$Na & $^{27}$Al(n,$\alpha$)$^{24}$Na   &Fuel Cladding, Structural Material& 2754.0      & 15.0\text{h}\\
   unkn. &&&5297.0&\\ 
    $^{55}$Fe & $^{54}$Fe(n,$\gamma$)$^{55}$Fe &Steel& 5507.5    &\text{fs}\\
    $^{16}$N & $^{16}$O(n,p)$^{16}$N &Water& 6128.6     &  7.2\text{s}\\
    $^{57}$Fe & $^{56}$Fe(n,$\gamma$)$^{57}$Fe &Steel& 6318.8    & \text{fs}  \\
    $^{16}$N & $^{16}$O(n,p)$^{16}$N &Water& 7115.2    &   7.2\text{s}\\
    $^{57}$Fe & $^{56}$Fe(n,$\gamma$)$^{57}$Fe &Steel& 7631.1    &  \text{fs} \\
    $^{57}$Fe & $^{56}$Fe(n,$\gamma$)$^{57}$Fe &Steel& 7645.5    &  \text{fs} \\
    $^{28}$Al & $^{27}$Al(n,$\gamma$)$^{28}$Al &Fuel Cladding, Structural Material& 7724.0   &   2.2\text{m}\\
    $^{16}$N & $^{16}$O(n,p)$^{16}$N &Water& 8869.0    &   7.2\text{s}\\
    $^{55}$Fe & $^{54}$Fe(n,$\gamma$)$^{55}$Fe &Steel& 8886.4    &  \text{fs} \\
    $^{55}$Fe & $^{54}$Fe(n,$\gamma$)$^{55}$Fe &Steel& 9297.8    & \text{fs}  \\
    \hline
   \end{tabular}
    \caption{Radionuclides identified as contributing to the $\g$-ray background at \NBSR{} and \ATR{}. Listed are likely production reactions, source materials, $\g$-ray energy, and half-life (for short-lived reactor-correlated products).  Many of these $\g$-rays should be expected at similar facilities, though relative line strengths could vary considerably.}
\label{tab:isotopes}
\end{center}
\end{table*}

\clearpage

\section{Background Measurement Results}
\label{sec:results}

In this section we describe the results obtained from measurement campaigns at each reactor site.

\subsection{High Resolution $\g$-ray Spectroscopy Results}
\label{sec:gammaResults}

High energy resolution $\g$-ray spectra were acquired at most of the locations of interest. At \HFIR{} measurements were not possible in the outdoor far location. Data taken at the HFIR near location with the reactor off are thought to provide a reasonable representation of what would be encountered there.  Similarly, no high energy resolution data was taken at either far site at \NBSR{}.  Again, reactor-off data taken inside the confinement building should provide a reasonable approximation of likely backgrounds.

\begin{figure}[tb]
\centering
\includegraphics*[width=0.48\textwidth]{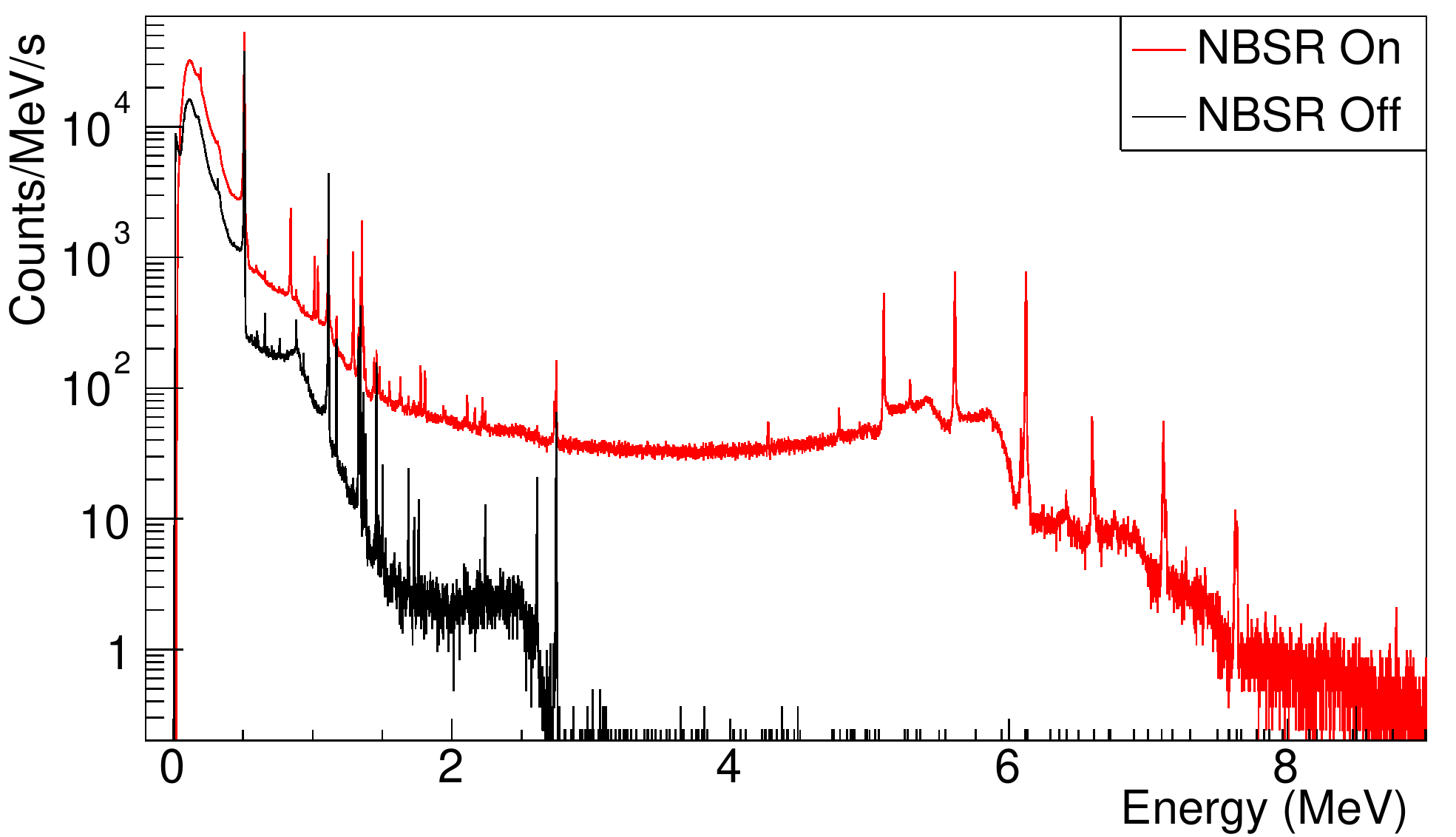}
\caption{Example HPGe $\g$-ray spectra taken with the NBSR on and off. Prominent lines, and associated escape peaks and Compton continua, are evident. The line sources are identified in Table~\ref{tab:isotopes}.
}
\label{fig:nistRawSpec}
\end{figure}

Typical raw HPGe spectra acquired at the \NBSR{} near location with both the reactor-on and the reactor-off are shown in Fig.~\ref{fig:nistRawSpec}. Clearly visible in the reactor-on spectrum are the prompt $\gamma$-ray lines associated with $^{16}$O(n,p)$^{16}$N (6128.6 keV) and $^{56}$Fe(n,$\gamma$)$^{57}$Fe (7631.1 keV and 7645.5 keV) as well as the associated single and double 511 keV escape peaks and Compton edges.  Much of the apparent continuum is a function of detector response and thus a full understanding of the source spectrum would require a full deconvolution.  However, it is evident that the high energy resolution of the HPGe instruments allows for identification of prominent lines contributing to the spectrum. It is also clear that there is a substantial increase in  $\g$-ray background when the reactor is operating, particularly at higher energies. Many short-lived isotopes, some with high energies in the range $6$--$9$~MeV are produced by a variety of mechanisms.  We have identified the isotopes that make the largest contributions to  $\g$-ray background at \NBSR{} and \ATR{} in Table~\ref{tab:isotopes}~\cite{Molnar2004,Chadwick20112887}.

\begin{figure}[tb]
\centering
\includegraphics*[width=0.48\textwidth]{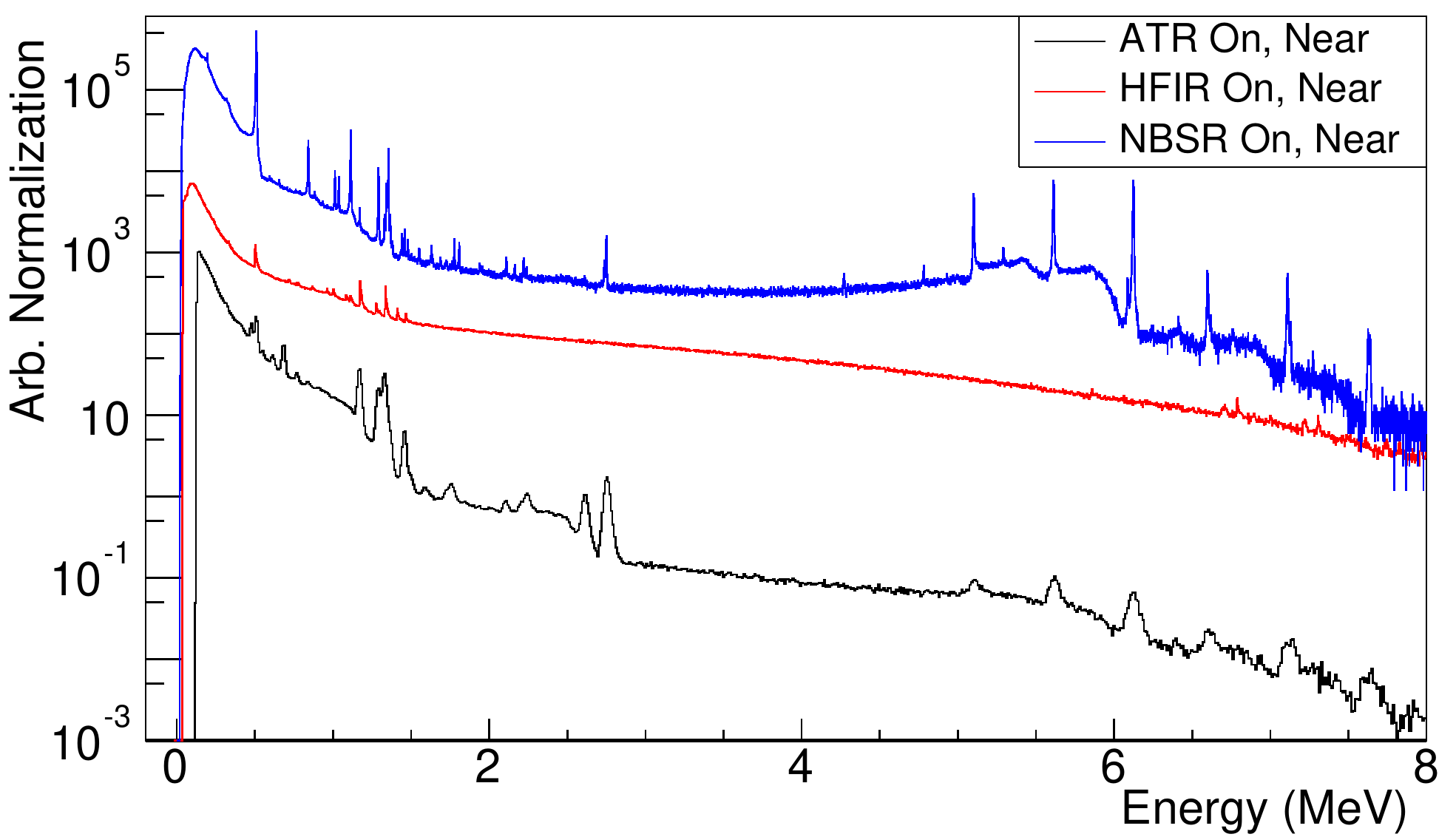}
\caption{Comparison of the high resolution $\g$-ray spectra measured at the near locations using different HPGe detectors at \HFIR{} and \NBSR{} and a LaBr$_3$ detector at \ATR{}. Spectra are collected with the reactors operating at nominal power. Note that spectra are offset in normalization for clarity and the detectors have different response functions, therefore this comparison only illustrates the general features of the $\g$-ray backgrounds in these locations.}
\label{fig:highResGammaComparison}
\end{figure}

Considering Table~\ref{tab:isotopes} more fully, the short half-life of many of the observed isotopes is notable. Those  with half-lives measured in seconds or less are likely produced by neutron interactions in the immediate vicinity of the measurement location. For example, neutron leakage fields can interact with Fe in structural steel components giving rise to the observed $^{55}$Fe and $^{57}$Fe $\g$-ray lines. Similarly, neutron interactions with water or HDPE shielding give rise to $^{16}$N and $^{2}$H emissions. Isotopes with half-lives measured in minutes-to-hours can also be produced in this way, but can additionally be produced in shielded regions with high neutron flux, e.g. in primary or secondary cooling loops, and then transported in plant piping to the measurement locations. One prominent example is $^{24}$Na which can be produced from trace amounts of dissolved $^{27}$Al in cooling water when it is exposed to the large neutron flux at or near the reactor core.  Note that all of these isotopes will have decayed substantially within $\approx1$~day of reactor shutdown. 

The high resolution  $\g$-ray spectra acquired at the three near locations are compared in Fig.~\ref{fig:highResGammaComparison}. Note that since each detector has different efficiency, it is not possible to draw conclusions about the relative intensity of the $\g$-ray flux at each location from this comparison. Instead, we can discern important differences relating to the background sources at the sites. The most obvious difference is the relatively featureless spectrum observed at \HFIR{}. We interpret this to mean that the primary source of $\g$-ray background at the HFIR near location is $\g$-rays that are down-scattered as they propagate from intense radiation sources through shielding material. Given that the broad continuum observed extends to high energies, neutron interactions on steel and water are the likely source for the majority of these $\g$-rays. As will be discussed in Sec.~\ref{sec:rxCharacteristics}, the emission of the down scattered $\g$-ray continuum is strongly correlated with the wall closest to the reactor, and there is evidence of a neutron capture $\g$-ray source within the \HFIR{} near location as well.

%\clearpage

The $\g$-ray background at both \ATR{} and \NBSR{} shows clear line structure implying that a significant fraction of that background is due to decays that occur locally with little intervening shielding. The high-energy features can be attributed to neutron interactions on steel and water, shedding light on local neutron backgrounds. As demonstrated by the spectral unfolding process to be described in Sec.~\ref{sec:unfold}, there is also a down-scattered continuum in these locations which can be attributed to locally produced $\g$-rays or the sort of incomplete shielding observed at \HFIR{}. 

\begin{figure}[tb]
\centering
\includegraphics*[width=0.48\textwidth]{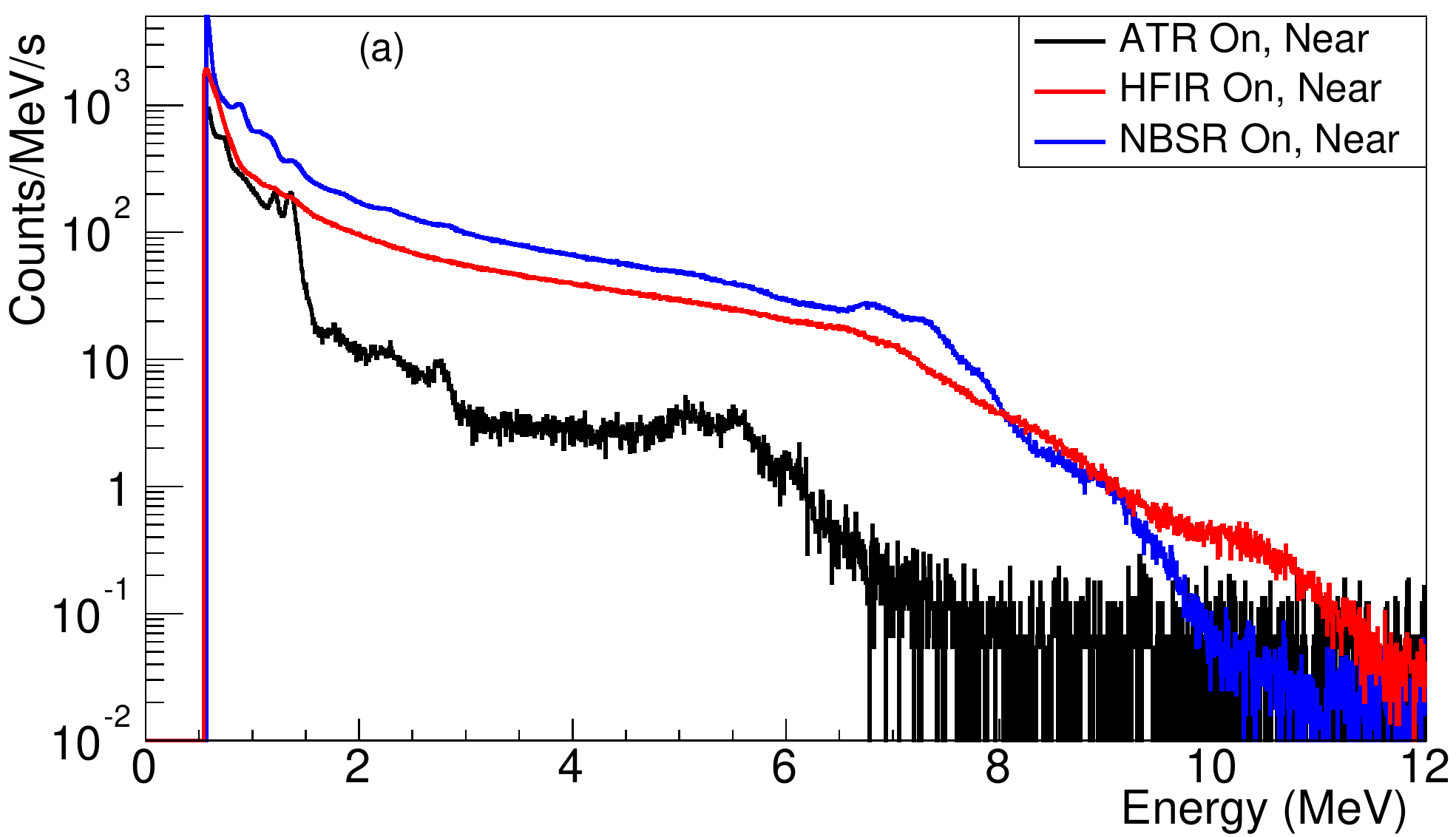}
\includegraphics*[width=0.48\textwidth]{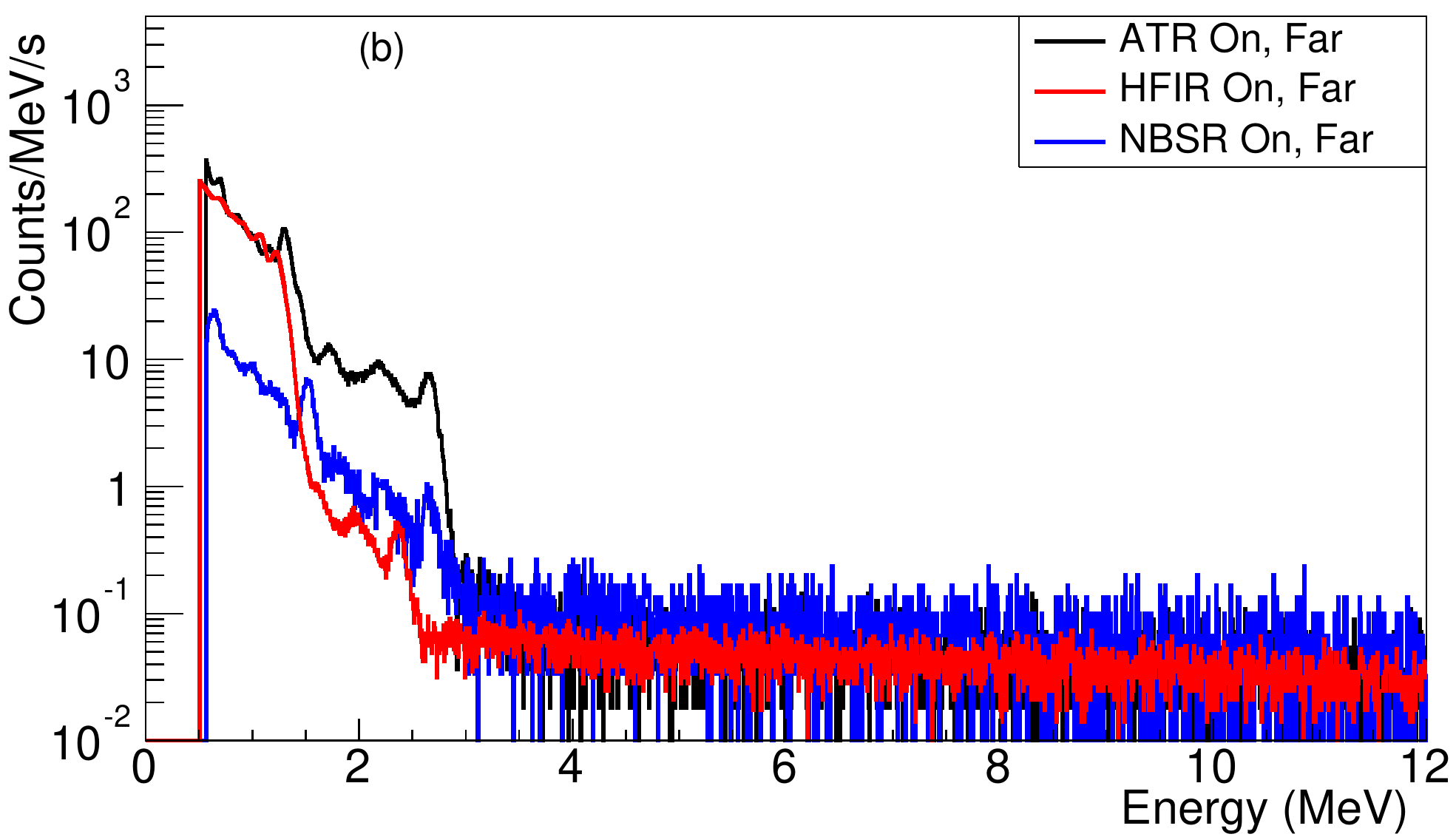}
\caption{Comparison of NaI(Tl) spectra acquired at each reactor site, in both the (a) near and (b) far locations. Spectra are collected with the reactors operating at nominal power. The near location data are averaged over several measurement positions representing the extent of the available deployment footprint to account for the effects of positional variations.}
\label{fig:NaI_Comparison}
\end{figure}

%\clearpage

\subsection{Moderate Resolution $\g$-ray Spectroscopy Results}

As described previously, a NaI(Tl) inorganic crystal spectrometer was used to obtain data for a relative comparison of the reactor sites. This device was calibrated using $^{60}$Co sources. Spectra representative of the near and far detector locations at each site are compared in Fig.~\ref{fig:NaI_Comparison}. Significant differences exist between the near and far locations, and among the sites themselves. The  spectral shapes observed at the near locations are consistent with those observed in the high resolution measurements (Fig~\ref{fig:highResGammaComparison}). Absolute $\g$-ray fluxes estimated using the unfolding procedure described in Sec.~\ref{sec:unfold} to account for the detector response are given in Table~\ref{tab:LowResGamma}. 

These lower resolution spectra display features similar to that observed in Fig.~\ref{fig:highResGammaComparison}. There is a considerable high-energy $\g$-ray background ($>3$~MeV) at every near location. This can attributed to short-lived isotopes produced by neutron reactions either at or nearby the near locations. Contrasting the potential near locations we observe variation in the total $\g$-ray flux and its character. The counts recorded above $7$~MeV at  \NBSR{} and \HFIR{} imply a larger thermal neutron background, leading to neutron capture on structural steel. The feature at $\approx2.7$~MeV observed at the \ATR{} far location is due to $^{24}$Na produced near the reactor and transported in piping. At all sites, there is an indication of a continuum background due to scattered $\gamma$-rays leaking through shielding walls, or local $\g$-ray scattering from surrounding material. Based upon the relative lack of peak structure, it appears that \HFIR{} has a more significant down-scattered component. 

\begin{table}[tb]{%
   \begin{tabular}{l|d{5.1}|d{5.1}} \hline
Location	& \multicolumn{1}{c|}{Flux $1$--$3$~MeV} 	&\multicolumn{1}{c}{Flux $3$--$10$~MeV} \\ 
	& \multicolumn{1}{c|}{(cm$^{-2}$s$^{-1}$)}	& \multicolumn{1}{c}{(cm$^{-2}$s$^{-1}$)}\\ \hline
ATR	near&$3.7$ &$0.3$ \\
HFIR	 near &$5.4$ &$4.3$\\
NBSR near&$11.7$&$7.7$\\ \hline
 ATR	far&$1.7$ &$-$ \\
HFIR	 far&$1.7$ &$-$\\
NBSR far&$0.1$&$-$\\ \hline
    \end{tabular}}
    \caption{Approximate $\gamma$-ray fluxes measured with the 2" NaI(Tl) detector at the three reactor sites. While statistical errors on these values range between $0.1$--$1$\%, a conservative $10\%$ relative systematic is assumed for the unfolding procedure used since an absolute efficiency calibration was not performed. No values for the far sites are reported in the upper energy range since there is no significant $\gamma$-ray contribution to the spectra in these cases. Note that the \NBSR{} far site represents a typical laboratory background spectrum dominated by naturally occurring radioactivity. }
 \label{tab:LowResGamma}
\end{table}

%\clearpage

\subsection{$\g$-ray Spectrum Unfolding}
\label{sec:unfold}
The measured $\g$-ray spectra are strongly dependent upon instrument response, as evidenced by the prominent escape peaks and Compton edge features visible in Fig.~\ref{fig:nistRawSpec}. To obtain an accurate representation of the absolute $\g$-ray flux for use in shielding studies, we must account for both the structure the response imprints upon the measured spectra and the energy-dependent detection efficiency of the $\g$-ray instruments. There exists a rich literature describing statistical unfolding or inversion techniques for problems such as this. Many difficulties can arise in applying unfolding algorithms, producing error estimates for unfolded quantities, and in selecting appropriate regularization parameters and/or convergence techniques. Furthermore, there is typically no guarantee that the solution obtained is unique. Nonetheless, this is still a useful exercise for our purpose: obtaining a reasonable estimate of the $\g$-ray source term for propagation through simulations of proposed detector shielding configurations. The ``reasonableness'' of an unfolded solution can be readily assessed by convolving it with the detector response function and making a qualitative comparison with the measured spectrum. Here we describe the method used to unfold the various $\g$-ray measurements, using the near location measurement at ATR with the LaBr$_3$(Ce) detector as an example. Since this detector was only used at ATR, and the measured $\gamma$-ray flux at ATR was the lowest of the three sites, this unfolding procedure was also performed for the NaI(Tl) measurements at all locations to estimate absolute fluxes for inter-comparison (Table~\ref{tab:LowResGamma}). While pileup has been neglected in what follows due to the relatively small $\gamma$-ray detectors used, it will clearly be an important consideration for cubic meter scale \nuebar{} detectors.

The following data processing steps were taken prior to performing the spectral unfolding. For the LaBr$_3$(Ce) detector, background due to internal La and $^{40}$K radioactivity was subtracted using a background run taken in a shielded enclosure. All measured spectra were calibrated using known line positions in the $479$--$7645.5$~keV energy range. This calibration also provided a measurement of the detector resolution as a function of energy.

Detector response functions were generated using a dedicated GEANT4 simulation. Electron energy depositions in the crystal volume of the detector package (including an aluminum casing and readout PMT) were recorded. Detector resolution effects were accounted for by convolving the simulation result with the energy resolution function determined during calibration. For the LaBr$_3$(Ce) detector the simulated response was validated against measurements in an \INL{} laboratory using $^{137}$Cs and $^{60}$Co sources. Furthermore, the simulation-predicted efficiency was in good agreement with tabulated values supplied by the detector manufacturer.  The LaBr$_3$(Ce) response was then simulated over the energy range of interest ($0$--$8$~MeV). Simulated $\gamma$-rays were propagated towards the detector model uniformly from all directions. The response function was generated with a bin size of $20$~keV for the incident $\g$-ray energy, and a non-linear binning matching the experimental data for the detected energy. 

\begin{figure}[tb]
\centering
\includegraphics*[width=0.48\textwidth]{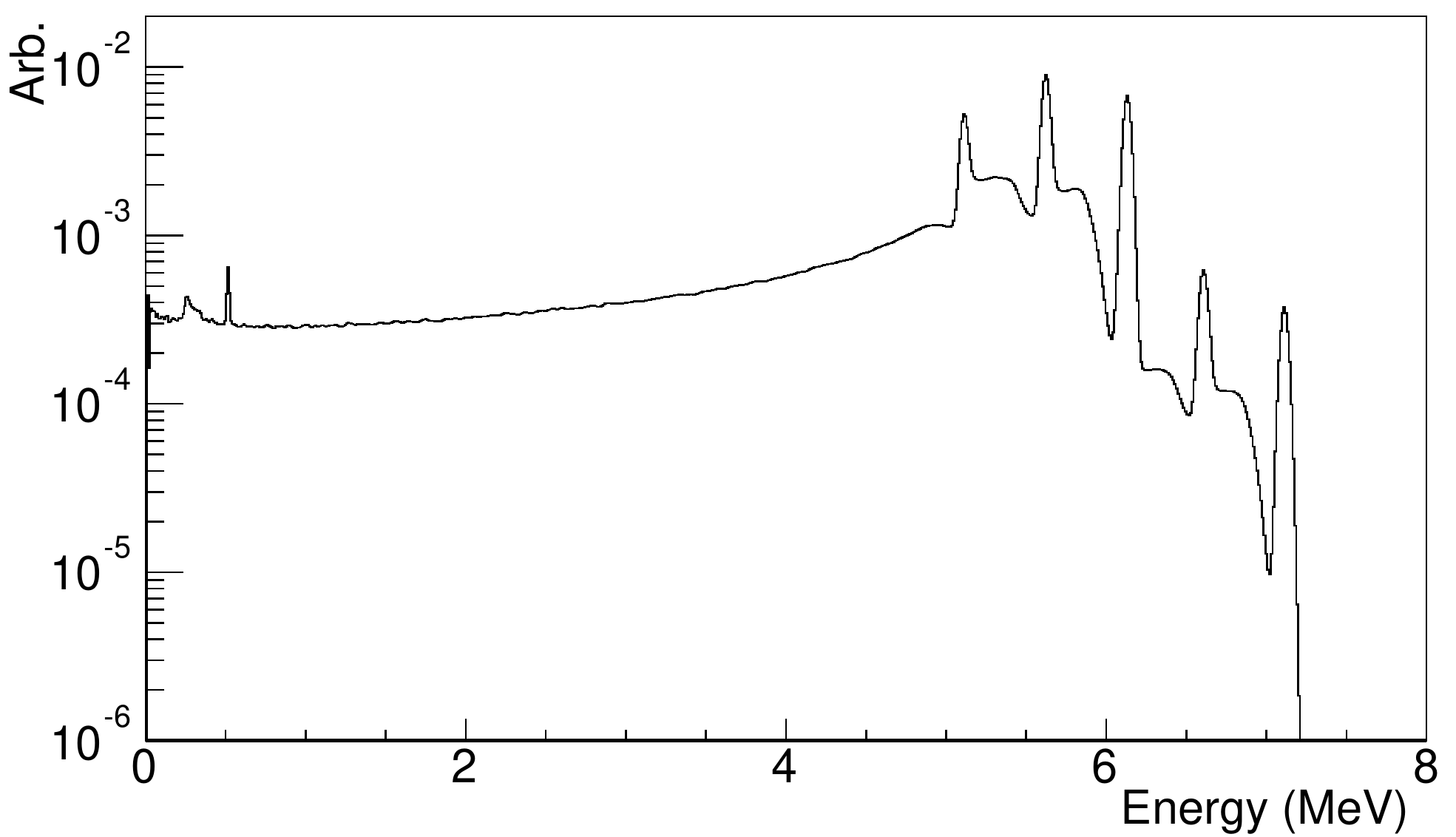}
\caption{Simulated response of the LaBr$_3$(Ce) detector to the $6.13$ and $7.12$~MeV  $\g$-rays emitted by $^{16}$N. The prominent features in this response are full absorption peaks, single and double escape peaks, and the summed Compton continuum from both lines. }
\label{fig:LaBrResponse}
\end{figure}

An example of the generated response function, convolved with the experimentally measured resolution, is shown in Fig.~\ref{fig:LaBrResponse} for $\g$-ray lines due to $^{16}$N.  These and 14 other prominent line responses were also generated so that contributions from monoenergetic lines could be directly subtracted from the measured spectra. Doing so leaves only the relatively smooth down-scattered continuum to unfold which presents an easier task. The monoenergetic line response was estimated by fitting the  sum of relevant line responses and a smooth background model to the data in several energy ranges. The background model was generated using the Sensitive Nonlinear Iterative Peak clipping algorithm (SNIP)~\cite{morhac1997} implemented in the TSpectrum class of the ROOT analysis package~\cite{root}.

\begin{figure}[tb]
\centering
\includegraphics*[width=0.48\textwidth]{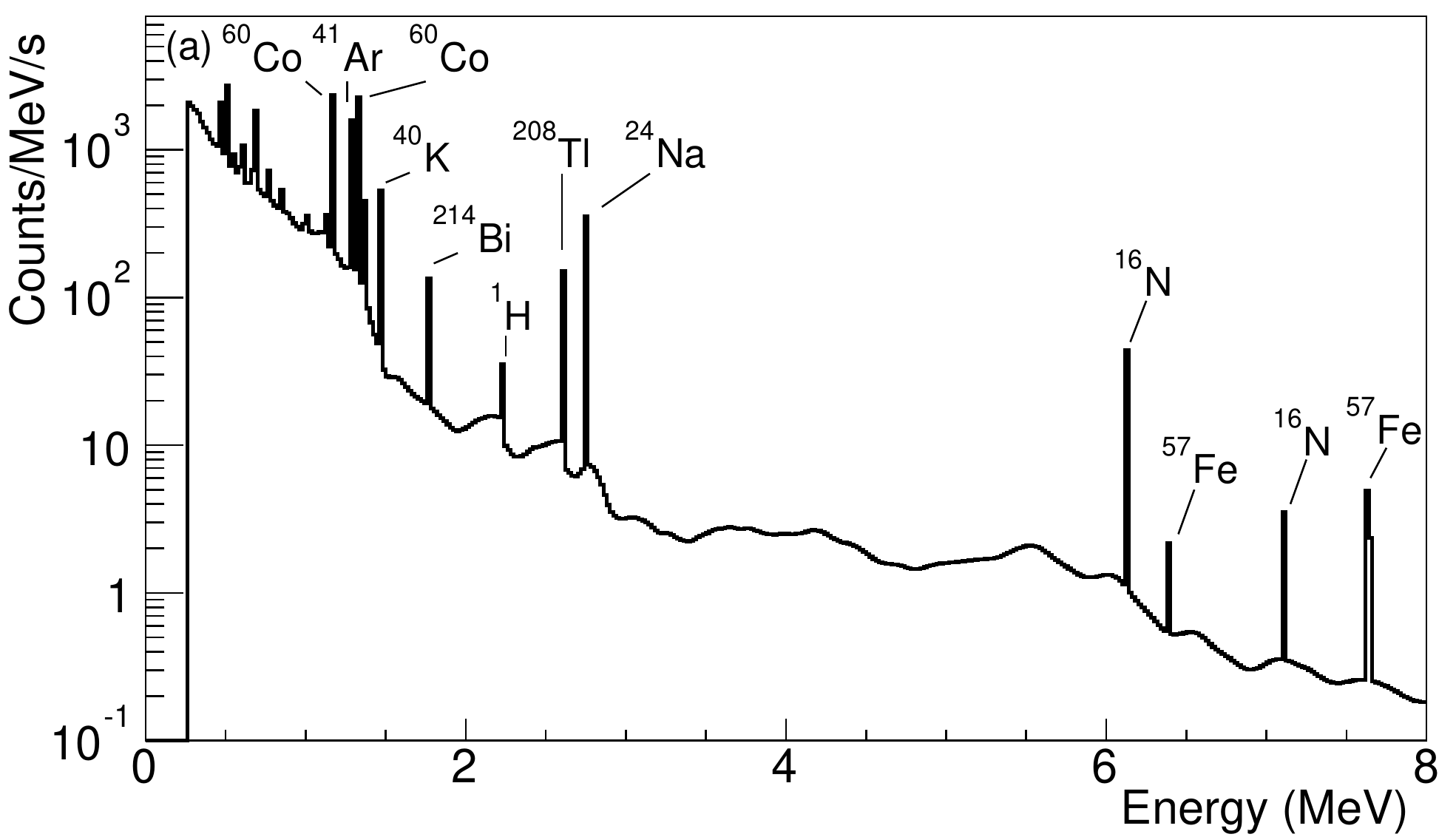}
\includegraphics*[width=0.48\textwidth]{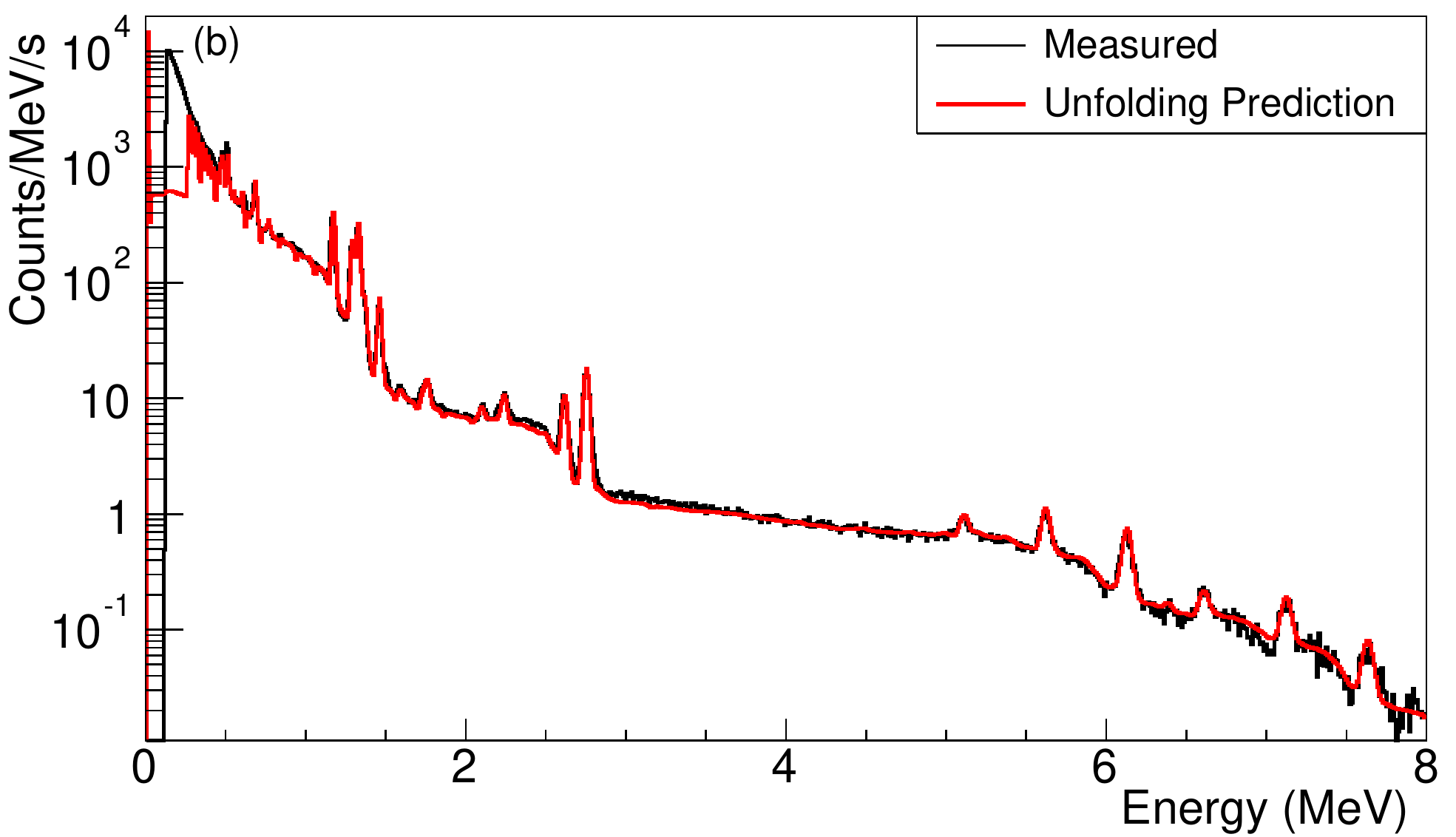}
\caption{(a) The $\g$-ray spectrum incident upon the LaBr$_3$(Ce) detector at the \ATR{} near location, as predicted by the unfolding of the measured spectrum. Prominent line sources are identified. (b) A comparison of the measured $\g$-ray spectrum with that predicted from the unfolded source term and the simulated detector response. Note that the unfolding procedure accounts for escape peaks and Compton scattering events in the measured spectrum. The residual continuum in (a) is due to $\g$-rays that have down-scattered in the surrounding environment interacting in the detector.}
\label{fig:unfoldingExample}
\end{figure}

Finally, an unfolding algorithm~\cite{Morhac2000108} is applied to the residual continuum. This is done within several energy ranges where the count rate is similar, to aid convergence. The predicted source term for the measured continuum is assembled piece-wise and the monoenergetic line contributions added. The results of this procedure for the LaBr$_3$(Ce) measurement taken at the \ATR{} near location are shown in Fig.~\ref{fig:unfoldingExample}. The detector response predicted from the unfolded source term is in good qualitative agreement with that measured and can be readily used for detector shielding simulation studies. 

\begin{figure}[tb]
\centering
\includegraphics[width=0.48\textwidth]{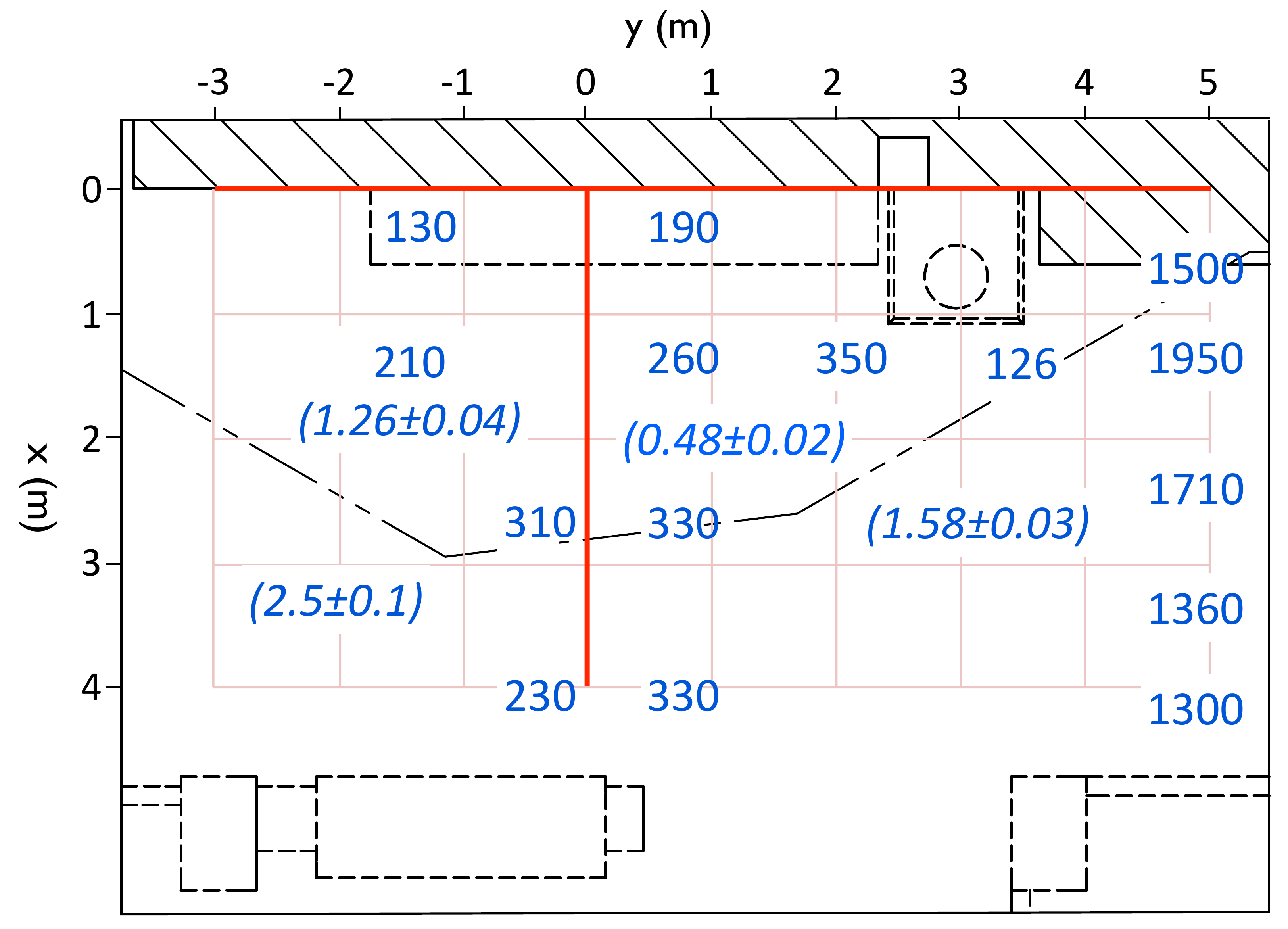}\hfil%
\caption{A pictorial representation of neutron dose rates (measured in nSv/h) and thermal neutron rates in italics (cm$^{-2}$s$^{-1}$) at the \HFIR{} near location roughly 15~cm (z = 0.15) above the floor. Measurements are plotted on a one meter square grid referenced to the reactor wall ($x=0$) and the smallest baseline ($y=0$). The reactor core is centered at $(x,y,z) = (-4.06,0,-3.85)$. }
\label{fig:hfirThermalNeutron}
\end{figure}

\subsection{Neutron Dose and Thermal Measurement Results}
\label{sec:thermalNeutronResults}

Shown in Fig.~\ref{fig:hfirThermalNeutron} 
are the results of neutron dose and thermal neutron flux (italics) measurements  taken in the \HFIR{} near location with the reactor  at the nominal operating thermal power of $85$~MW.  The neutron dose data represent two data sets taken roughly six months apart.  These measurements generally agreed at the 10\% level, except where large gradients were observed.  In these cases the disagreement is likely due to inaccuracies in locating the instruments.  As is evident in the figure, considerable spatial variation was observed, in particular, a strong increase in neutron rate to both the left and right of the proposed detector location. The likely cause of this spatial variation is a large shielding structure on the level below that terminates in approximately this area (the dotted polygon shaped region in the Fig.~\ref{fig:hfirLayout}). This structure is probably shielding the central area from scattered neutrons originating from the neutron beamlines on the lower level, while in other regions of this space they can propagate through the floor.  The effect is particularly pronounced on the right side above the cold neutron source and guides.  Consistent with this hypothesis, dose measurements taken above the cold-neutron beamline shielding, but below the experimental level floor, were 2.35~$\mu$Sv/h.  Similarly, the dose rate at  $(x=1$~m, $y=1$~m)  dropped by a factor of two when the HB3 beamline shutter was closed (lower floor as indicated in Fig.~\ref{fig:hfirLayout}).  In this scenario, localized shielding would be  difficult,  but may still be possible since relatively thin layers of borated materials can be very effective for thermal neutron suppression.

Measurements were taken at the \NBSR{} near location and in a lab space far from the confinement building as a reference point.  The near location measurements were taken multiple times, approximately 2 m from the face of the reactor biological shield indicated in Fig.~\ref{fig:nistNearLocation},  with adjacent instruments on.  Dose rates (as described above) were 1.44~$\mu$Sv/h which, assuming a spectrum centered around energies close to the maximum detector efficiency, corresponds to an approximate flux of $2$--$3$~cm$^{-2}$ s$^{-2}$.  This is four times the rate observed with the adjacent instruments off.  For context, the rate in the far lab space was 22~nSv/h, consistent with natural backgrounds.  These rates were fairly constant between measurements.  A bare BF$_3$ tube was used to measure the thermal flux in the same location.  The flux of $2$~cm$^{-2}$s$^{-1}$ indicates that the spectrum is likely peaked at lower near-thermal energies. 
Such measurements were not taken at \ATR{}, but the relatively low flux of neutron-capture $\g$-rays  observed at that site implies the thermal neutron flux is also low.

\subsection{Fast Neutron Measurement Results}

\subsubsection{Fast Neutron Spectrum Measurements with FaNS-1}

\begin{figure}[tb]
\begin{center}
\includegraphics[width=0.48\textwidth]{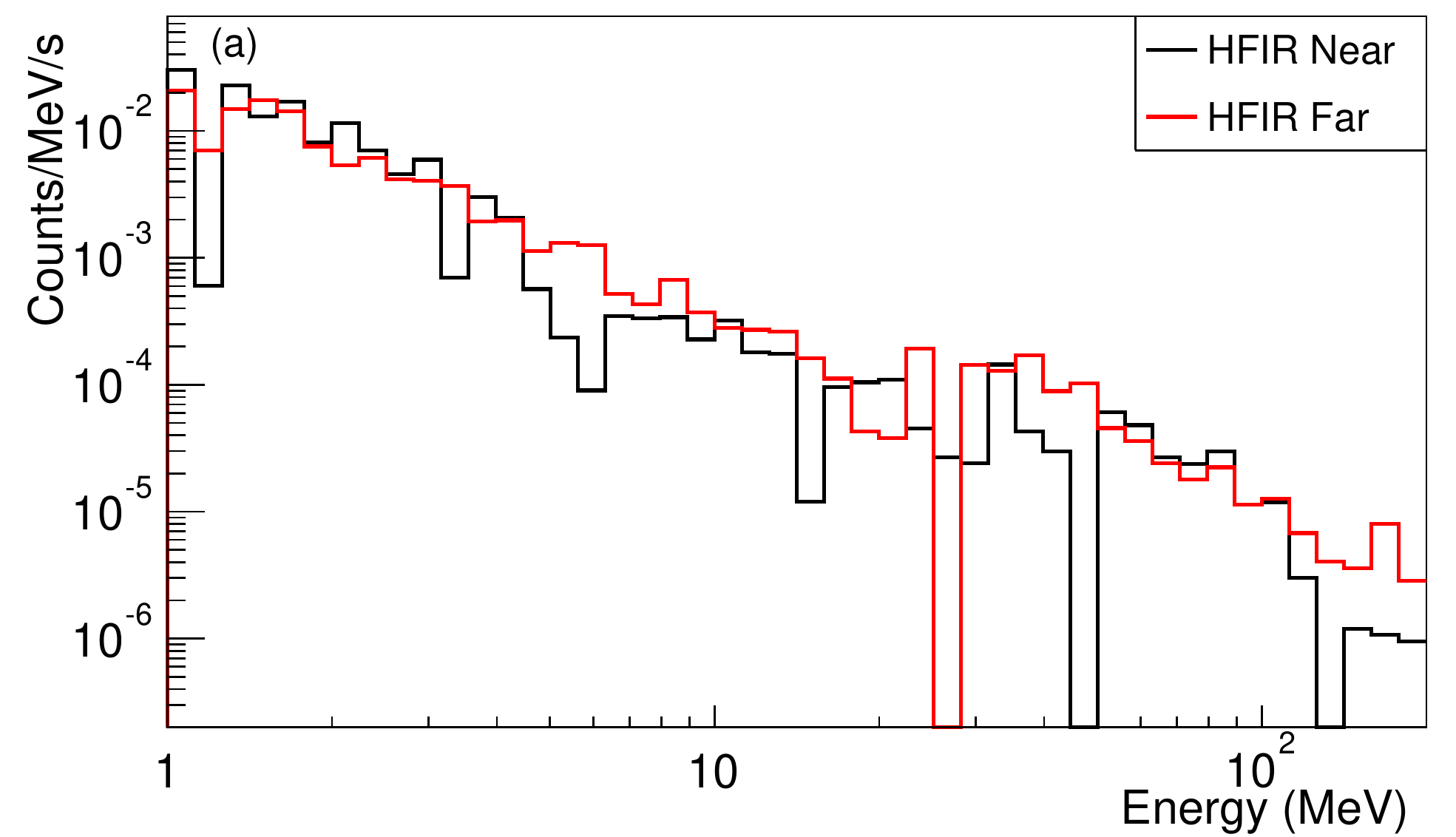}
\includegraphics[width=0.48\textwidth]{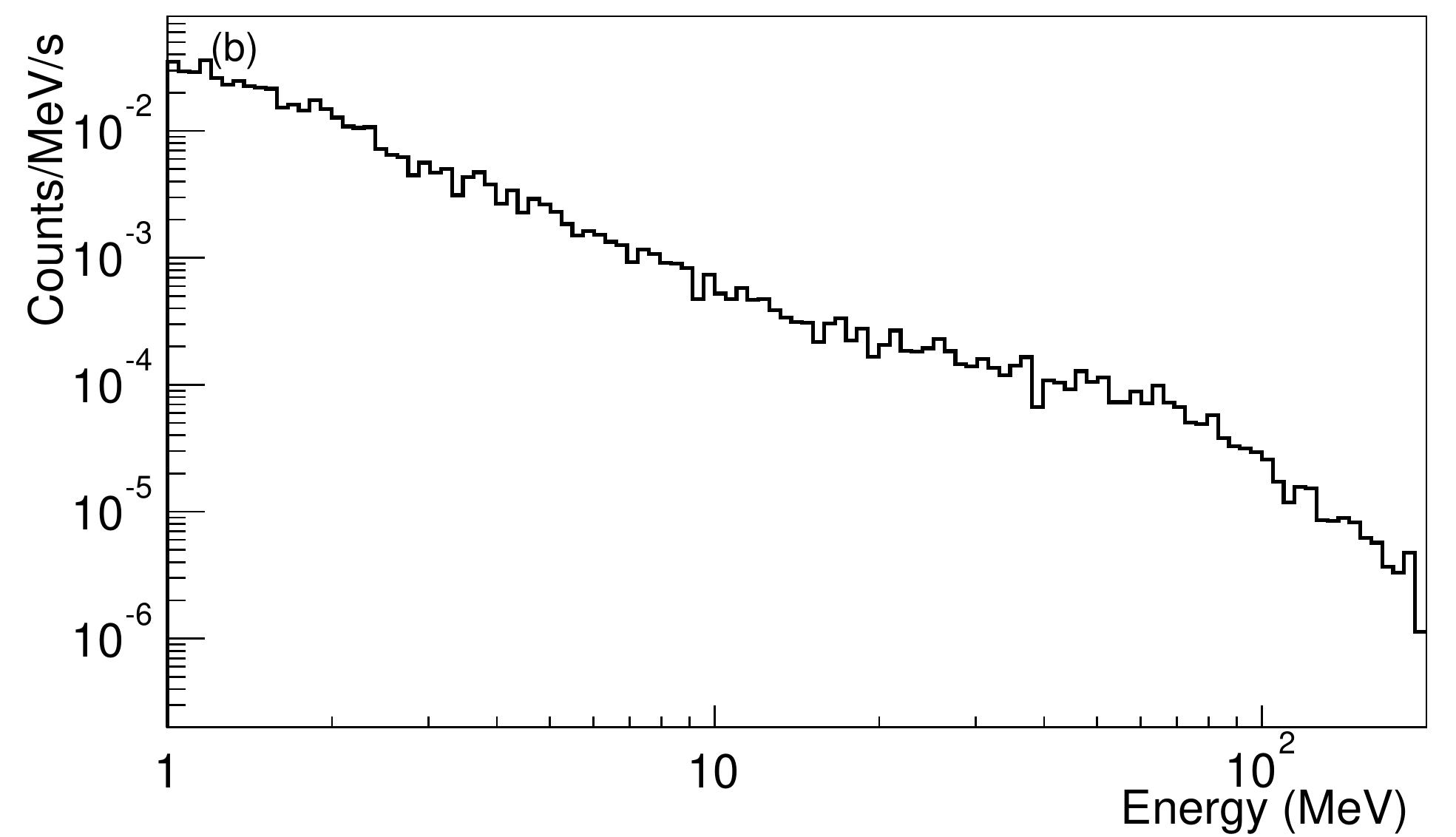}
\caption{The cosmogenic neutron induced energy spectrum recorded at the (a) \HFIR{} near and far locations  and (b) \NBSR{} far location.}
\label{fig:FaNSEnSpec}
\end{center}
\end{figure}

Measurements of the cosmogenic neutron spectra (Fig.~\ref{fig:FaNSEnSpec}) and fluxes (Table~\ref{tab:FaNSCosmogenics}) at \NBSR{} and \HFIR{} were performed using FaNS-1. Reactor off measurements were taken at \HFIR{} at the near and far locations and the \NBSR{} far location. The sensitivity of FaNS-1 to cosmogenic neutrons has been simulated using MCNPX. An isotropic distribution of neutrons following the JEDEC standard spectrum was launched at the detector and the sensitivity, in neutrons detected per incident neutron fluence, for neutron energies above 1~MeV was determined to be 10.3$\pm$2.5~(n/(n/cm$^2$))~\cite{Langford2013}. This is akin to the efficiency weighted by the cosmogenic spectrum times the cross-sectional area. This sensitivity is then used to convert a measured count rate in s$^{-1}$ into the incident flux in cm$^{-2}$s$^{-1}$. 

Comparing the \NBSR{} measurement with the \HFIR{} far location we see a slight deficit in the \HFIR{} flux which can possibly be explained by the presence of a large ($10$~m--$12$~m) concrete wall that shadows the location.  The  \HFIR{} far and near measurements are comparable.  Note the similarity of the spectra shape between each site. This similarity reinforces the previously discussed notion that the spectrum of cosmogenic neutrons does not vary significantly between sites. It is important to note that these fluxes have not been corrected for fluctuations in the barometric pressure and solar cycle. These environmental parameters are known to influence the total flux by 10--20\% (-0.73\% per millibar change in pressure)~\cite{Langford2013, Paschalis2013}. However, from a qualitative point of view, we find that the difference between \NBSR{} and \HFIR{} is minimal.  

\begin{table}[tb]
\begin{center}
\begin{tabular}{l|d{1}|c}
\hline
Location & \multicolumn{1}{c|}{Exposure} & \multicolumn{1}{c}{Flux (E$_n >$ 1~MeV)} \\
&\multicolumn{1}{c|}{(h)} &\multicolumn{1}{c}{(cm$^{-2}$s$^{-1}$)}\\
\hline
HFIR near 	 & 12 &	$(4.1\pm0.3)\times10^{-3}$\\ 
HFIR far  	 & 8 &	$(4.4\pm0.3)\times10^{-3}$\\
NBSR far & 156 & 	$(5.6\pm0.1)\times10^{-3} $\\    
\hline
\end{tabular}
\caption{Cosmogenic neutron background measurements conducted with FaNS-1 at the \HFIR{} near location and the \NBSR{} and \HFIR{} far locations. Quoted uncertainties are statistical only.}
\label{tab:FaNSCosmogenics}
\end{center}
\end{table}%

\subsubsection{Fast Neutron Relative Rate Measurements with a Portable Stilbene Detector}
\label{sec:stilbene}

The stilbene detector system records list mode data for each event. A digital filtering algorithm that mimics the function of an analog constant fraction discriminator is applied to a stream of waveform samples to derive a trigger. Two integrals of waveform samples are acquired relative to the trigger time: a ``full'' integral summing the total PMT charge resulting from an interaction in the crystal and a ``tail'' integral summing charge produced primarily by the slow component of the scintillator response. Since more heavily ionizing particles, like recoil protons, preferentially excite long-lived states in the scintillator, the ratio of the ``tail'' to ``full'' integrals can be used to distinguish particle type. With this DAQ setup it was not possible to record the full waveforms corresponding to each event, only these integrals. Subsequently there is a potential for misidentification if the triggering algorithm is disrupted by pulse pileup or baseline disturbances caused by relatively high interactions rates.

\begin{figure}[tb]
\centering
\includegraphics*[width=0.48\textwidth]{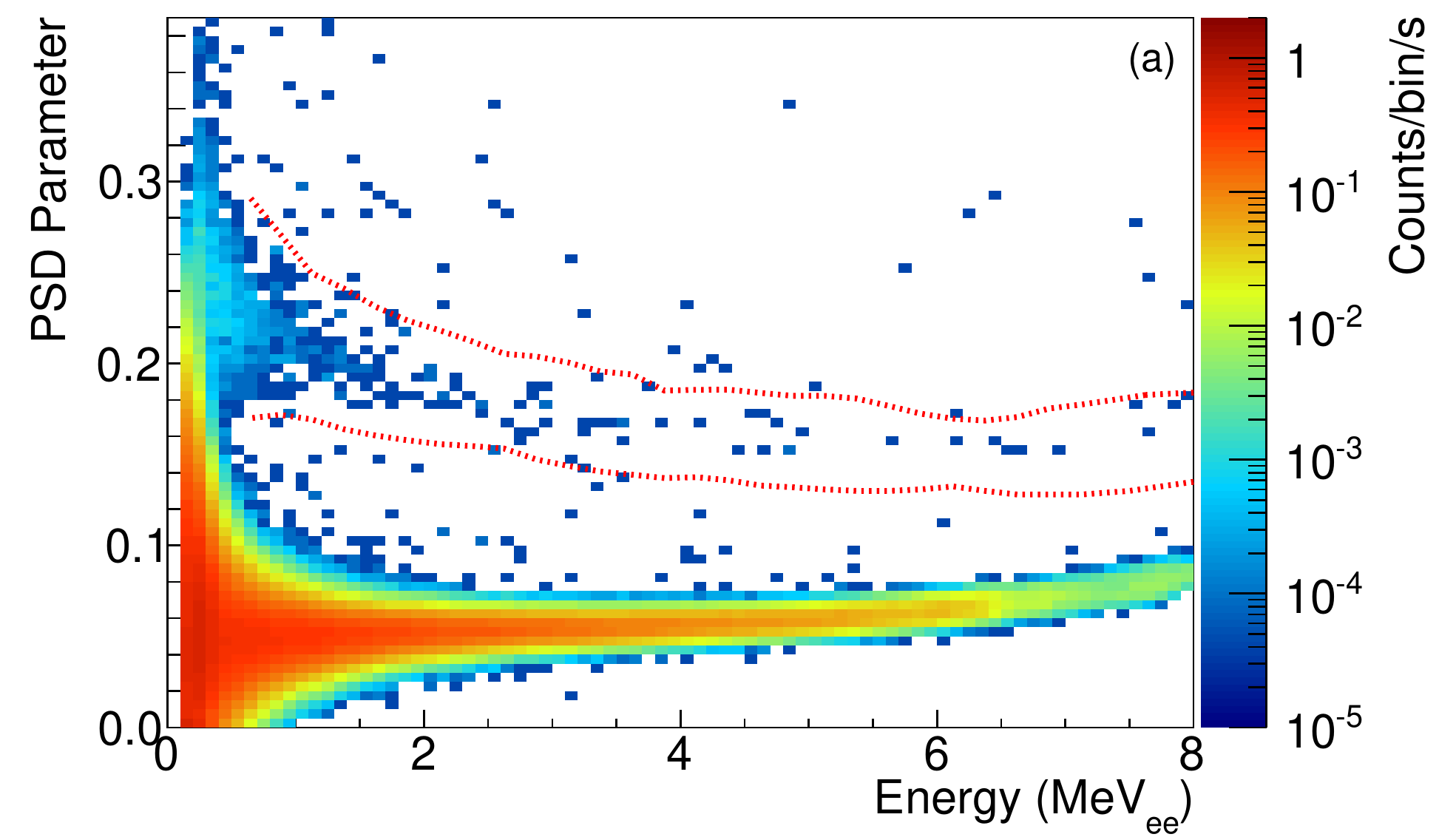}
\includegraphics*[width=0.48\textwidth]{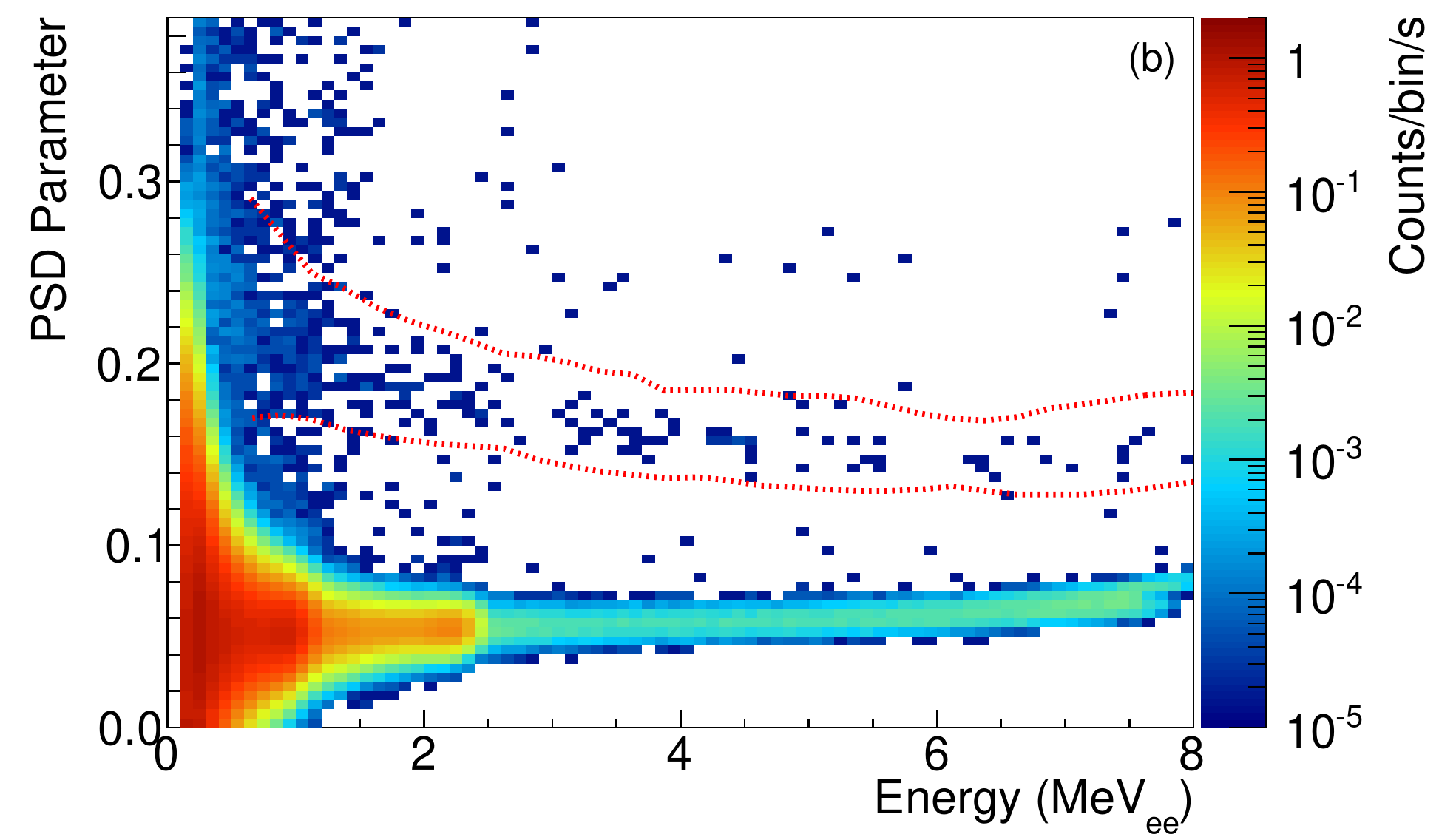}
\caption{Stilbene PSD measurements performed at the (a) \HFIR{} and (b) \ATR{}  far locations. Spectra are collected with the reactors operating at nominal power. The two horizontal bands correspond to $\g$-rays (lower) and fast neutron recoil interactions (upper). The region used for fast neutron rate measurements is indicated by dashed lines. The higher $\g$-ray rate encountered at \ATR{} due to $^{24}$Na causes particle misidentification at energies $<1.5$~MeV$_{ee}$.} 
\label{fig:stilbeneExample}
\end{figure}

Example fast neutron measurements using the stilbene detector are shown in Fig.~\ref{fig:stilbeneExample} for the \ATR{} and \HFIR{} far locations. As is conventional for fast neutron PSD measurements, the electron equivalent energy (MeV$_{ee}$) of an event is plotted against the ``tail'' to ``full'' ratio PSD parameter. This allows the energy dependence of the PSD parameter to be readily observed. The energy scale is determined using calibration sources and background $\g$-rays observed at each site. Two clear horizontal bands are observed in the data, the lower corresponding to electron depositions (primarily from $\g$-rays) and the upper corresponding to neutron induced recoil protons.  As the total event energy decreases the width of these bands increases due to worsening photo-statistics causing direct spread in the ratio and larger jitter in the trigger time determination. The PSD parameter range used for neutron counting was determined by performing a Gaussian fit to the neutron band as a function of total energy using a high statistics background dataset taken at \LLNL{}. The selection band is set $\pm3\sigma$ about the mean PSD parameter value found for a  particular energy range. 

Comparing the \HFIR{} and \ATR{} measurements we see that the larger $\g$-ray background at the \ATR{} far location causes misidentification at energies $<1.5$~MeV$_{ee}$. Subsequently, this is the lower threshold implemented for all site-to-site comparisons. At some locations, notably the \NBSR{} near location and the \HFIR{} near location close to the reactor wall, this misidentification was significant across almost all of the stilbene detector energy range. These measurements were therefore not included in the comparison. Values reported for \HFIR{} are the average of measurements taken at the middle and rear of the potential detector location, while values for \NBSR{} are based on data taken in a nearby laboratory. 

The conversion from electron equivalent deposition energy  to recoil proton energy (denoted MeV$_{nr}$) for this material is obtained from~\cite{Robinson2011404}. The $1.5$~MeV$_{ee}$ lower threshold corresponds to a proton energy of $4$~MeV$_{nr}$, while the dynamic range of the eMorpho DAQ implies an upper limit of $14.5$~MeV$_{nr}$. As well as giving the rate in the full range of comparable sensitivity, we also calculate the rate in the range $10$--$14.5$~MeV$_{nr}$ where there will be effectively no contribution from reactor-correlated fission spectrum neutrons, allowing us to make a comparison based only on cosmogenic fast neutron interactions. 

The measured fast neutron rates for the near and far locations at all reactor sites are given in Table~\ref{tab:stilbeneFastNeutron}.  All measurements were performed with the reactors operating. The values for NIST were recorded in a nearby laboratory, since the $\g$-ray background encountered in the near location was too high across much of the energy range of interest. The rates given in Table~\ref{tab:stilbeneFastNeutron} for \NBSR{} are therefore a lower bound in the  $4$--$14.5$~MeV$_{nr}$ range since any possible fission neutron contribution is not included, and an upper bound in the $10$--$14.5$~MeV$_{nr}$ range since the attenuating affect of the reactor confinement building is not included. 

\begin{table}[tb]{%
   \begin{tabular}{l|c|c} \hline
Location	& \multicolumn{2}{c}{Rate ($\times10^{-3}$s$^{-1}$)} \\
&$4$--$14.5$~MeV$_{nr}$ 	& $10$--$14.5$~MeV$_{nr}$     \\ \hline
ATR	near&$4.7 \pm 0.3$ &$1.0 \pm 0.1$ \\
HFIR near&$2.2 \pm 0.2$ &$0.3 \pm 0.1 $ \\ \hline 
ATR	far&$1.8 \pm 0.2$ &$0.4 \pm 0.1$ \\
HFIR far&$3.5 \pm 0.2$ &$0.6 \pm 0.1$ \\
NBSR far&$2.8 \pm 0.2 $&$0.8 \pm 0.1$ \\ \hline
    \end{tabular}}
    \caption{Relative fast neutron rates and associated statistical uncertainties measured at the three reactor sites. See text for additional comments.}
 \label{tab:stilbeneFastNeutron}
\end{table}

The \ATR{} near location experiences the highest fast neutron rate, presumably due to the higher elevation of that site which is not entirely offset by the overburden provided by the building structure. Conversely, the relatively deep \ATR{} far location has the lowest fast neutron rate. Comparing the \HFIR{} near and far locations, we see that the near location has a lower rate which is presumably due to the greater overburden provided by the reactor confinement building at the near location relative to the (effectively) outdoor far location. The \NBSR{} result is consistent with that at the \HFIR{} far location, which has similar elevation and overburden. 

\begin{figure}[tb]
\centering
\includegraphics*[width=0.48\textwidth]{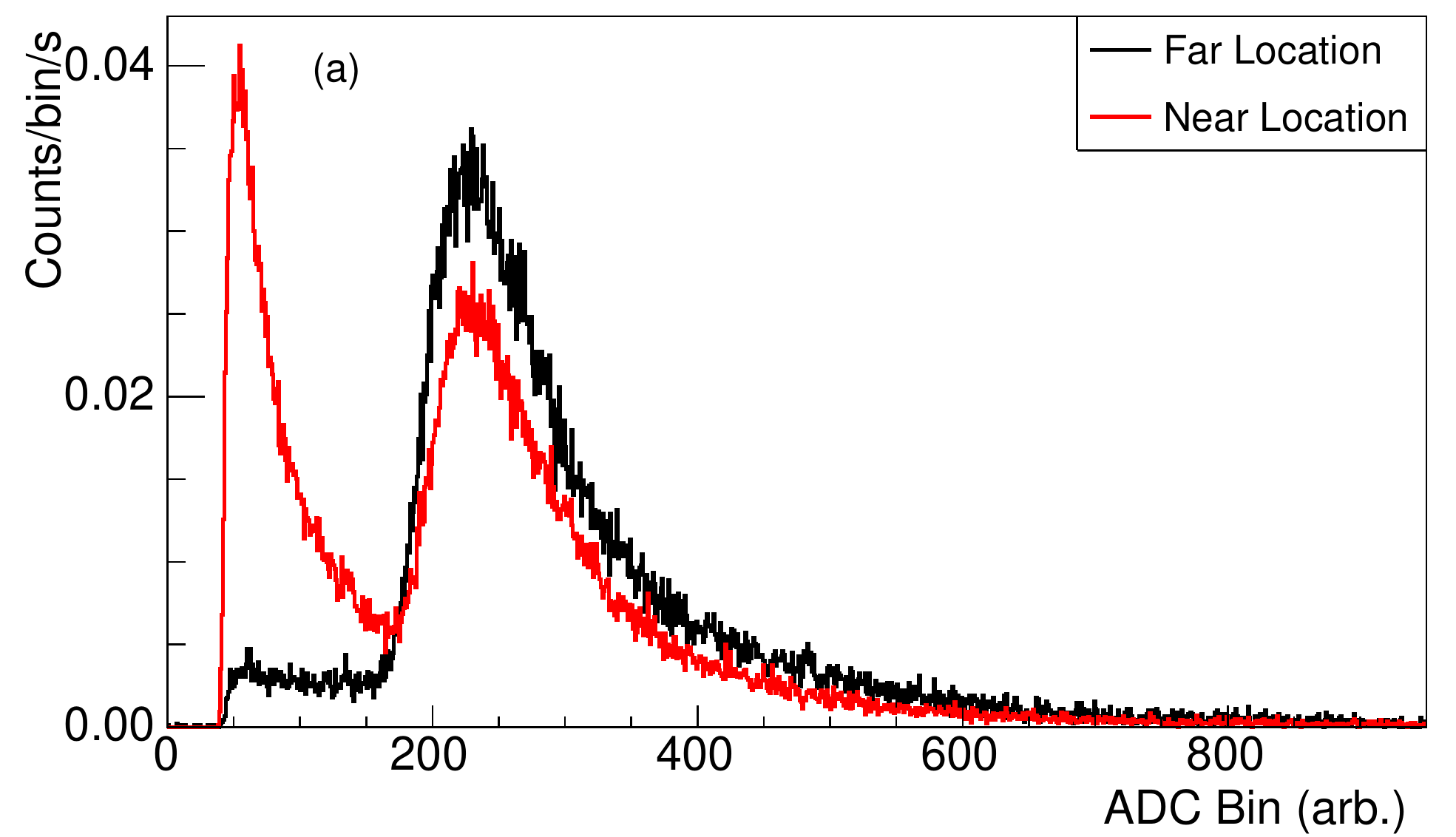}
\includegraphics*[width=0.48\textwidth]{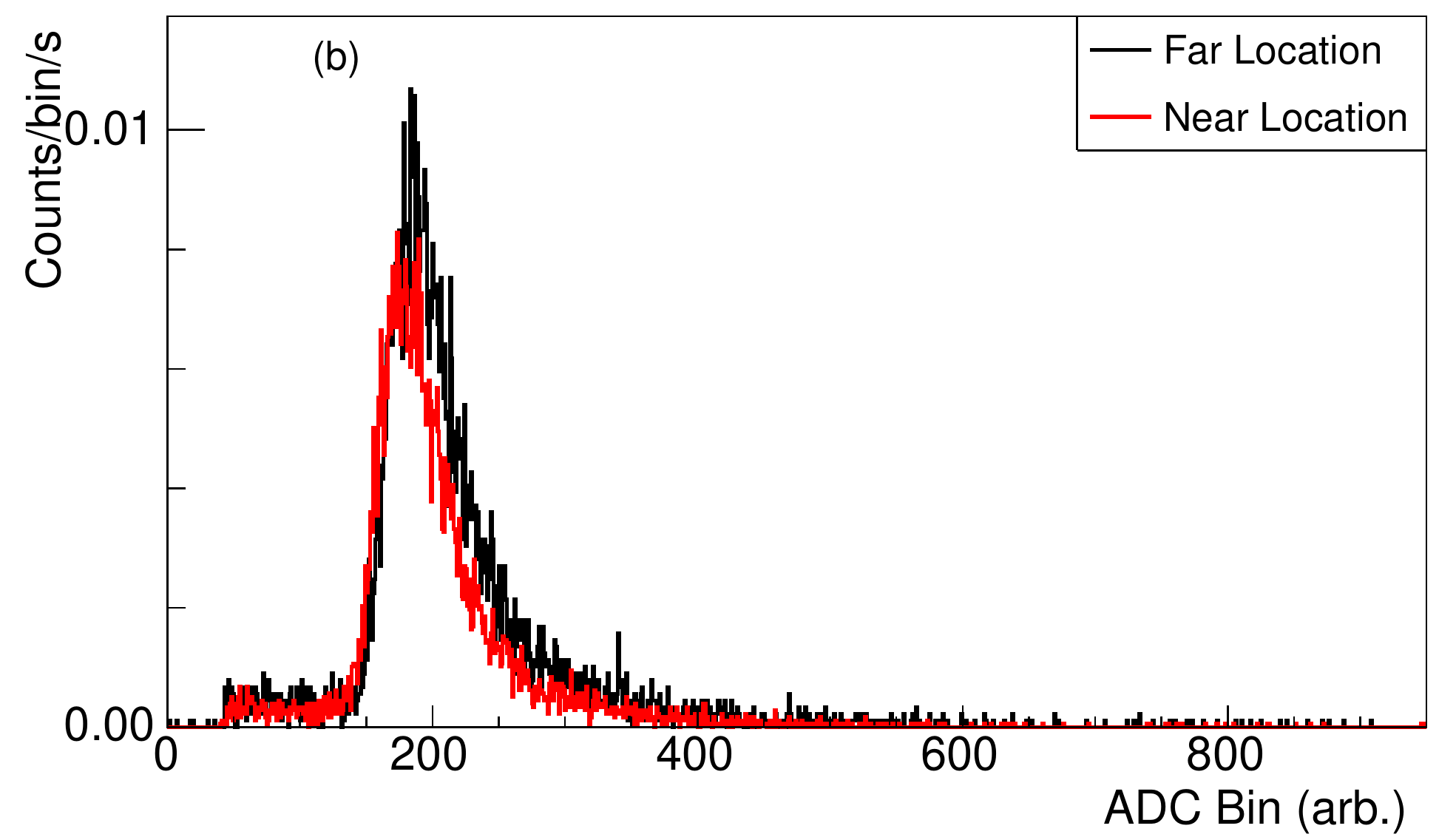}
\caption{(a) Channel 1 spectrum requiring a twofold coincidence (Ch0*Ch1) at the \NBSR{} near (red) and far  (black)  locations . A large correlated background from the high background $\gamma$-ray rate in the near location is apparent. (b) The same comparison for Channel 3 requiring a threefold coincidence (Ch0*Ch1*Ch3). No additional background is observed. }
\label{fig:muon_Compare}
\end{figure}

\subsection{Muon Measurement Results}
\label{sec:muon}
For this study cumulative histograms were recorded for two different trigger conditions: a twofold coincidence between two of the lower paddles $(\approx\pm55^{\circ} \times ~\pm65^{\circ}$ angular range) or a threefold coincidence where the topmost paddle was also required 
$(\approx\pm10^{\circ} \times  \pm15^{\circ}$ angular range). 

With a trigger threshold of $\approx1$~MeV and a coincidence window of $250$~ns, accidental coincidences were found to be negligible. Given a $2$~MeV/cm energy deposition for minimum ionizing particles, a typical signal deposition in the paddles is $\approx5$~MeV, a value higher than most background $\gamma$-rays. In the  high $\gamma$-ray fluxes encountered at some sites paddle singles rates were $< 250$~s$^{-1}$, yielding a twofold accidental rate of $<0.05$~s$^{-1}$. This should be compared to the measured muon signal rate of $4$--$7$~s$^{-1}$.  The energy spectra for the twofold coincidence requirement are shown in Fig.~\ref{fig:muon_Compare}a at \NBSR{}. The minimum ionizing peak is evident in far location data. Muons clipping the edge of the scintillator paddle  produce the flat shape in the bins lower than the Landau peak in simulation studies. Inside the reactor confinement building at the near location a large background was observed in the twofold coincidence spectra at low energy. This background is probably due to multiple scatter $\gamma$-ray interactions, since the accidental background previously calculated is too small to account for this feature. Requiring a threefold coincidence strongly suppresses  this source of background (Fig.~\ref{fig:muon_Compare}b). The coincidence spectra shape is consistent  in the data taken inside and outside  the \NBSR{} confinement building, despite a factor of 200 increase in the singles rate inside the building.

Since the scintillator spectra  requiring a threefold coincidence were consistent with clean muon signals, these measurements are used for the site comparison. Due to equipment damage during transport between the sites, the paddle separations were not identical for all measurements which had a small effect on the telescope acceptance efficiency. A geometry dependent correction factor was estimated via a simple Monte Carlo simulation. The resulting threefold coincidence rates are given in Table~\ref{tab:Muon}. Of the near locations, \NBSR{}  and \HFIR{} have similar rates, while the higher rate observed at \ATR{} is presumably due to the greater elevation at that site and the modest overburden provided by the crane access hatch at that location. The measured far location rates are very similar. At \HFIR{} and \NBSR{} these measurements were taken outside of the reactor confinement structures with reduced overburden relative to their respective near locations. Conversely, at \ATR{} the far location is in a deeper basement level providing more overburden relative to the near location. For comparison, the rate measured at grade level at \ATR{} was $0.85$~s$^{-1}$.

\begin{table}[tb]{%
   \begin{tabular}{l|c|c} \hline
Reactor	& Rate at near location 	& Rate at far location \\ 
	& (s$^{-1}$)	&  (s$^{-1}$)\\ \hline
ATR	&$0.78 \pm 0.03$ &$0.68 \pm 0.02$  \\
HFIR	&$0.59 \pm 0.02$ &$0.71 \pm 0.03$ \\
NBSR&$0.56 \pm 0.01 $&$0.69 \pm 0.01$ \\ \hline
    \end{tabular}}
    \caption{Muon rates measured at the 3 possible near and far locations for the three-fold telescope. The far location at \NBSR{} was a lab space whose rate should approximate the \NBSR{} far location.}
 \label{tab:Muon}
\end{table}

Measurements at azimuthal angles of 0$^\circ$, 45$^\circ$, and 90$^\circ$  in different orientations relative to the reactor core were made at each of the sites. The measured rate at 90$^\circ$  was essentially zero. The data are consistent within errors with the expected $\rm{cos}^2\theta$ dependence.  At 45$^\circ$, the measured rates were lower by $10$--$40$\% when the telescope was oriented towards the more massive shielding structures surrounding the reactor cores.

Translating the rates in Table~\ref{tab:Muon} to an absolute muon flux requires a correction for the trigger efficiency and for the solid angle acceptance. The trigger efficiency was measured with data using all four paddles stacked on top of each other.   An efficiency of 98.7\% is assumed for all paddles.  The solid angle acceptance of the coincidences was calculated from a simple simulation. The average threefold acceptance with the extended paddle is 0.189 sr.  The twofold acceptance is 3.16 sr. Thus the fluxes at the near location obtained from the threefold measurements are $79.5~\rm{m^{-2}sr^{-1}s^{-1}}$ at \NBSR{}, $84.9~\rm{m^{-2}sr^{-1}s^{-1}}$ at \HFIR{}, and $111.4~\rm{m^{-2}sr^{-1}s^{-1}}$ at \ATR{}.

\section{Characteristics of Reactor-Correlated Background}
\label{sec:rxCharacteristics}

As noted above, there are three important sources of background encountered in research reactor facilities: naturally occurring radioactivity in facility structures, cosmogenic background, and emissions correlated with reactor operations. Extensive discussion of naturally occurring and cosmogenic background can be found elsewhere (e.g.~\cite{Gordon2004,Kowatari2005}). In this section we use the measurements described in Sec.~\ref{sec:results} to examine the production mechanisms and other pertinent characteristics of reactor-correlated backgrounds. In particular, in reference to the $\gamma$-ray lines identified in Table~\ref{tab:isotopes}, it is apparent that reactor-produced neutrons play a crucial role in elevated reactor-correlated $\gamma$-ray fluxes at the locations examined. Furthermore, elevated neutron rates at the locations are themselves a source of background for many experiments. In this section we describe how the physical characteristics of a reactor facility influence the observed background and make a qualitative comparison of the three facilities examined here.

The observation of significant spatial and/or temporal variations in reactor-correlated background rates at each near location  further illustrates the mechanisms at work. Here we give several indicative examples of the spatial variation encountered at each site. These spatial variation studies were somewhat ad-hoc, being dependent upon the particular configurations of detectors and shielding materials  available at each site at the time of the measurements. We qualitatively associate these variations  with the following characteristics of reactor facilities:
\begin{itemize}
\item \textbf{Local concentrations of water, polyethylene, or iron.}  In locations with thermal neutron leakage from the core or beam lines, neutron interaction with these materials will produce prompt high-energy  $\g$-rays;
\item \textbf{Plant piping carrying water that has been exposed to high neutron fluxes.}  Activated $^{16}$O or trace impurities in water can be transported outside of shielding walls. We do not believe significant $^{16}$N was observed via this pathway due to the relatively low flow in visible pipes and the short $7.3$~s half-life involved. This mechanism can cause activity to be transported a considerable distance from the reactor, depending upon the details of the site configuration;
 \item \textbf{Shielding walls or penetrations between the reactor and the measurement location.}  The shielding between a location of interest and a high intensity background source (e.g. pipe carrying a large amount of primary coolant, or indeed the reactor core itself) may not attenuate the emitted $\g$-ray flux to levels comparable with natural background. Seams or piping penetrations in shielding walls may allow a scattering path for $\g$-rays that results in a localized ``hot-spot'';
 \item \textbf{Experiments or other devices attached to neutron beamlines.}  In facilities that support neutron scattering experiments, beamlines or experiments themselves can be significant sources of scattered neutrons and/neutron capture $\g$-rays. Large time variation can be expected from such sources during reactor on periods as experiments are reconfigured.
\end{itemize}
While the measurements here support the mechanism described above, we note that a more detailed survey at the site selected to host an experiment would be required to fully characterize the $\g$-ray background fields in order to optimize a shielding configuration.

\begin{figure}[tb]
\centering
\includegraphics*[width=0.48\textwidth]{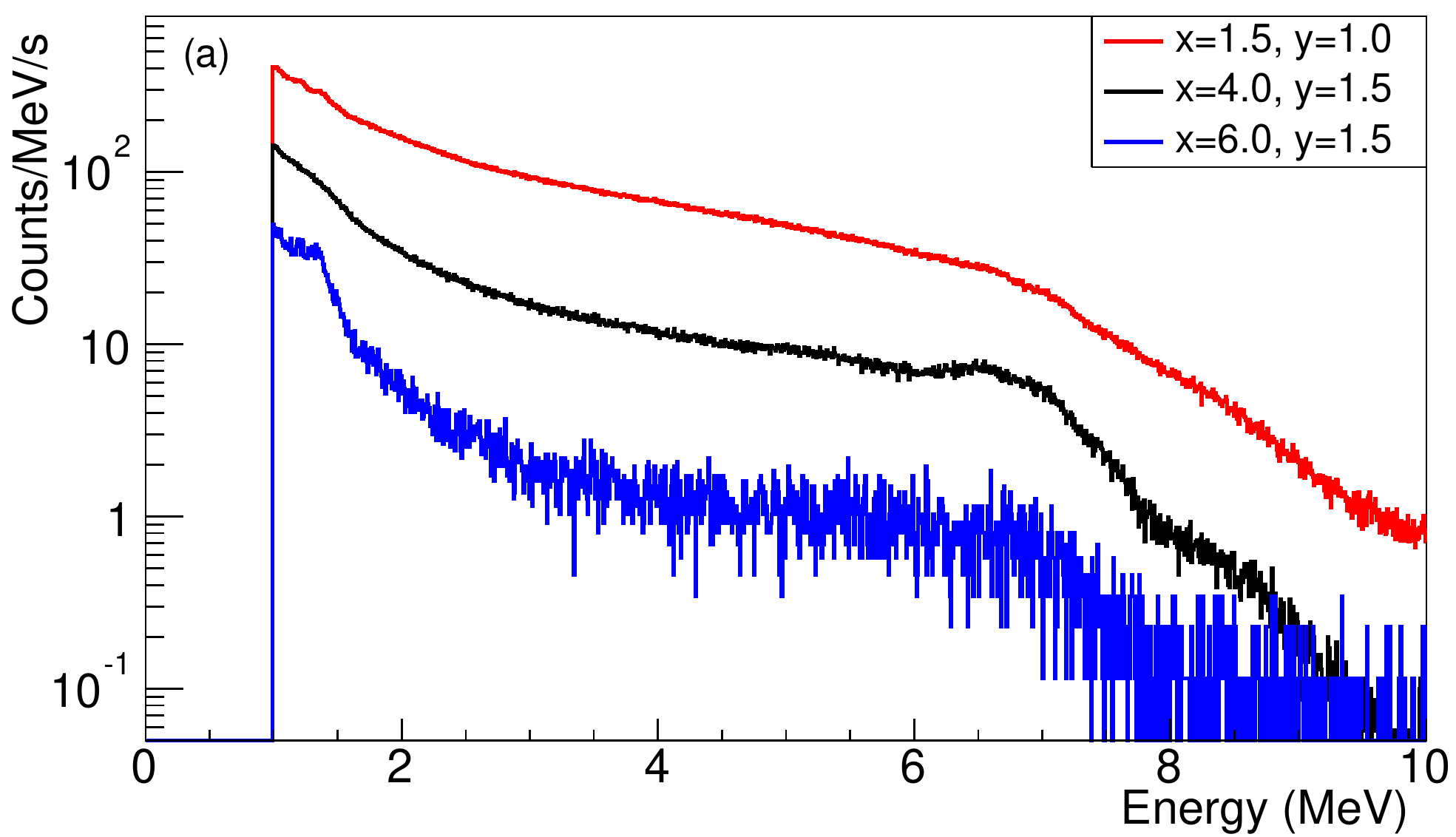}
\includegraphics*[width=0.48\textwidth]{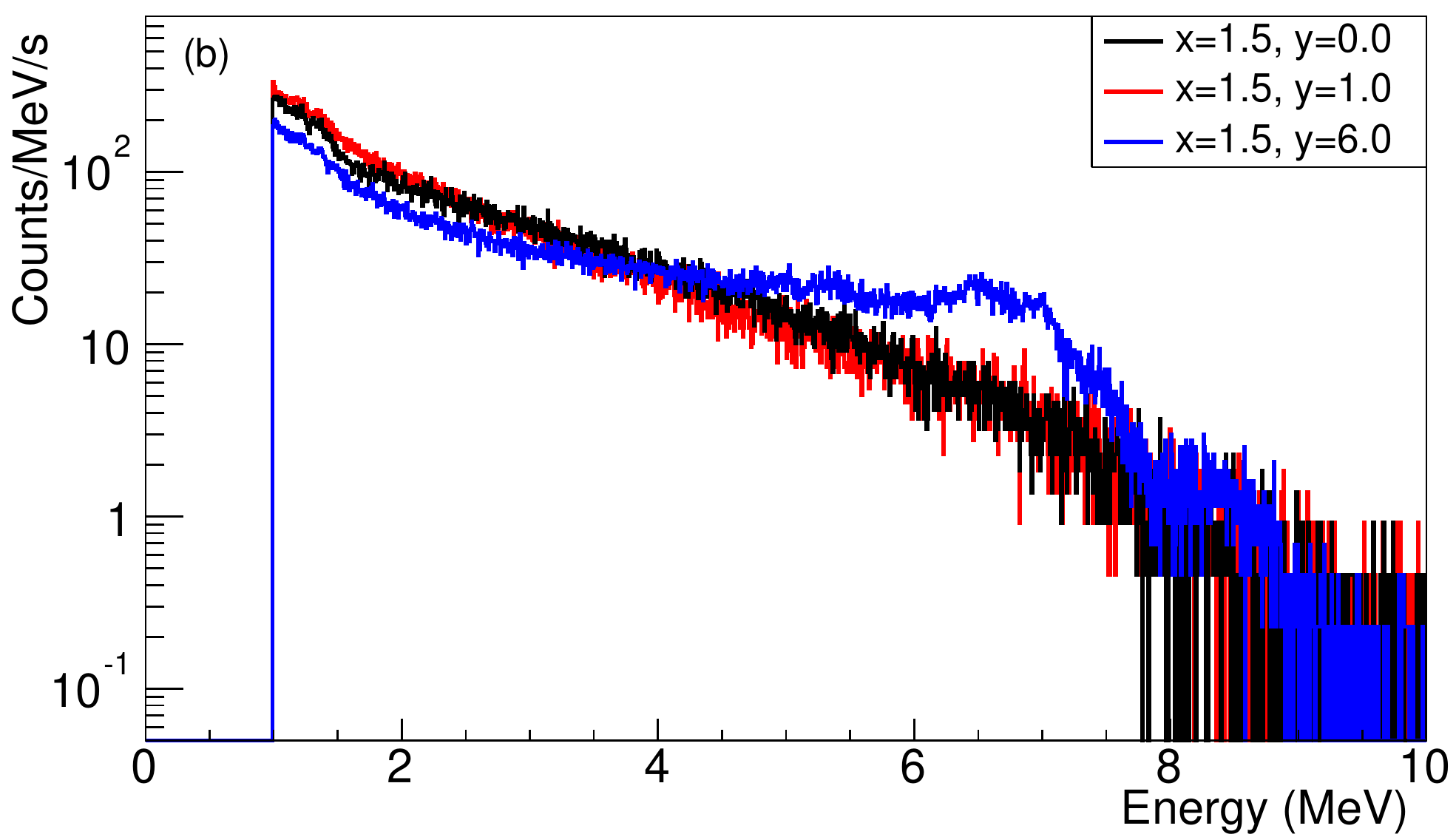}
\caption{The $\g$-ray background at the \HFIR{} near location has significant spatial variations when the reactor is operating at nominal power. These include variation with  (a) distance from the wall closest to the reactor and (b) position along that wall. 
}
\label{fig:hfirWall}
\end{figure}

\begin{figure}[tb]
\centering
\includegraphics*[width=0.48\textwidth]{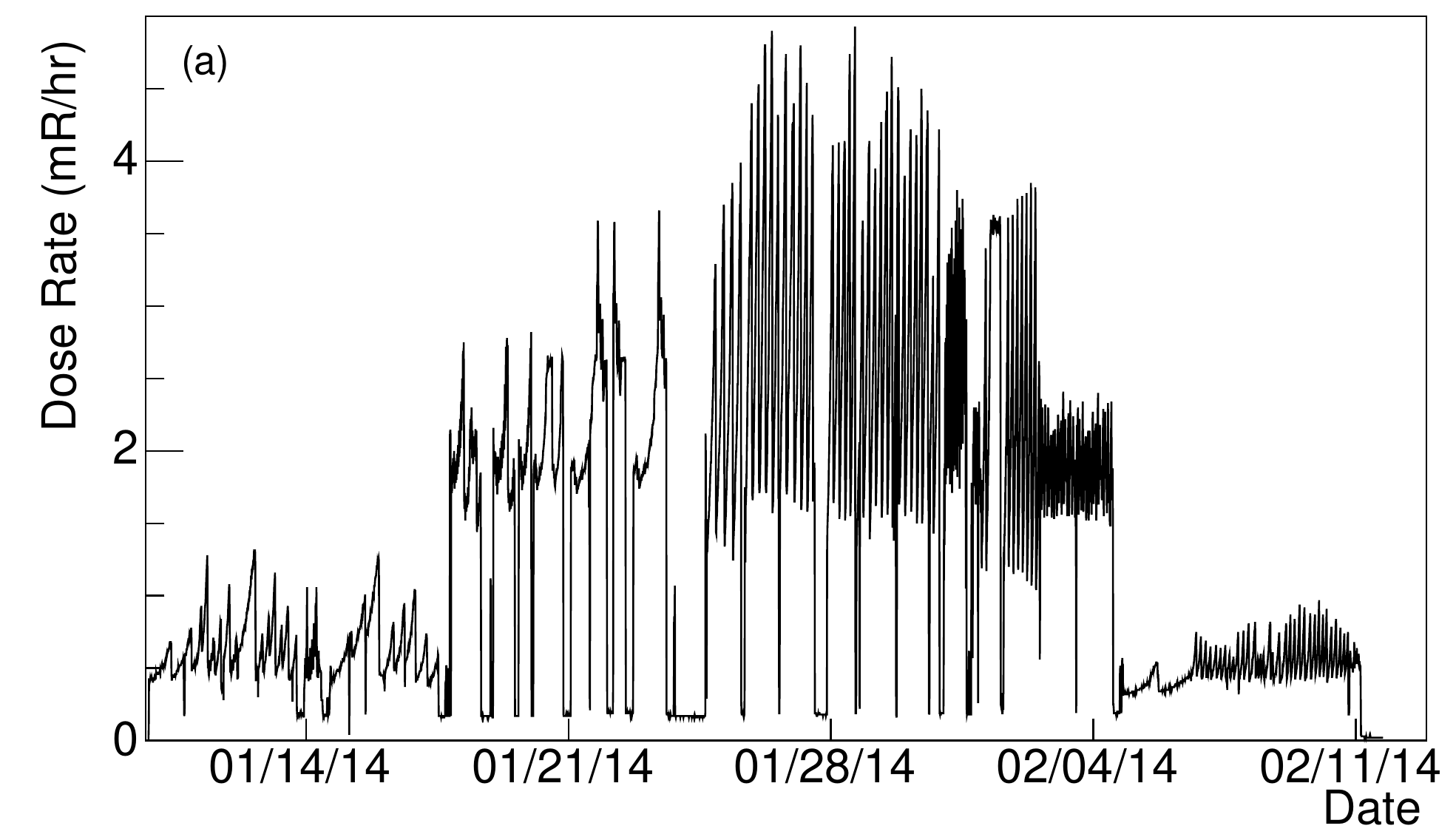}
\includegraphics*[width=0.48\textwidth]{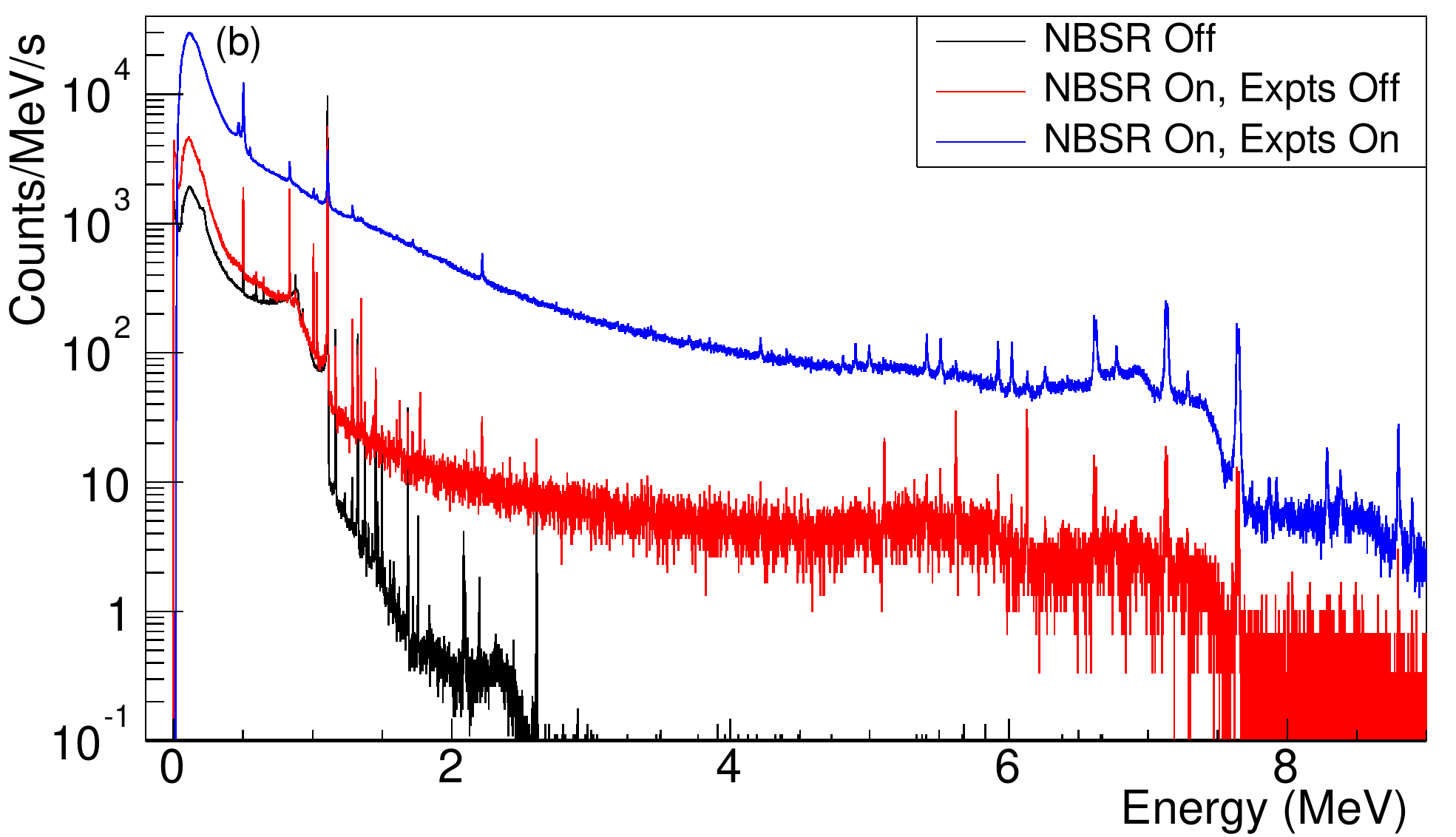}
\caption{(a) The radiation dose rate due to background $\g$-rays as a function of time at the \NBSR{} near location. (b) $\g$-ray spectra measured with the NIST reactor-off, with the reactor-on, and with the reactor-on and an adjacent neutron scattering experiment operating.}
\label{fig:nistTimeVary}
\end{figure}

At \HFIR{} considerable variation was found in the $\g$-ray flux with respect to proximity to the wall nearest the reactor (Fig.~\ref{fig:hfirLayout} and Fig.~\ref{fig:hfirWall}). This wall contains several penetrations which might be the source of the observed  increase, or may simply not be sufficiently thick to completely attenuate emissions from activated water in the reactor pool.  Collimated measurements indicate a higher  flux in the direction of the wall, but not other directions. Removal of lead shielding about the NaI(TI) detector in the vertical direction has little effect on the observed rates while removal of shielding in the direction of the wall closest to the reactor increased the rate by a factor of~$4.5$. Similarly, the detection rate measured with the unshielded NaI(Tl) detector exhibits a steep fall off as the distance from this wall was increased (Fig.~\ref{fig:hfirWall}a).

\begin{figure}[tb]
\centering
\includegraphics*[width=0.48\textwidth]{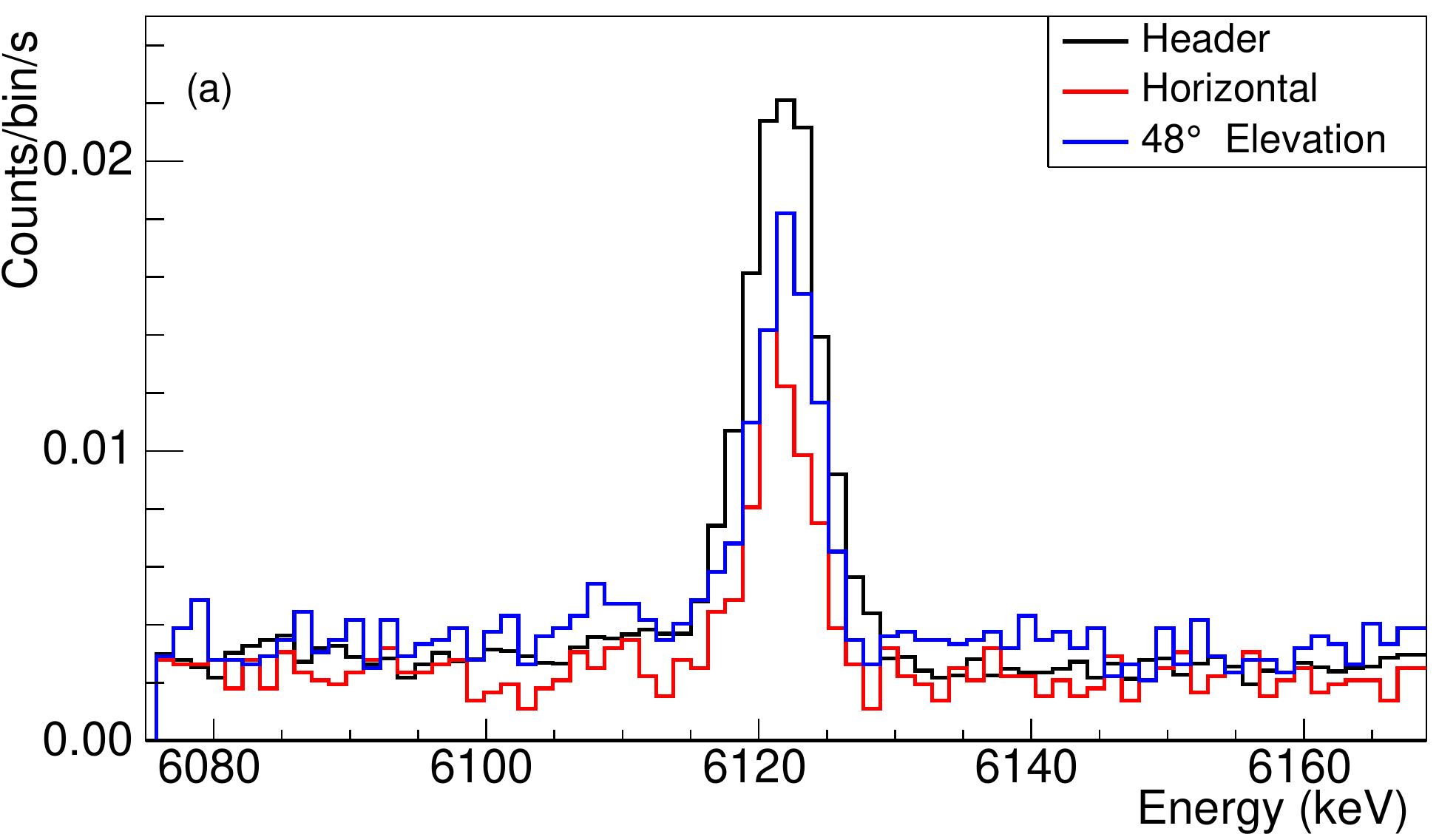}
\includegraphics*[width=0.48\textwidth]{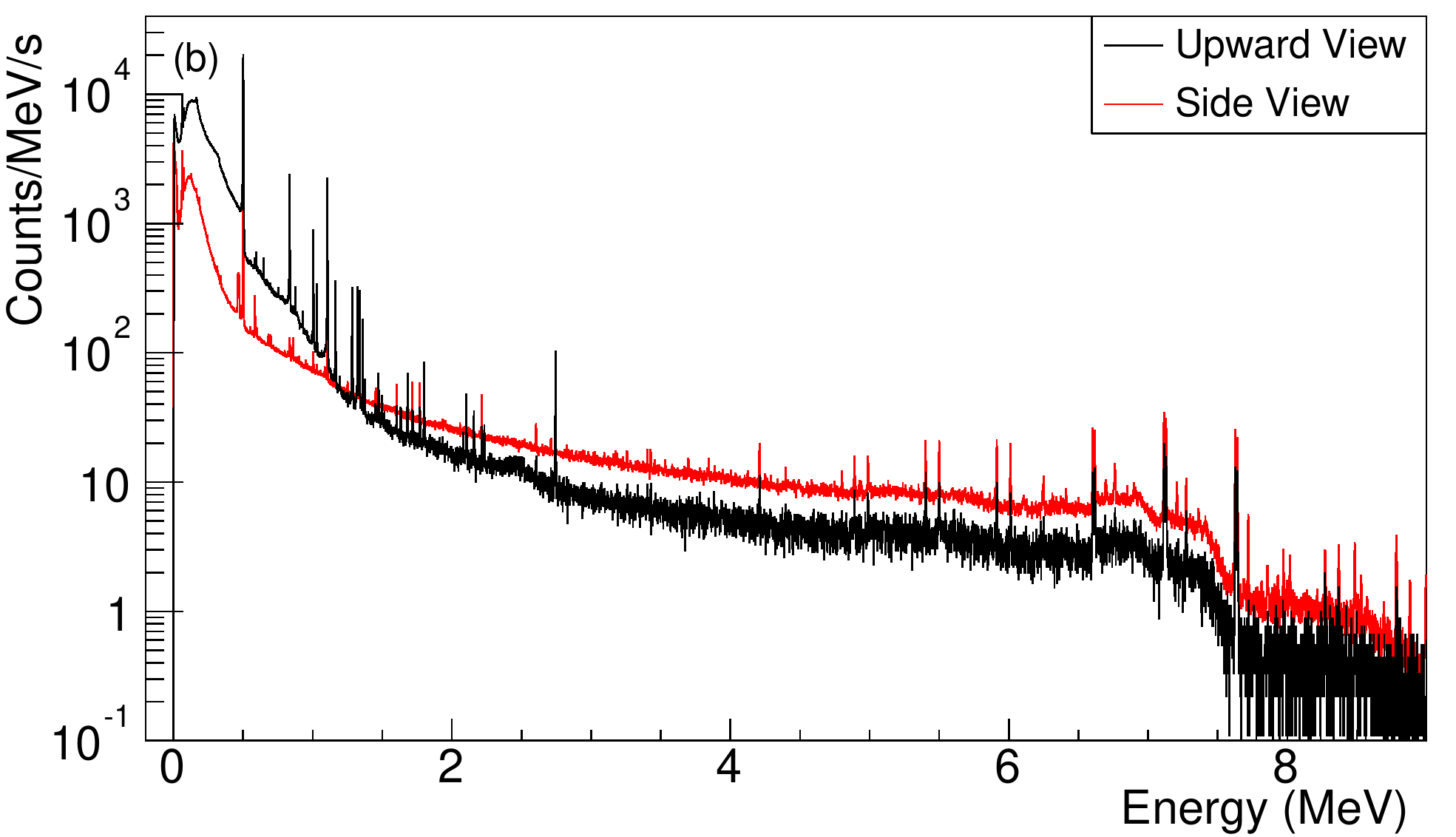}
\caption{Examples of spatial variations observed at the \NBSR{}. (a) The intensity of the observed $^{16}$N line is correlated with header pipes carrying cooling water in the reactor thermal shield. (b) The intensity of the $^{56}$Fe lines depends on the field of view with higher rates observed in the horizontal plane.}
\label{fig:nistSpatial}
\end{figure}

As described in Sec.~\ref{sec:thermalNeutron}, an increased thermal neutron flux was observed to one side of the \HFIR{} near location. One effect of this can be observed in measurements made with the unshielded NaI(Tl) detector. At the locations with elevated thermal neutron flux, high-energy capture $\g$-rays are more prominent (Fig.~\ref{fig:hfirWall}b). Examination of the continuum portion of these spectra also sheds light on the source of this $\g$-ray background. The similarity in the intensity and shape of this continuum at locations along the wall closest to the reactor suggest that the source is not localized to a single penetration or narrow leakage path. Instead, it appears likely that the entire length of the wall is emitting downscattered $\g$-rays from the reactor pool.

\begin{figure}[tb]
\centering
\includegraphics*[width=0.48\textwidth]{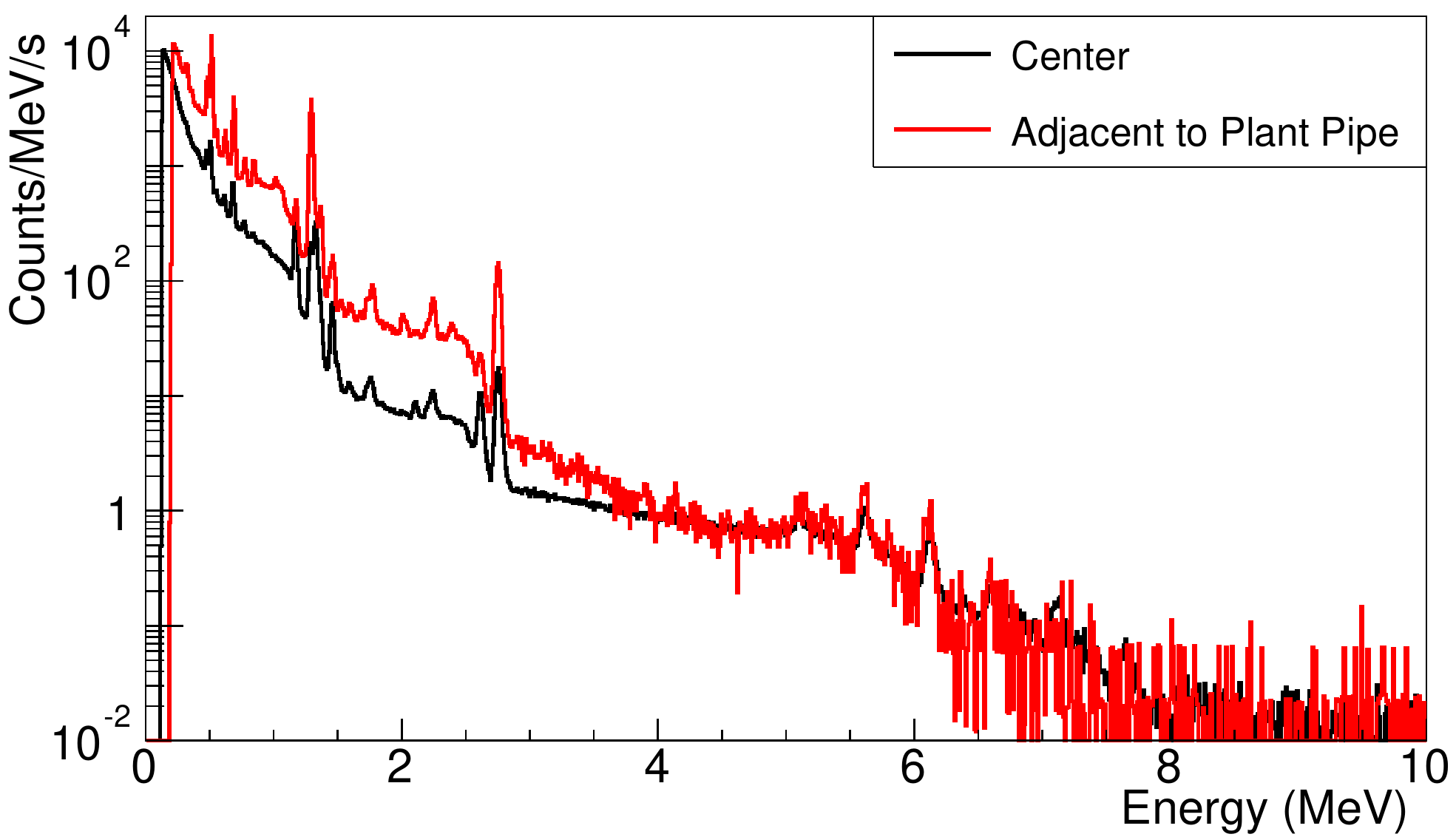}
\caption{The $\g$-ray flux due to $^{24}$Na ($\approx1.3$ and $\approx2.7$~MeV) increases substantially when the LaBr$_3$(Ce) detector is moved from the center of the \ATR\ near location to be adjacent to plant piping in the area and the reactor is operating at nominal power.}
\label{fig:atrNearPipe}
\end{figure}

At \NBSR{} a number of spatial and temporal variations were observed. The variation in $\g$-ray background due to the operation of an adjacent neutron scattering instrument (MACS) is displayed in Fig.~\ref{fig:nistTimeVary}a showing results from a logging dosimeter~\cite{gammaTracer} that was used to record the $\g$-ray dose rate in the near location over a $1$~month period. During the period spanning 1/18/14--1/31/14 the instrument was operated with cadmium thermal neutron shielding which increased the background rate (use of boron thermal neutron shielding should result in a substantial decrease in $\gamma$-ray dose).  The detailed time structure seen in Fig.~\ref{fig:nistTimeVary}a is the result of different configurations of the neutron scattering instrument. The average dose during the period shown was $13$~$\mu$Sv/h.  The effect of the operation of the adjacent instruments on the $\g$-ray energy spectrum is also displayed in Fig.~\ref{fig:nistTimeVary}b. Large increases in downscattered continuum background and $^{57}$Fe emissions due to thermal neutron capture on structural steel are observed.

Significant spatial variation of the $\g$-ray background was also observed at NBSR.  
A primary source of background was identified as coming from the thermal shield cooling-water lines located above the proposed near location as indicated in Fig.~\ref{fig:nistLayout}a.  The dominant 6.128 MeV line is clearly seen in Fig.~\ref{fig:nistRawSpec}.  Measurements taken with $5~$cm thick lead apertures that restricted the detector field of view to approximately 30$^\circ$ demonstrated qualitatively that these lines are originating in the header assembly.  This is shown in Fig.~\ref{fig:nistSpatial}a.   As can be seen in Fig.~\ref{fig:nistNearLocation}, this source of background is partially shielded by the biological shield and illuminates roughly half of the potential near location.   Fig.~\ref{fig:nistSpatial}b  compares data taken with two different apertures at a position roughly 50 cm from the face of the reactor biological shielding: a 2$\pi$ upward view and an arrangement that views primarily the horizontal plane.  The spatial dependence and lack of significant downscattering in these data suggest that the dominant source of higher-energy $\gamma$-rays are thermal neutron capture on the steel shielding surrounding the adjacent beamlines, consistent with the interpretation of Fig.~\ref{fig:nistTimeVary}b. 
They also indicate that the low energy part of the spectrum is dominated by overhead sources.  The fact that the $\g$-ray backgrounds at NIST are highly directional, and in some cases, well localized, suggests that targeted shielding may be particularly effective.

At \ATR{}, both the near and far locations have line-of-sight to piping carrying small amounts of water that has been in close proximity to the core. The length of these pipes and the relatively low flow rates they carry result in there being little $^{16}$N activity observed from them. However, as demonstrated in Fig.~\ref{fig:atrNearPipe}, proximity to these pipes results in a substantial increase in the observed $^{24}$Na $\g$-ray flux, most notably the line at $2.754$~MeV. At the near location a piping manifold, used for monitoring water flows near control devices in the core, on the wall closest to the reactor is therefore the likely source of the observed $^{24}$Na activity. At the far location, a small ceiling mounted pipe carrying primary coolant diverted to a power monitoring system is the $^{24}$Na source. 

\section{Case Study: the PROSPECT Experiment at the High Flux Isotope Reactor}
\label{sec:prospect@hfir}

After an assessment process that considered the background characteristics described here in addition to logistical and engineering considerations, the PROSPECT collaboration decided to pursue PROSPECT Phase~I at \HFIR{}~\cite{Ashenfelter:2013oaa}. Therefore the background characteristics of the \HFIR{} near location were examined in greater detail. Described here are $\gamma$-ray surveys to more fully determine spatial variations in the background $\gamma$-ray flux, studies to develop detector shielding appropriate for the background encountered at this location, and results from a prototype detector deployment to demonstrate background reduction. 

\subsection{Detailed Spatial $\gamma$-ray Surveys}
\label{sec:hfirMap}

The previously described NaI(Tl) measurements used for  comparing reactor sites were unshielded measurements made within the expected near detector footprint. To better identify $\gamma$-ray background sources at \HFIR{}, it was necessary to explore a wider range of positions with unshielded and shielded detectors. Over 200 NaI(Tl) measurements were made during three background measurement campaigns at \HFIR{} with the reactor was operating at a thermal power of $85$~MW. The grid shown in Fig.~\ref{fig:hfirThermalNeutron}  provides a convenient reference for comparison of different positions. The $y$-axis measured distance along the wall surrounding the reactor water pool, the $x$-axis measured the distance from the wall and the $z$-axis measured the height above the floor, with $y = 0.0$ being in line with the reactor core. The complex spatial variations observed in Sec.~\ref{sec:rxCharacteristics} indicate that multiple sources contribute to the background at any given location. Lead shielding was used  to restrict the angular acceptance of the NaI(Tl)  during some measurements to indicate the spatial distribution of these background $\gamma$-ray sources. 

\begin{figure}[tb]
\centering
\includegraphics[clip=true, trim=5mm 0mm 10mm 10mm,width=0.48\textwidth]{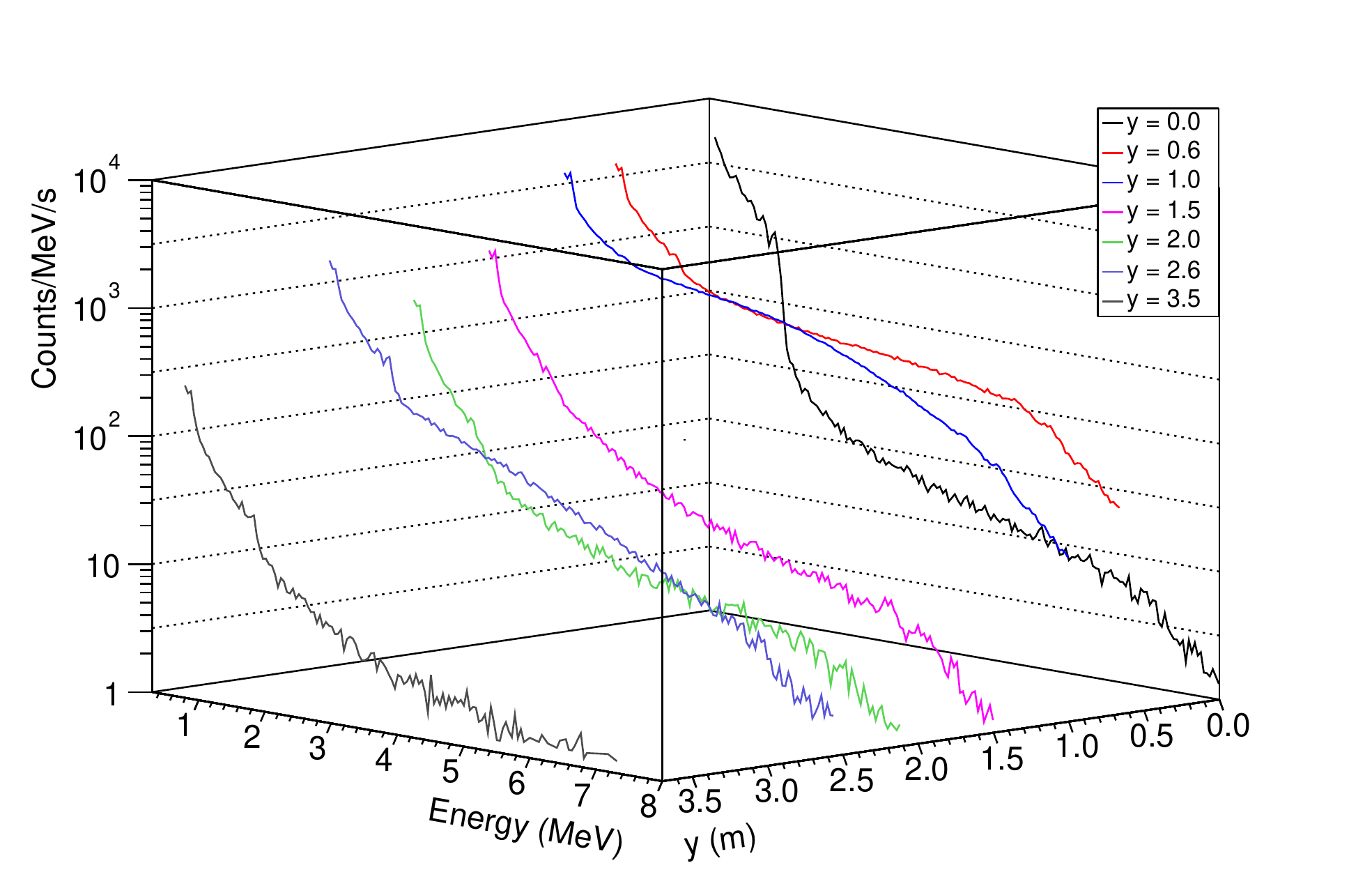}\hfil%
\caption{Measured energy spectra for an unshielded NaI(Tl) detector at different $y$ locations along the wall surrounding the reactor pool, with the reactor operating at nominal power. All data are taken $0.1$~m from the wall and $1$~m above the floor ($x=0.1$, $z=1.0$). Several locations exhibit significantly higher rate, associated with penetrations in the wall described in the text.  }
\label{fig:unshielded_xscan}
\end{figure}

%\clearpage

The differing rate and energy spectra of the background sources along the reactor pool wall are illustrated in Fig.~\ref{fig:unshielded_xscan}, where measured $\gamma$-ray spectra taken at different $x$~positions are plotted. Two prominent Òhot spotsÓ are evident. A pipe directly through the concrete wall to the reactor water pool near $x= -0.04$~m is an intense source of lower-energy $\gamma$-rays ($\leq 1.5$~MeV). An unused beam tube between $y = 0.66$--$1.0$~m, pointing almost directly back to the reactor core, is the dominant source of higher-energy $\gamma$-rays ($\geq 2$~MeV) despite being filled with a concrete plug.  Less prominent Òhot spotsÓ interrupt the general reduction in rate with increasing $y$ at $y = 2.56$~m (a notch in the wall) and $y= 3.0$~m (above another unused beam tube in the floor).

\begin{figure}[tb]
\centering
\includegraphics*[clip=true, trim=0mm 20mm 10mm 10mm,width=0.48\textwidth]{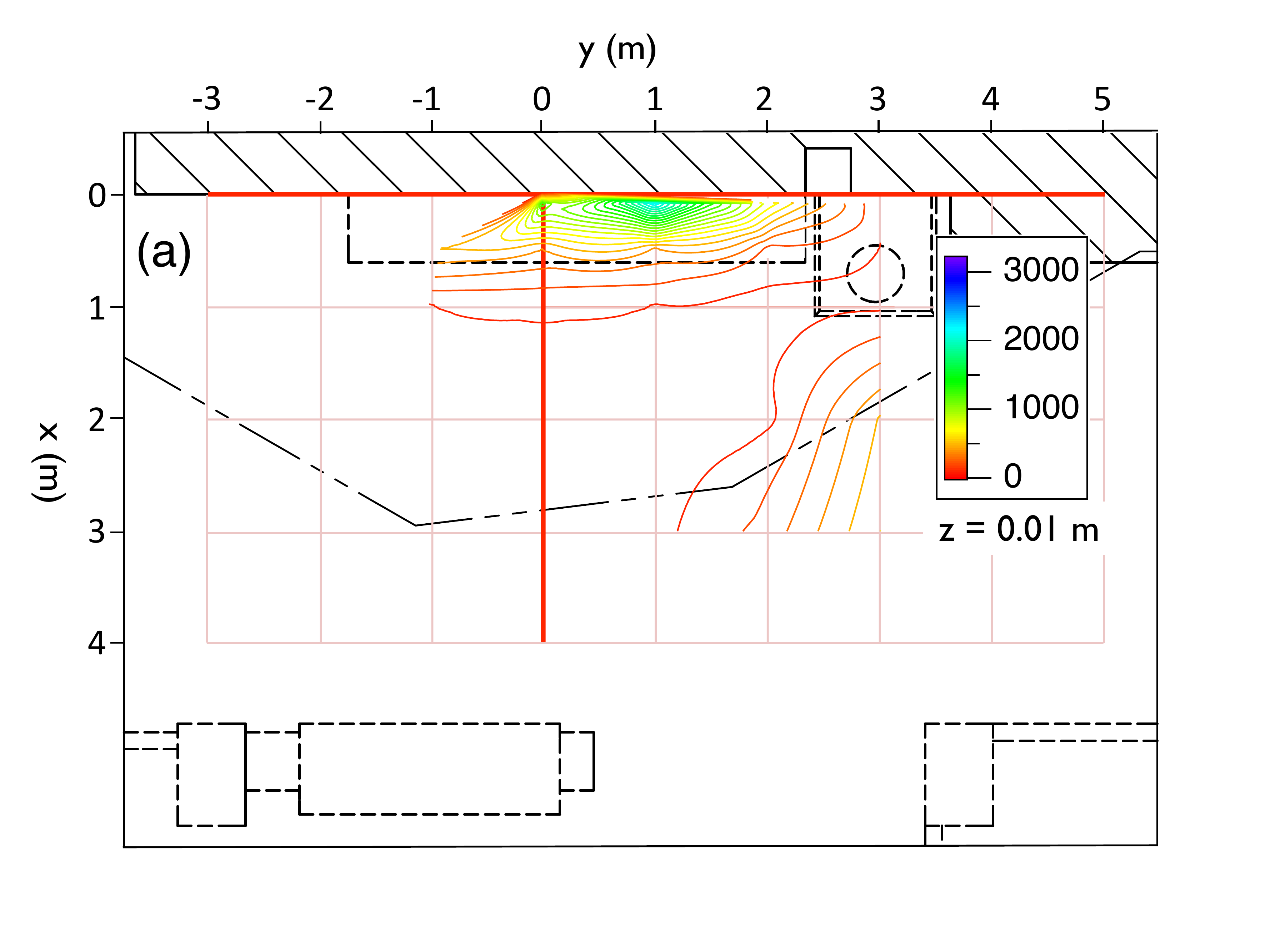}
\includegraphics*[clip=true, trim=0mm 20mm 10mm 10mm,width=0.48\textwidth]{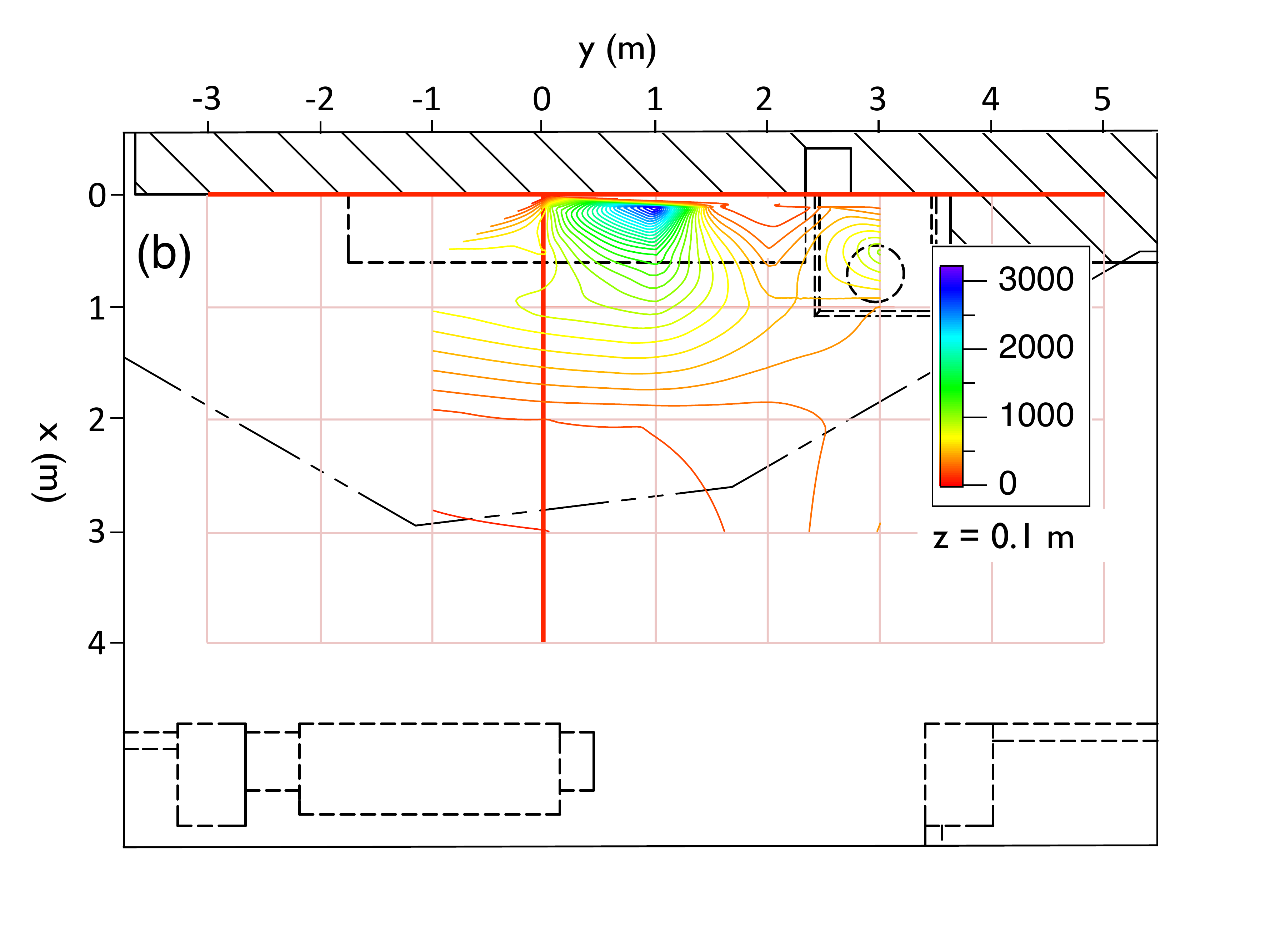}
\caption{Measured count rates (s$^{-1}$) over the energy range of $1$--$10$~MeV, for an unshielded NaI(Tl) detector either 10 cm above the floor (a) or 100 cm above the floor (b). The reactor was operating at nominal power. The reactor core is centered at $(x,y,z) = (-4.06,0,-3.85)$.}
\label{fig:V_scan}
\end{figure}

General trends in the spatial variation of $\gamma$-ray backgrounds can be seen in Fig.~\ref{fig:V_scan}, which displays integrated $\gamma$-ray counting rates between $1$--$10$~MeV as a function of position. 
Contour plots at two different heights above the floor are shown: (top) z = 0.1 m and (bottom) z = 1.0 m.
Variation along the $y$-axis close to the wall (x = 0.1 m) follows the trends seen in Fig.~\ref{fig:unshielded_xscan}. 
Integrated rates decrease  along the $y$-axis  as the distance from the reactor increases, consistent with  the spectra shown in top of Fig.~\ref{fig:hfirWall}. 
The variation is most pronounced close to the floor as can be seen comparing Fig.~\ref{fig:V_scan}a and Fig.~\ref{fig:V_scan}b.
This large reduction in background rate is attributed to the large  concrete support monolith under this level whose outline can 
be seen as a dashed line in  Fig.~\ref{fig:V_scan} or in the elevation view of Fig.~\ref{fig:hfirLayout}.  
Backgrounds from the water pool much below the level of the floor are strongly suppressed.

Close to the reactor wall both the average $\gamma$-ray energy and rate are significantly lower 2 meters above the floor than at 1 meter.
Rates below 1.5 MeV are a factor of 10 lower while rates $\approx$3-6 MeV are nearly 100 times lower.
However, further from the wall (x~$\geq$~0.7 meter), the spectra at z = 1 and 2 meters are similar while rates 
just above the floor (monolith) are very low.  These distributions imply that higher-energy $\gamma$-rays from the wall are emitted roughly
at 45$^{\circ}$ to the vertical i.e. along the unused beam tube.   

\begin{figure}[tb]
\centering
\includegraphics[width=0.45\textwidth]{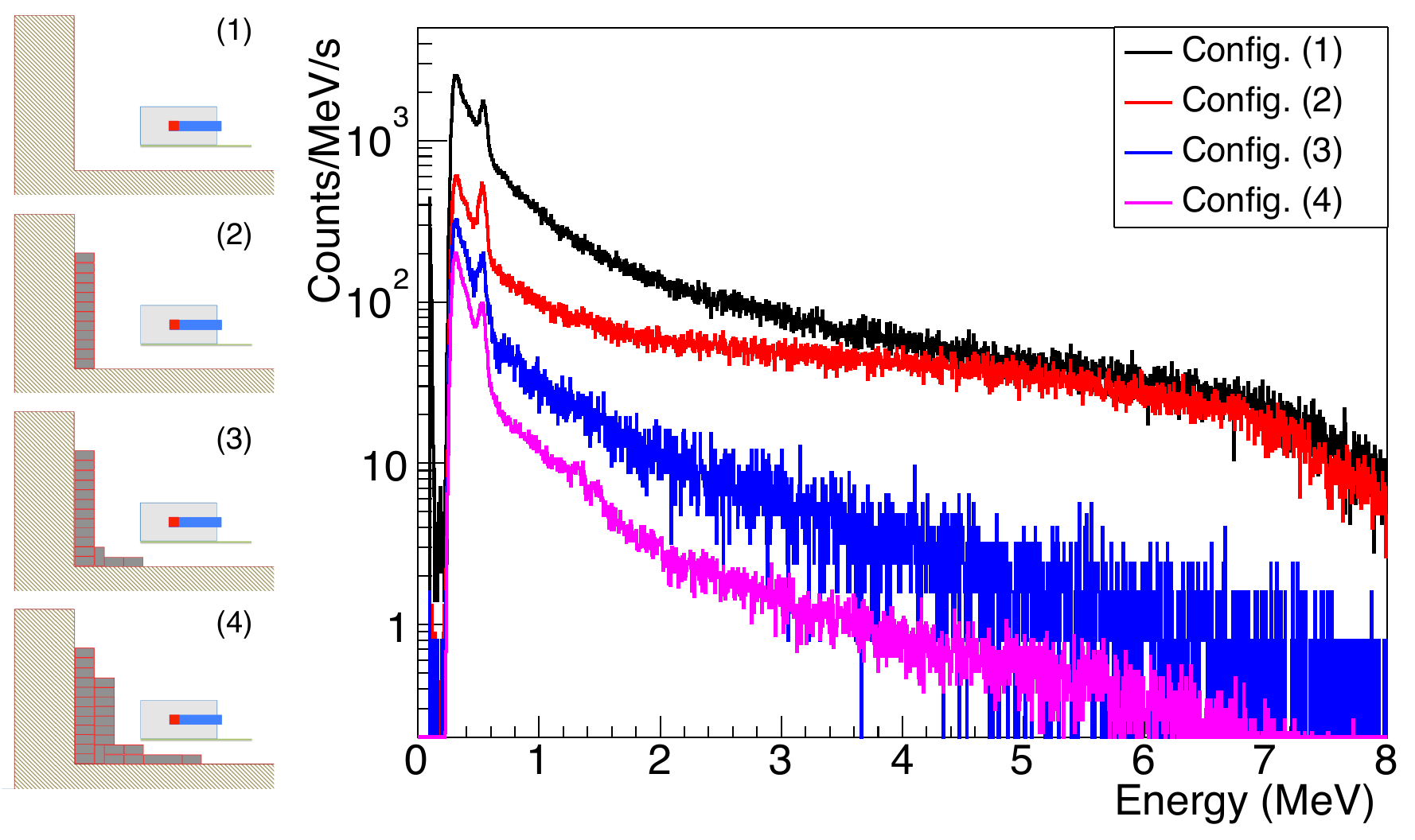}\hfil%
\caption{Measured energy spectra for a NaI(Tl) detector inside a horizontal lead collimator placed at $x = 0.5$~m, $z = 0.2$~m for different configurations of a 102~cm wide ($y$-axis in Fig.~\ref{fig:V_scan}) lead wall, with the reactor operating at nominal power. This location is in front of a localized $\g$-ray background source. }
\label{fig:Wall}
\end{figure}

\begin{table}[tb]%
\begin{center}
\begin{tabular}{l|d{1}|d{1}l} 
\hline
Configuration	& \multicolumn{2}{c}{Rate (Hz)} \\ 
(dimensions in cm)  			& \multicolumn{1}{c|}{$1$-$3$~MeV} 	&\multicolumn{1}{c}{$3$-$10$~MeV} \\ \hline
No Wall (1) 							&$512.4$ &$246.3$ \\
Add wall: $10\times102\times51$ (2)		&$169.8$ &$168.4$\\
Add floor: $25\times102\times5$ (3)		&$52.5$&$15.2$\\ 
Add to floor: $25\times102\times10$		&$32.2$ &$10.5$ \\
Add to wall: $20\times102\times51$		&$28.5$ &$12.7$ \\
Extend floor: $30\times20\times5$ (4)	&$15.5$ &$3.0$ \\
\hline
 \end{tabular}
    \caption{Integrated background rates for energy ranges $1$--$3$~MeV and $3$--$10$~MeV for sequential augmentation of a lead shielding wall. The shielding spans the range $y = 0.6$--$1.0$~m against the wall indicated in at $x=0$ in Fig.~\ref{fig:V_scan}. Wall dimensions are given as $x\times y \times z$ values, with $(x,y,z)$ directions also as indicated in Fig.~\ref{fig:V_scan}. Several configurations are pictorially represented in Fig.~\ref{fig:Wall}.}
 \label{tab:Wall}
    \end{center}
\end{table}

Measurements were taken with the NaI(Tl) detector inside  a 10 cm thick rectangular lead well, intended to attenuate all $\gamma$-rays not coming from directly beneath the detector.  An intense local hot spot is observed near $y = -0.2$~m near the wall closest to the reactor. Away from the wall rates were uniformly  low over the shielding monolith. Background rates increased with the detector over the relatively thin 15~cm concrete floor outside of the monolith footprint ($x\gtrsim2$~m). The level beneath the location being examined contains multiple neutron beam lines. Scattered beam neutrons interacting with structural materials in that level or the floor itself are thought to be cause of the increased $\g$-ray background rates observed past the shielding monolith.

A study of shielding effectiveness was conducted by varying the configuration of a lead wall in front of the beam tube at $y = 0.6$--$1.0$~m and measuring background $\g$-ray rates (Fig.~\ref{fig:Wall}). The NaI(Tl) detector was placed between two 10~cm thick lead walls oriented perpendicular to the lead wall, thus limiting the detector acceptance in the horizontal plane in directions other than the wall. Table~\ref{tab:Wall} gives the background rates summed over the energy ranges $1$--$3$~MeV and $3$--$10$~MeV for each wall configuration. Fig.~\ref{fig:Wall}  shows the background energy spectra at selected configurations. 

With the detector $0.5$~m from the wall, a $10$~cm thick lead wall reduced the $\g$-ray detection rate at energies below $3$~MeV by a factor of~3. Extending the wall onto the floor by $25$~cm  significantly reduced the rate of higher-energy $\g$-rays by as much as a factor of ten.  Doubling the thickness of the floor layer further reduced rates, while doubling the thickness of the vertical wall had little effect. Extending the floor bricks another $20$~cm lowered the high-energy $\g$-ray rate by an additional factor of four. 

Both  background sources and shadows were observed during these studies. The solid concrete monolith effectively blocks any background sources directly beneath the location under consideration. Penetrations or relatively thin sections in concrete structures were associated with higher backgrounds. In particular the beam tube near $y = 1.0$~m was the dominant source of high-energy background.  Less intense sources of higher-energy $\g$-rays were likely to be associated with higher neutron fluxes at large $y$ ($y\gtrsim2$ m) or off the monolith ($x\gtrsim3$ m).  Accordingly, PROSPECT aims to build a localized lead shielding structure against the wall and floor closest to the reactor and then remeasure these background distributions before designing detector shielding.

\begin{figure}[tb]
\centering
\includegraphics[clip=true, trim= 00mm 0mm 0mm 90mm,width=3.0in]{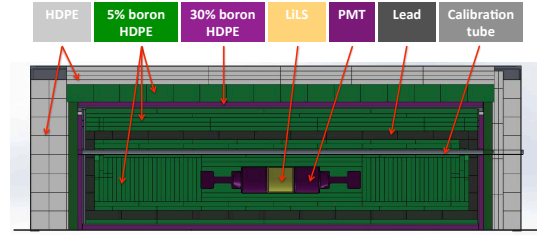}
\caption{PROSPECT2 as installed at HFIR. The 5" cylindrical LS detector (yellow), PMTs and HV bases (purple) are surrounded by 
 5\% borated polyethylene sheets (green),  lead (dark grey),  more 5\% borated polyethylene sheet, an Al containment box ( grey), 30\%  borated polyethylene sheet (purple), more 5\% borated polyethylene sheet, and  polyethylene sheet (light grey). }
\label{fig:p2Shield}
\end{figure}

\subsection{Deployment of the PROSPECT2 Prototype at the HFIR Near Location}

To test the efficiency of shielding and provide data for simulation validation, a prototype detector  
was deployed at the \HFIR{} near location. The detector is a right cylindrical acrylic vessel with an internal diameter of $12.7$~cm containing $1.7$~liters of organic liquid scintillator doped with $0.1\%$ by weight $^6$Li (LiLS). Since the active volume is almost $2$~liters, the device is denoted as \ptwo{}; later prototypes of larger size follow a similar naming convention. Optical readout was via two $5$~inch PMTs (ET9823KB~\cite{et}) coupled directly to each face of the vessel with EJ550 optical grease~\cite{eljen550}. All other sides are covered with a diffuse reflective TiO$_2$ paint. Each PMT is readout using a CAEN~V1720 waveform digitizer~\cite{caen1720} sampling at 250~MHz with 12 bits per sample.

\begin{figure}[tb]
\centering
\includegraphics*[width=0.48\textwidth]{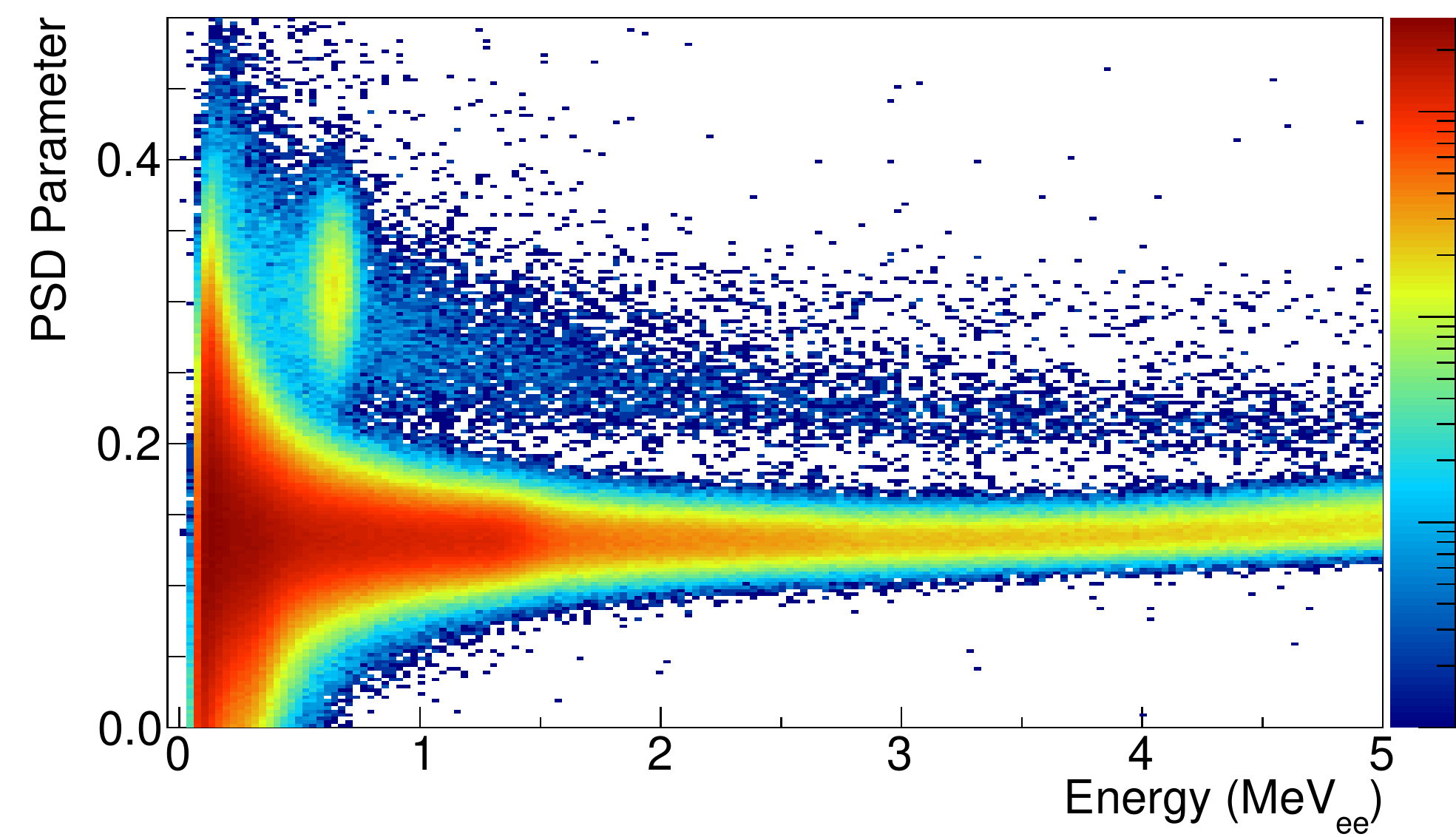}
\caption{PSD parameter vs. electron equivalent energy for the shielded \ptwo{} detector operated at \HFIR{} with the reactor off. }
\label{fig:p2_psd}
\end{figure}

The \ptwo{} detector was deployed within a multilayer shield enclosure designed to reduce both $\g$-ray and neutron fluxes. 
A diagram of the shielding configuration is shown in Fig.~\ref{fig:p2Shield}. 
The $\geq50$~cm thick shield consists of  (from the outside in) 10-20~cm of high-density polyethylene, $\approx$20~cm of 5\% borated polyethylene, 
2.5~cm of 30\% borated polyethylene, 5-10~cm of lead, and finally 10~cm more of borated polyethylene.
In addition, a 10-cm-thick lead shield was placed over the beam port describe in Sec.~\ref{sec:hfirMap} to locally shield that intense background source.

The guiding concepts behind this design are:
\begin{itemize}
\item{Thermalize and capture low-energy neutrons in an outer layer of borated polyethylene to reduce high-energy capture $\gamma$-rays;}
\item{Use a layer of high-Z material to stop external $\gamma$-rays as well as those produced from neutron capture in the outer borated polyethylene later;}
\item{Thermalize and capture any neutrons produced from cosmic rays interactions in the high-Z material in a second layer of borated polyethylene.}
\end{itemize}

The combination of $^6$Li doping and PSD in the \ptwo{} detector allows the same device to simultaneously measure $\gamma$-ray, fast neutron recoil, and neutron capture rates. As with the stilbene detector described in Sec.~\ref{sec:stilbene} a PSD parameter is determined by taking the ``tail'' to ``full'' ratio of the PMT pulse. As demonstrated in Fig.~\ref{fig:p2_psd}, interactions of each of these particle types fall in a different region of an PSD parameter vs. energy plot. We use this capability to assess the effectiveness of the shield enclosure at reducing reactor generated backgrounds from each of these particle types. Data sets totaling $109$~hours with the reactor operational at a thermal power of $85$~MW and $348$~hrs with the reactor off were collected.

\begin{figure}[tb]
\centering
\includegraphics*[width=0.45\textwidth]{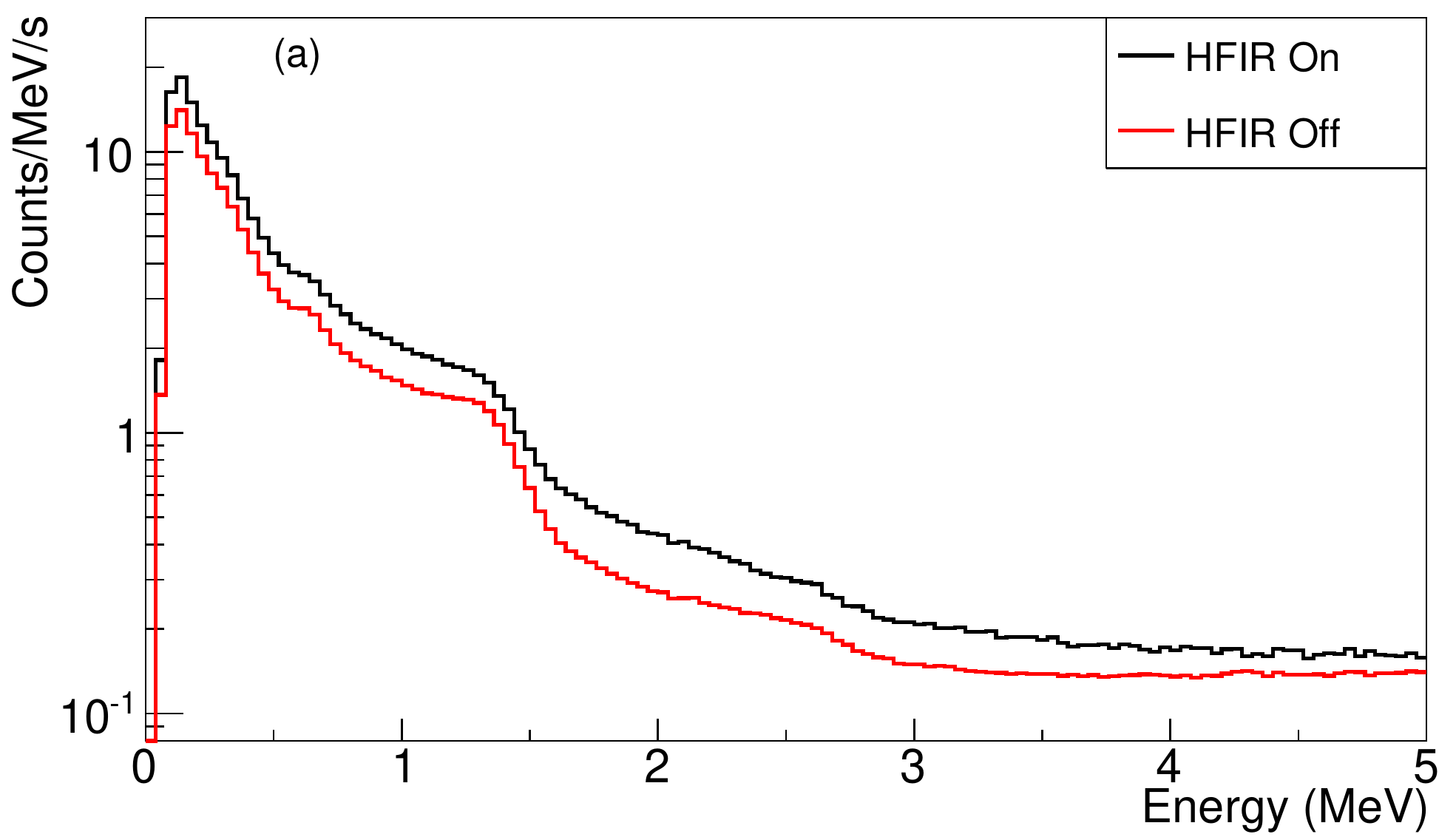}
\includegraphics*[width=0.45\textwidth]{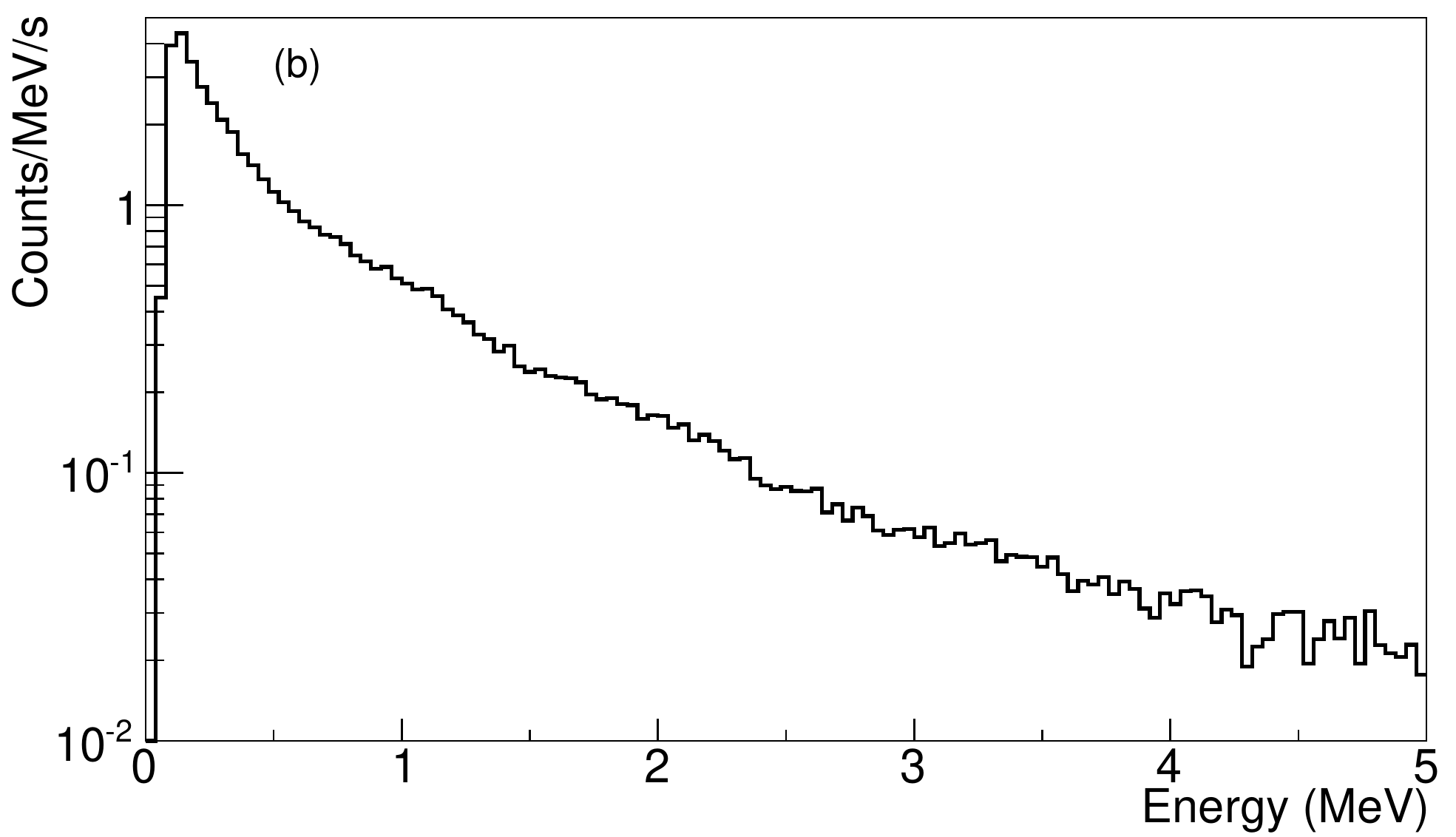}
\caption{(a) The energy spectra of depositions in the \ptwo{} detector are compared with the reactor on and off. (b) The residual after subtraction of the reactor-off spectrum from the reactor-on. }
\label{fig:p2_energy_spectra}
\end{figure}

\begin{figure}[tb]
\centering
\includegraphics*[width=0.45\textwidth]{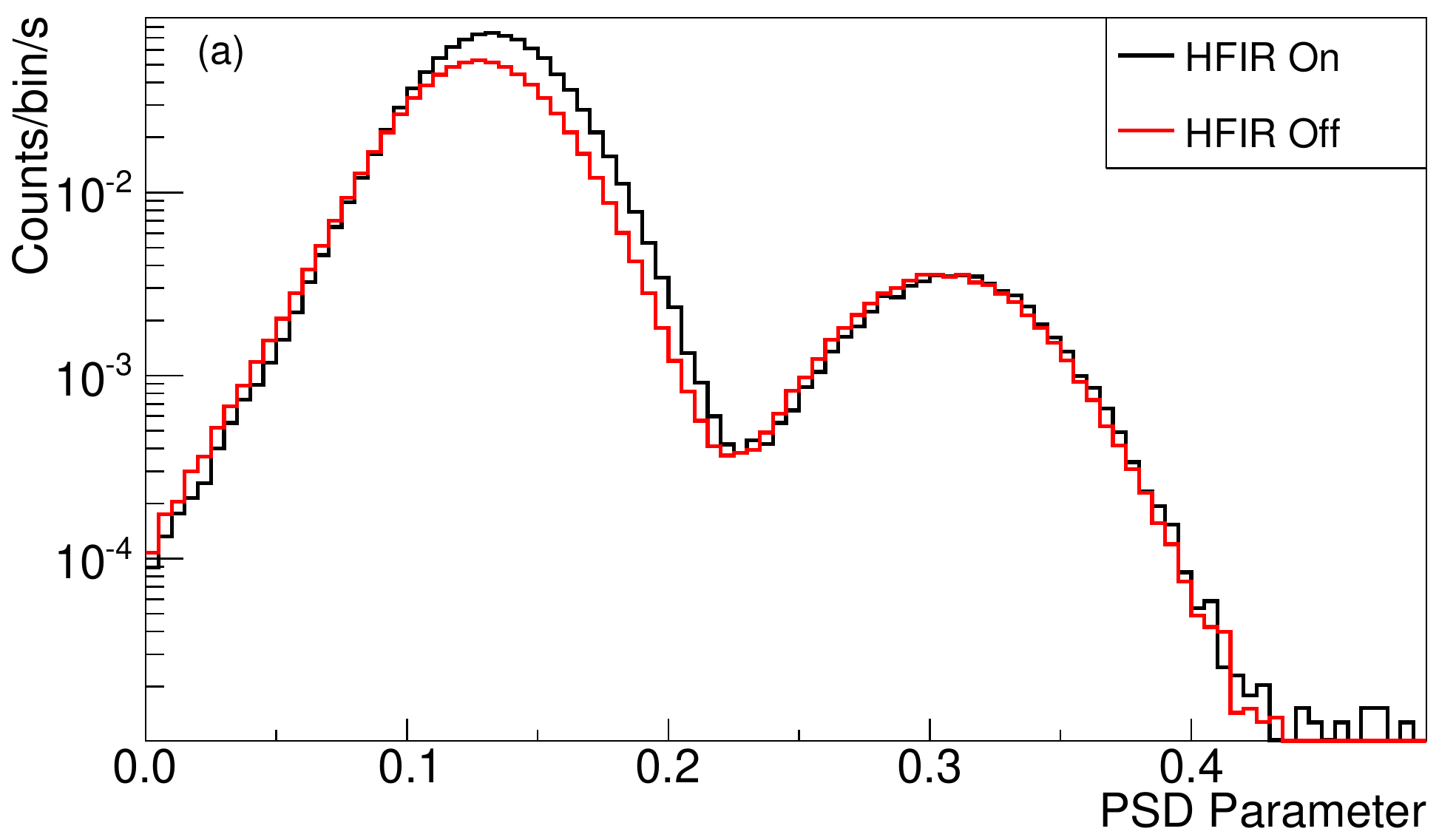}
\includegraphics*[width=0.45\textwidth]{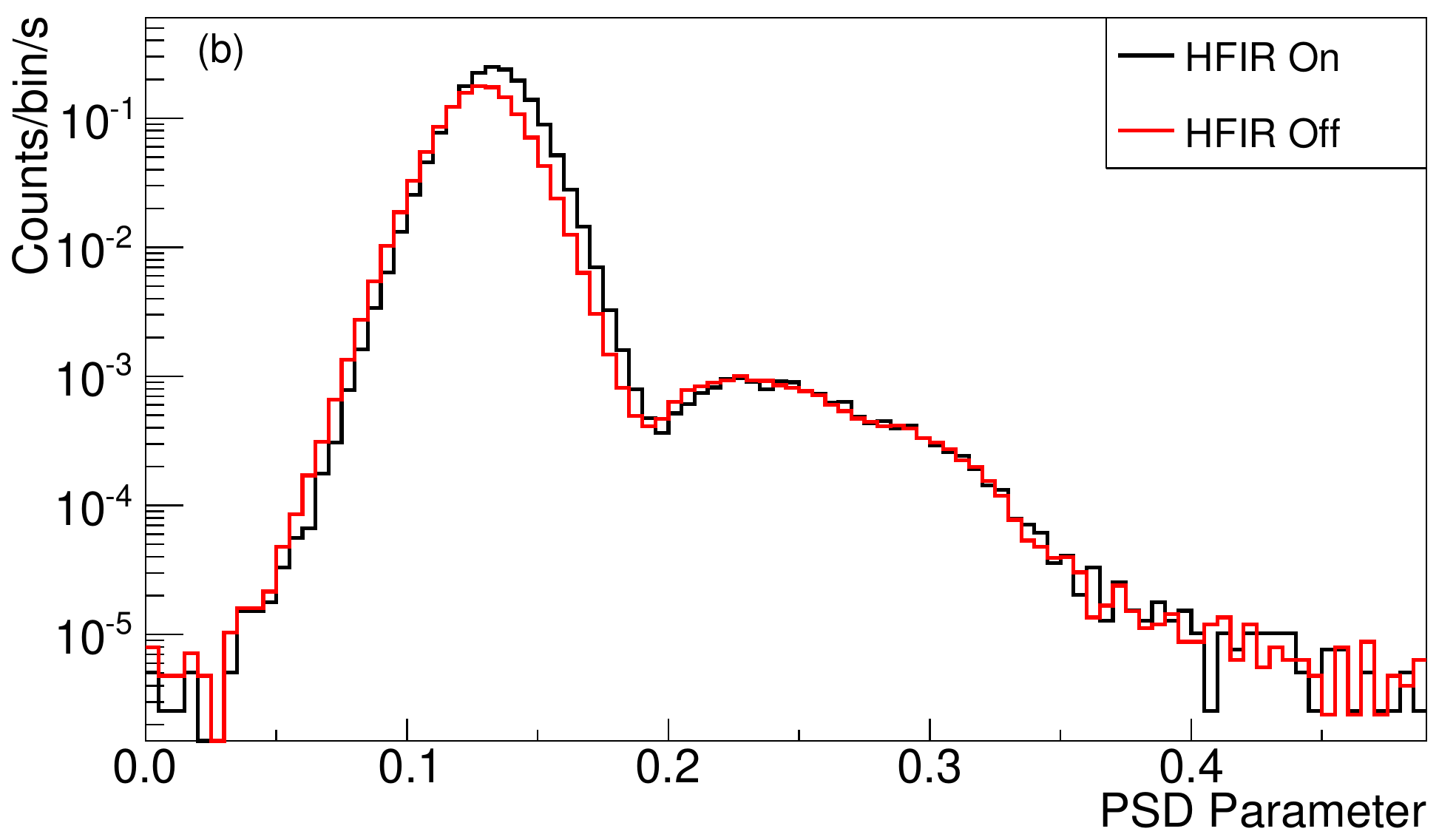}
\caption{PSD parameter distributions are compared for reactor-on and reactor-off data for two energy ranges in the \ptwo{}  detector: (a) $0.5$--$0.8$~MeV, corresponding to neutron capture on $^6$Li, and (b) $1$--$5$~MeV. The similarity of the distributions at high PSD parameter values indicates the detection no reactor-correlated thermal or fast neutrons. } 
\label{fig:p2_psd_proj}
\end{figure}

The electron-equivalent energy spectrum of all depositions in the \ptwo{} detector is shown in Fig.~\ref{fig:p2_energy_spectra} for reactor-on and reactor-off data. Clearly, there is an increase in the detector interaction rate that can be attributed to reactor generated particles. However, a considerable reduction in background is achieved compared to what would be expected with no shielding. Using the reactor-correlated $\g$-ray fluxes given in Sec.~\ref{sec:gammaResults}, the unshielded \ptwo{} detector would be estimated to have reactor related excess count rates of $\approx4\times10^3$~s$^{-1}$ and $\approx3\times10^3$~s$^{-1}$ in the $1$--$3$~MeV and $3$--$10$~MeV energy ranges, respectively. Instead rates of $1.38$~s$^{-1}$ and $1.44$~s$^{-1}$ are observed in these respective ranges using the shielded detector.

The increased background occurs at energies $\lesssim7$~MeV, with the greatest enhancement occurring at low energies. We can use the PSD ability of the detector to infer the relative contribution of this increase background from different particle types. PSD spectra are displayed in Fig.~\ref{fig:p2_psd_proj} for two energy ranges: $0.5$--$0.8$~MeV, corresponding to the $^6$Li neutron capture feature, and $1.0$--$5.0$~MeV, corresponding to the region where positrons from the inverse beta-decay \nuebar{} interaction would be observed. In these projections, the features at lower values of the PSD parameter correspond to electromagnetic interactions (predominately Compton scattering of $\g$-rays), while those at higher values correspond to thermal neutron capture (Fig.~\ref{fig:p2_psd_proj}a) or fast neutron recoils (Fig.~\ref{fig:p2_psd_proj}b).

It is apparent from these figures and the integrals of the two PSD regions given in Table~\ref{tab:p2PSD} that the increased background is due to $\g$-ray interactions in the \ptwo{} detector. That is, the shielding surrounding the detector effectively eliminates any reactor generated thermal or fast neutrons. The reactor-correlated background observed can be attributed to $\g$-rays produced outside and transported through the shield, or to $\g$-rays produced by neutron capture interactions within the shield. Since care was taken not to include materials that produce high-energy $\g$-rays within the shield, the high-energy excess observed in the \ptwo{} detector is attributed to external production and transport. 
While the $\gamma$-ray interaction rate changes with the reactor status, the rate of fast neutrons and neutron captures in the \ptwo{} detector are unchanged. This is an indication that any reactor-correlated neutron flux is highly suppressed by the shielding package. 

\begin{table}[tb]{%
\begin{tabular}{l|d{1.1}|d{1.1}l} 
\hline
Parameter space region			& \multicolumn{2}{c}{Background rate ($\times10^3$s$^{-1}$)} \\ 
							& \multicolumn{1}{c|}{Reactor-on}	&\multicolumn{1}{c}{Reactor-off} \\ \hline
$0.5$--$0.8$~MeV, $\gamma$-like	&960.1 \pm 1.5 & 700.6 \pm 0.7 \\
$0.5$--$0.8$~MeV, n-like		&58.0 \pm 0.4		&58.6 \pm 0.2 \\
$1.0$--$5.0$~MeV, $\gamma$-like		&1719.5\pm2.1&1261.2 \pm 1.0 \\
$1.0$--$5.0$~MeV, n-like			&15.2\pm0.2&15.5 \pm 0.1 \\
\hline
 \end{tabular}}
    \caption{Integrated rates for $\gamma$-like and neutron-like events in the \ptwo{} detector for reactor-on and reactor-off conditions. The $1$--$5$~MeV energy range approximately corresponds to inverse beta decay positrons, while the $0.5$--$0.8$~MeV energy range corresponds to neutron capture on $^6$Li . Quoted uncertainties are statistical only.}
 \label{tab:p2PSD}
\end{table}

\section{Conclusion}
\label{sec:conc}

The background characteristics of three research reactor facilities have been measured. Both significant similarities as well as important differences between the sites were encountered, and thus it is expected that these measurements will inform work at research reactor sites generally. Features common to all sites, include:
\begin{itemize}
\item{significant spatial variations in $\g$-ray and neutron backgrounds due to irregular shielding, localized leakage paths through shielding, or the presence of piping carrying activated materials. Detailed site-specific characterization of background is therefore essential to optimize a shielding design. In some cases, localized shielding applied to compact background sources could be a  cost and weight efficient approach to reducing detector backgrounds;}
\item{higher reactor-correlated background rates are encountered at potential near detector locations, when compared to far detector locations. This is not surprising, considering the near locations are closer to the reactors and therefore have less shielding from that intense source, and/or are more likely to be proximate to plant systems or other experiments that can transport radiation from the reactor to a detector location. A far detector may therefore require less shielding than a near detector;}
\item{neutron leakage and/or scattering is a significant background source, via neutron interactions on water, steel, or other structural materials.  The resulting high-energy  $\g$-rays are relatively difficult to shield. Application of relatively light neutron absorbing shielding to localized neutron sources could therefore be a cost and weight efficient approach to reducing $\g$-ray backgrounds.}
\end{itemize}

Features particular to \ATR{} include:
\begin{itemize}
\item{the lowest near site $\g$-ray background, due to relatively low thermal neutron leakage and good shielding from the reactor, and few nearby plant systems. This is offset by the highest cosmogenic background flux (muon and fast neutron), due to the high site elevation;}
\item{the lowest cosmogenic background flux (muon and fast neutron) of any location at the potential far detector location. This is due to the location being $\approx12$~m below grade in a basement. The far location $\g$-ray background is the highest of any far site, but still significantly lower than the near locations;}
\item{no expected or observed time variation of reactor-correlated $\g$-ray or neutron backgrounds.}
 \end{itemize}

Features particular to \HFIR{} include:
\begin{itemize}
\item{a large down-scattered $\g$-ray background coming from the entire length of the wall closet to the reactor at the near location. This implies that an intense radiation source (likely the reactor pool) is being only partially shielded. However, the flux falls rapidly as the distance to this wall increases, suggesting that localized shielding applied along the length of the wall may be able to attenuate this flux in a cost and weight effective manner. }
 \end{itemize}
 
Features particular to \NBSR{} include:
\begin{itemize}
\item{both large spatial and temporal variations of $\g$-ray and thermal neutron backgrounds at the near location. This is due to both the facility design and the operation of nearby experiments. Localized shielding may therefore be able to attenuate these sources in a cost and weight effective manner. The $\g$-ray background encountered at \NBSR{} is similar to that at \HFIR{}.}
\end{itemize}

While the background surveys reported here should be useful in the preliminary design of an experiment, given the considerable variation in background sources and intensity observed, a primary conclusion of this paper is that any sensitive experiment intending to operate in such facilities must perform detailed assessment of  the background in the particular location of interest. Such detailed measurements conducted by the PROSPECT collaboration at \HFIR{} have illustrated the complex nature of the background fields in that facility as well as the ability to strongly suppress backgrounds with well placed shielding. Deployment of the \ptwo{} detector in a shielding enclosure verified this conclusion, and importantly indicated that  reactor-correlated neutron backgrounds can be essentially completely suppressed.

\section*{Acknowledgements}

We gratefully acknowledge the support and hospitality of the Advanced Test Reactor at the Idaho National Laboratory; the High Flux Isotope Reactor and the Physics Division at Oak Ridge National Laboratory; and the National Bureau of Standards Reactor at the National Institute of Standards and Technology. The Oak Ridge National Laboratory is managed by UT Battelle for the U.S. Department of Energy. 

This material is based upon work supported by the U.S. Department of Energy Office of Science. Additional support for this work is provided by Yale University. 

LLNL-JRNL-669986. Prepared by LLNL under Contract DE-AC52-07NA27344.

\bibliographystyle{elsarticle-num} 
\bibliography{ReactorBackgroundPaperArxivRev1}% Produces the bibliography via BibTeX.%

\begin{thebibliography}{10}
\bibitem{disclaimer}
Certain trade names and company products are mentioned in the text or
  identified in illustrations in order to adequately specify the experimental
  procedure and equipment used. In no case does such identification imply
  recommendation or endorsement by the National Institute of Standards and
  Technology, nor does it imply that the products are necessarily the best
  available for the purpose.
\end{thebibliography}

\end{document}